\documentclass[a4paper,11pt]{article}
\usepackage{jheppub} 
\pdfoutput=1
\usepackage{amsmath,amssymb,amsfonts}
\usepackage{graphicx}
\usepackage[normalem]{ulem}
\usepackage{slashed}
\usepackage{color}
\usepackage{braket}
\usepackage{tikz}
\usepackage{url}
\usepackage{subcaption}

\usepackage{hyperref}
\newcounter{qnumber}

\newcommand{\beq}{\begin{equation}}
\newcommand{\eeq}{\end{equation}}

\newcommand{\colmatf}[1]{\left(\begin{array}{cccc}#1\end{array}\right)}

\newcommand{\exps}[1]{\exp{\left\{#1\right\}}}

\usepackage{stackrel}

\makeatletter\g@addto@macro\bfseries{\boldmath}\makeatother
\DeclareMathSymbol{\shortminus}{\mathbin}{AMSa}{"39}
\def\figureautorefname~#1\null{fig.\,#1\null}
\def\equationautorefname~#1\null{eq.\,(#1)\null}

\makeatletter
\DeclareFontFamily{OMX}{MnSymbolE}{}
\DeclareSymbolFont{MnLargeSymbols}{OMX}{MnSymbolE}{m}{n}
\SetSymbolFont{MnLargeSymbols}{bold}{OMX}{MnSymbolE}{b}{n}
\DeclareFontShape{OMX}{MnSymbolE}{m}{n}{
    <-6>  MnSymbolE5
   <6-7>  MnSymbolE6
   <7-8>  MnSymbolE7
   <8-9>  MnSymbolE8
   <9-10> MnSymbolE9
  <10-12> MnSymbolE10
  <12->   MnSymbolE12
}{}
\DeclareFontShape{OMX}{MnSymbolE}{b}{n}{
    <-6>  MnSymbolE-Bold5
   <6-7>  MnSymbolE-Bold6
   <7-8>  MnSymbolE-Bold7
   <8-9>  MnSymbolE-Bold8
   <9-10> MnSymbolE-Bold9
  <10-12> MnSymbolE-Bold10
  <12->   MnSymbolE-Bold12
}{}

\let\llangle\@undefined
\let\rrangle\@undefined
\DeclareMathDelimiter{\llangle}{\mathopen}%
                     {MnLargeSymbols}{'164}{MnLargeSymbols}{'164}
\DeclareMathDelimiter{\rrangle}{\mathclose}%
                     {MnLargeSymbols}{'171}{MnLargeSymbols}{'171}
\makeatother

\begin{document}

\title{The Quantum Spectral Method: From Atomic Orbitals to Classical Self-Force}
\author[a]{Majed Khalaf,}
\emailAdd{khalafmajed@gmail.com}
\affiliation[a]{Racah Institute of Physics, Hebrew University of Jerusalem, Jerusalem 91904, Israel}

\author[a]{and Ofri Telem}
\emailAdd{t10ofrit@gmail.com}

\abstract{
Can classical systems be described analytically at all orders in their interaction strength? For periodic and approximately periodic systems, the answer is yes, as we show in this work. Our analytical approach, which we call the \textit{Quantum Spectral Method}, is based on a novel application of Bohr's correspondence principle, obtaining non-perturbative classical dynamics as the classical limit of \textit{quantum matrix elements}. A major application of our method is the calculation of self-force as the classical limit of atomic radiative transitions. We demonstrate this by calculating an adiabatic electromagnetic inspiral, along with its associated radiation, at all orders in the multipole expansion. Finally, we propose a future application of the Quantum Spectral Method to compute scalar and gravitational self-force in Schwarzschild, analytically.}

\maketitle
\section{Introduction}

The advent of gravitational wave signals \cite{Abbott2016} from black hole inspirals sparked a remarkable worldwide theoretical effort to reliably and efficiently compute inspiral trajectories and their associated waveforms. The latter require some of the most accurate numerical and analytical computations ever done in general relativity.
Current state-of-the-art approaches, such as the effective-one-body approach \cite{Albanesi2023,Meent2023,Albertini2022,Albertini2022a,Nagar2022,Albanesi:2021rby,Damgaard:2021rnk,Bini2016,Damour:2016abl,Damour2016a,Damour:2015isa,Damour:2009sm,Damour:2009wj,Buonanno:1998gg},  synthesize data from numerical relativity \cite{Pretorius2005,Boyle2019}, Post-Newtonian (PN) \cite{Blanchet:2013haa,Foffa:2012rn,Foffa:2019hrb}, Post-Minkowskian (PM) \cite{Bjerrum-Bohr:2014zsa,Bjerrum-Bohr:2014zsa,Arkani-Hamed:2017jhn,Guevara:2017csg,Cachazo:2017jef,Bjerrum-Bohr:2018xdl,Cheung:2018wkq,Arkani-Hamed:2019ymq,Cristofoli:2019neg,Chung:2019duq,Bern:2019crd,Bern:2019nnu,Bjerrum-Bohr:2019kec,Cheung:2020gyp,Cheung2020,Bern:2020buy,Bern:2020gjj,Bjerrum-Bohr:2021din,Bjerrum-Bohr:2021vuf,Herrmann2021,Herrmann2021a,Chen2022,Bern:2021dqo,Bern2022a,Bern2022,Bern2023a,Bern2023,Cristofoli:2019neg,Bini2018a,Damour:2019lcq,Damour:2020tta,Bini2022,Damour2023,Dlapa2024,Driesse2024}, and self-force \cite{Mino:1996nk,Quinn:1996am,Rosenthal2006,Gralla:2008fg,Pound2010a,Detweiler2012,Pound2012,Gralla2012,Miller2016,Barack:2018yvs,Pound2020,Pound2021,Upton2021,Warburton2021,Miller2021,Albertini2022,Albertini2022a,Spiers2023,Spiers2023a,Meent2023,Wardell2023,Bini:2024icd,Long:2024ltn} computations, and synergize them into a comprehensive description of the black hole dynamics from the early moments of the inspiral, through the plunge and into the ringdown phase. Among these complementary approaches, PM perturbation theory has seen a remarkable renaissance in recent years due to its inherent ability to leverage high-loop-order \textit{quantum} scattering results to describe \textit{classical} PM and PN inspirals \cite{Kosower:2018adc,Kalin:2020fhe,Mogull:2020sak,Kalin:2020lmz,Kalin:2020mvi,Dlapa:2021npj,Jakobsen:2021smu,Dlapa2022,Jakobsen2023,Dlapa2023a,Kaelin2023,Dlapa2024,Driesse2024,Kalin:2019rwq,Kalin:2019inp,Cho2022a,Gonzo2023,Goldberger:2004jt,Foffa:2011ub,Porto:2016pyg,Levi:2018nxp,Foffa2017,Foffa:2019yfl,Foffa2019,Foffa2021,Goldberger2022,Bjerrum-Bohr:2014zsa,Bjerrum-Bohr:2014zsa,Arkani-Hamed:2017jhn,Guevara:2017csg,Cachazo:2017jef,Bjerrum-Bohr:2018xdl,Cheung:2018wkq,Arkani-Hamed:2019ymq,Cristofoli:2019neg,Chung:2019duq,Bern:2019crd,Bern:2019nnu,Bjerrum-Bohr:2019kec,Cheung:2020gyp,Cheung2020,Bern:2020buy,Bern:2020gjj,Bjerrum-Bohr:2021din,Herrmann2021,Herrmann2021a,Chen2022,Bern:2021dqo,Bern2022a,Bern2022,Bern2023a,Bern2023,Albanesi2023,Meent2023,Albertini2022,Albertini2022a,Nagar2022,Albanesi:2021rby,Damgaard:2021rnk,Bini2016,Damour:2016abl,Damour2016a,Damour:2015isa,Damour:2009sm,Damour:2009wj,Buonanno:1998gg}. This is done either by setting up a PM 2-body effective field theory (EFT) or by directly mapping scattering data to inspiral data, leading to highly non-trivial results up to 5PM order.

A common feature of most of the quantum-to-classical methods mentioned above is that they are all inherently perturbative in the gravitational constant $G$, i.e. their quantum-to-classical features are inherently bundled with their PM nature. This is not surprising, as the main workhorse to compute quantum scattering amplitudes are Feynman diagrams and their on-shell equivalents. This has lead some recent authors \cite{Adamo2022,Kosmopoulos2023,Cheung2023,Adamo2023} to stress the importance of (at least a partial) resummation of PM perturbation theory. The possibility of such a resummation is particularly appealing in the context of the self-force/post-adiabatic (PA) expansion in the mass ratio $m_1/m_2$ of the black holes. The latter is the main framework for computing gravitational wave signals from extreme-mass-ratio inspirals, relevant for the LISA future space-based gravitational wave detector \cite{Babak2017,Barack:2018yly,Barausse2020}. 

In this paper, we propose the first-ever quantum-to-classical method, dubbed the \textit{Quantum Spectral Method} (QSM), which is inherently \textit{non-perturbative} in the coupling $G$ (yet perturbative in the mass ratio $m_1/m_2$). As such, it is particularly suited for self-force calculations. Rather than resumming perturbative amplitudes, our method starts right away with \textit{quantum bound states} \'a la the quantum hydrogen-like atom. As such it is already a non-perturbative resummation of scattering amplitudes, in the sense of the famous Bethe-Salpeter formalism \cite{Salpeter1951}. Rather than performing any resummation of amplitudes, we compute quantum matrix elements among bound wavefunctions - the same way one would compute spontaneous emission in the hydrogen atom. Taking the classical limit $\hbar\rightarrow 0$, we are able to reproduce classical dynamics within the self-force/PA expansion, to all orders in the coupling. For example, in the case of the hydrogen atom and its classical limit -- Keplerian motion, we find that the $\Delta n_{th}$ Fourier coefficient of any classical observable $\mathcal{O}$ is given by the $\hbar\rightarrow 0$ limit of the $\Delta n_{th}$ transition $\left\langle n-\Delta n|\mathcal{O}|n\right\rangle$. This simple yet deep relation could have been derived early in the 20th century, but we could not find previous accounts for it. It can be seen as a direct corollary of Ehrenfest's theorem when applied to periodic systems, or as a result of the classical saddle-point behavior of coherent states. As a first paper outlining the method, we \textit{prove} that our method works for motion in a Schwarzschild background, but only \textit{apply} it explicitly to derive a fully analytical expression of the first-order electromagnetic self-force in flat space.

Our method gives exact analytical expressions, even in cases where only numerical integration has been available before (see, e.g. \cite{Fujita2009,Hopper2015} for the case of gravitational emission from a quasi-periodic orbit). We explicitly demonstrate our method by (a) reproducing time-dependent Keplerian motion; (b) deriving a first all-multipole analytical expression for the electromagnetic (EM) radiation from a classical electron in Keplerian motion; and (c) using our all-multipole analytical expression to calculate the adiabatic EM inspiral of a classical electron, and its associated waveform. Remarkably, we are able to show analytically how the classical EM self-force emerges as the classical limit of quantum \textit{spontaneous emission}.

The first three applications of the QSM presented in this paper concern the motion of a bound classical electron in a Coulomb potential, with or without radiation-reaction (c.f. the study of the dissipative motion of a \textit{scattering} electron in \cite{Manohar2022}). Nevertheless, a major future application of the QSM is within the PA approach \cite{Hinderer2008,VanDeMeent2018} for \textit{gravitational} inspiral problems in the extreme-mass-ratio regime. In this regime, the computation of the gravitational self-force (GSF) tables at first \cite{Mino:1996nk,Quinn:1996am,Gralla:2008fg,Pound2010a} and second-order \cite{Rosenthal2006,Detweiler2012,Pound2012,Gralla2012,Miller2016,Pound2020,Upton2021,Warburton2021,Miller2021,Albertini2022,Albertini2022a,Spiers2023,Spiers2023a,Meent2023,Wardell2023} required for the accurate calculation of the inspiral \cite{Isoyama2013,Burko2013} is a major computational bottleneck \cite{Upton2021}. This is in contrast with the online step of waveform generation, which is largely a solved problem \cite{Meent2016,Meent2018,Hughes2021}. Using the QSM might allow to replace some or all of the numerical source integrals used in the calculation of self-force tables with analytical expressions (c.f. \cite{Bini:2024icd,Long:2024ltn} for the unbound case), potentially allowing for a speedup of the computation at first and even second order in the mass ratio. As a first step in this direction, we present in section~\ref{sec:Schwarz} a proof of the QSM for the motion on an osculating geodesic in an ambient Schwarzschild metric. Furthermore, we show that the full analytical calculation of scalar self-force in Schwarzschild is equivalent to one single integral over the product of three confluent Heun functions. We expect the same conclusion to hold for GSF as well. Computing the latter integral would allow for a fully analytical expression for 1st-order self-force (1SF) in a black hole background, and would enable a direct comparison with multi-loop PM results, e.g. \cite{Driesse2024}.

The paper is structured as follows. In Section~\ref{eq:Periodic} we set-up the classical Hamilton-Jacobi theory for a point-particle moving in a spherically symmetric potential. Section~\ref{sec:QtoC} is a first glance into the details of the quantum-to-classical correspondence, in which we prove the link between the classical fundamental frequencies and the differences between quantum eigenenergies in the classical limit. Section~\ref{sec:QSM} is a complete proof of the QSM for motion in spherically-symmetric potentials. The proof simply uses the WKB approximation, which becomes exact in the $\hbar\rightarrow 0$ limit. In Section~\ref{sec:Coul} we specialize to a Coulomb potential, in which case the problem degenerates and there is only one fundamental frequency. In Section~\ref{sec:Kep} we use the QSM to reproduce the time-dependent position of Keplerian motion. Section~\ref{sec:EM} is a step-by-step derivation of the retarded potential generated by a Keplerian electron, while Section~\ref{sec:ESF} links the classical 0PA energy and angular momentum loss rates to the classical limit of quantum spontaneous emission. Section~\ref{sec:Schwarz} is a generalization of the QSM to relativistic motion in a Schwarzschild background. Finally Section~\ref{sec:future} outlines the future application of the QSM to gravity, focusing first on a toy scalar-force model in Schwarzschild. To make the paper more readable, we relegate some of the more technical aspects of our computations to our detailed appendices. Appendix~\ref{app:quantumtoclassicalladscape} highlights the place of the QSM within the landscape of quantum-to-classical methods; Appendix~\ref{app:WKB} contains an identity relevant to the WKB proof of the QSM; In Appendix~\ref{app:momWKB}, we generalize the WKB proof of the QSM to momentum-dependent operators; Appendix~\ref{0PA:SE} supplements the derivations in section~\ref{sec:ESF}; in Appendix~\ref{app:quant} we present a compendium of quantum matrix elements for hydrogen-like atoms; the classical limits of these matrix elements are given in Appendix~\ref{app:clas}. Finally, Appendix~\ref{app:aux} contains auxiliary computations.

\section{Classical Multiply Periodic Motion and Action-Angle Variables}\label{eq:Periodic}
Consider a classical particle of mass $\mu$ undergoing non-relativistic conservative motion in flat space. The particle moves in a time-independent external potential. The position of the particle is denoted by $\vec{r}$ and its momentum by $\vec{p}$, while the Hamiltonian is denoted $H(\vec{r},\vec{p})$. By Noether's theorem, the time translation symmetry of the Hamiltonian tells us that the energy of the particle, $E$, is conserved. If, in addition, $H(\vec{r},\vec{p})$ is rotationally invariant, the system also conserves angular momentum $\vec{L}$. In particular, this means that the motion derived from $H(\vec{r},\vec{p})$ is integrable, i.e. there are as many conserved quantities as coordinates. Here and below we assume the simple spherically symmetric form for $H$:
\begin{eqnarray}\label{eq:Ham}
H(\vec{r},\vec{p})=\frac{\vec{p}^2}{2\mu}+V(r)\,,
\end{eqnarray}
where $V(r)$ is a spherically symmetric potential. Though our method is useful for any integrable motion, not necessarily spherically symmetric, we specialize to this case for simplicity. For spherically symmetric motion, one conventionally defines the \textit{Hamilton-Jacobi} action,
\begin{eqnarray}\label{eq:sHJ}
S_{HJ}(r,\theta,\varphi)=S^r(r)+S^\theta(\theta)+L_z\varphi
\end{eqnarray}
where the radial action $S^r(r)$ and the angular action $S^\theta(\theta)$ are given by
\begin{eqnarray}\label{eq:radialangular}
S^r(r)=\pm\int_{r_{min}}^r\,dr'\,\sqrt{U^r(r')}~~~~~~~~~~~~&,&~~~~S^\theta(\theta)=\int_{\theta_{min}}^\theta\,d\theta'\,\sqrt{U^\theta(\theta')}\nonumber\\[5pt]
U^r(r')=2\mu\left(E-V(r')-\frac{L^2}{2\mu r'^2}\right)~~~&,&~~~U^\theta(\theta')=L^2-\frac{L^2_z}{\sin^2\theta'}\,.
\end{eqnarray}
We select the $+$ sign for $S^r(r)$ when we use it to compute variables over the first half of each radial cycle (when $r$ grows from $r_{\rm min}$ to $r_{\rm max}$). The $-$ sign, on the other hand, is used when computing variables over the second half of the radial cycle (when $r$ decreases from $r_{\rm max}$ to $r_{\rm min}$).
Since we are considering motion in a spherically symmetric potential, the motion is planar, and without loss of generality we can focus on motion in the XY plane where $\theta=\pi/2$ and $L_z=L\equiv|\vec{L}|$. The equations of motion of $r$ and $\varphi$ then read:
\begin{eqnarray}\label{eq:EOMrvarphi}
\dot{r}=\frac{p_r}{\mu}=\frac{1}{\mu}\frac{\partial S_{HJ}(r,\theta\varphi)}{\partial r}=\pm\frac{\sqrt{U^r(r)}}{\mu}~~~,~~~\dot{\varphi}=\frac{p_\varphi}{\mu r^2}=\frac{1}{\mu r^2}\frac{\partial S_{HJ}(r,\theta\varphi)}{\partial \varphi}=\frac{L}{\mu r^2}\,.
\end{eqnarray}
We also focus on the case where $E<0$, so that the motion is bound and (multiply-) periodic. By multiply periodic, we mean that there could be different time periods for the radial and angular motions, and so the motion is not truly periodic unless the radial and angular periods are commensurate \cite{Goldstein2011}. For bound motion derived from the spherically symmetric \eqref{eq:Ham}, we can choose to express all conserved quantities as functions of the constant \textit{action variables}\footnote{Our definition differs by a factor of $2\pi$ from \cite{goldstein2002classical}.},
\begin{eqnarray}
J_i\equiv\frac{1}{2\pi}\oint\,dq_i\,p_i~~~,~~~i\in\{r,\varphi\}\,.
\end{eqnarray}
Here we ignore the $\theta$ action variable since we assume that the motion is in the XY plane, without loss of generality.
As a matter of fact, explicit computation gives \cite{goldstein2002classical}
\begin{eqnarray}\label{eq:aa1}
J_r=\frac{1}{\pi}\,S^r(r_{max})~~,~~J_{\varphi}= L\,.
\end{eqnarray}
Over each radial period, the body undergoes a \textit{libration} between $r_{min}$ and $r_{max}$, two consecutive real roots of $r'^2U^r(r')$ separated by a minimum. 
Since $H=E$ (we will use the Hamiltonian $H$ and the energy $E$ interchangeably) can be expressed in terms of the $J_i$, it is only natural to treat the $J_i$ as new conjugate momenta, whose corresponding coordinates $\alpha^i$ are cyclic -- i.e. absent from the Hamiltonian. Hamilton's equations then give 
\begin{eqnarray}\label{eq:aldef}
\alpha^i\equiv\Upsilon^{i}_{J_r,L}\,t~~~,~~~\Upsilon^i_{J_r,L}\equiv\frac{\partial H(J_r,J_\varphi)}{\partial J_i}\,,~~~i\in\{r,\varphi\}.
\end{eqnarray}
Here the constant $\Upsilon^i_{J_r,L}$ are the \textit{fundamental frequencies} of the system, which are in-and-of themselves functions of $J_r$ and $L$. Explicitly,
\begin{eqnarray}\label{eq:Upcalc}
\frac{2\pi}{\Upsilon^{r}_{J_r,L}}=T^r_{J_r,L}=2\int_{r_{min}}^{r_{max}}dr\,\frac{dt}{dr}=2\int_{r_{min}}^{r_{max}}dr\,\frac{\mu}{\sqrt{U^r(r)}}=2\pi\frac{\partial J_r}{\partial E_{J_r,L}}\,,
\end{eqnarray}
where $dt/dr$ is computed from the EOM \eqref{eq:EOMrvarphi}. Similarly,
\begin{eqnarray}\label{eq:Upcalc2}
\Upsilon^{\varphi}_{J_r,L}&=&\Upsilon^{r}_{J_r,L}\frac{\varphi(r_{\rm max})-\varphi(r_{\rm min})}{\pi}\,\nonumber\\[5pt]
\varphi(r_{\rm max})-\varphi(r_{\rm min})&=&\int_{r_{min}}^{r_{max}}dr\,(\dot{\varphi}/\dot{r})=\int_{r_{min}}^{r_{max}}dr\,\frac{L}{\mu r^2}\frac{\mu}{\sqrt{U^r(r)}}\,.
\end{eqnarray}
For future reference, we can also express the $r$ equation of motion (EOM) \eqref{eq:EOMrvarphi} in terms of $\alpha^r$,
\begin{eqnarray}\label{eq:EOMral}
\frac{dr}{d\alpha^r}=\frac{1}{\Upsilon^{r}_{J_r,L}}\dot{r}=\pm\frac{T^r_{J_r,L}}{2\pi}\frac{\sqrt{U^r(r)}}{\mu}\,,
\end{eqnarray}
where again the $+\,(-)$ sign is for $0\leq\alpha^r<\pi\,(\pi\leq\alpha^r<2\pi)$.
\subsection{Doubly Periodic Motion}
From the theory of action-angle variables, the $r(t)$ which solves the EOM \eqref{eq:EOMrvarphi} is periodic in $\alpha_r$,
\begin{eqnarray}\label{eq:rper}
r(t)=r[\alpha^r(t)]=\sum_{\Delta j_r}\,r_{\Delta j_r}\,\exps{-i\Delta j_r\alpha^r(t)}\,,
\end{eqnarray}
where the coefficients are related to $r[\alpha^r]$ via
\begin{eqnarray}\label{eq:rper2}
r_{\Delta j_r}=\int_{0}^{2\pi}\frac{d\alpha^r}{2\pi}\,r[\alpha^r]\,\exps{i\Delta j_r\alpha^r}\,.
\end{eqnarray}
Of course, to explicitly determine the coefficients one usually has to explicitly solve the classical EOM \eqref{eq:EOMrvarphi}. In the bulk of this paper we present an alternative route for this, obtaining $r_{\Delta j_r}$ as the classical limit of quantum matrix elements.

To describe the solution for $\varphi(t)$ which solves \eqref{eq:EOMrvarphi}, it is customary to express it as
\begin{eqnarray}\label{eq:phidel}
\varphi(t)\equiv\chi(t)+\alpha^\varphi(t)\,.
\end{eqnarray}
The variable $\chi(t)=\chi[\alpha^r(t),\alpha^\varphi
(t)]$ is \textit{doubly periodic} in $\alpha^r$ and $\alpha^\varphi$, 
\begin{eqnarray}\label{eq:rper3}
\chi(t)=\chi[\alpha^r(t),\alpha^
\varphi(t)]=\sum_{\Delta j_r}\,\sum_{\Delta l}\,\chi_{\Delta j_r,\Delta l}\,\exps{-i\Delta j_r\alpha^r(t)-i\Delta l\alpha^\varphi(t)}\,,
\end{eqnarray}
where the coefficients are related to $\chi[\alpha^r,\alpha^\varphi]$ via
\begin{eqnarray}\label{eq:dvarphiper}
\chi_{\Delta j_r,\Delta l}=\int_{0}^{2\pi}\frac{d\alpha^r}{2\pi}\,\int_{0}^{2\pi}\frac{d\alpha^\varphi}{2\pi}\,\chi[\alpha^r,\alpha^\varphi]\,\exps{i\Delta j_r\alpha^r+i\Delta l\alpha^\varphi}\,.
\end{eqnarray}
Here $\Delta j_r,\,\Delta l$ are integers which label the Fourier harmonies. The reason that we chose the labels $\Delta j_r,\,\Delta l$ for our Fourier harmonies will be made clear in the next Section~\ref{sec:QtoC}. 

Similarly, we can consider any other physical observable of the system $\mathcal{O}_{J_r,L}(t)=\mathcal{O}_{J_r,L}[r(t),\chi(t)]$. From the fact that $r(t)$ is periodic in $\alpha^r(t)$ and $\chi(t)$ is doubly-periodic in $\alpha^r(t)$ and $\alpha^\varphi(t)$, we see that $\mathcal{O}_{J_r,L}(t)$ has to be doubly periodic as well. We put the label $J_r,L$ on $\mathcal{O}_{J_r,L}(t)$ to emphasize that the value of any observable of the system depends on the numerical values of the variables $J_r$, $L$, or equivalently $E$ and $L$.
Explicitly, the doubly periodic representation of $\mathcal{O}_{J_r,L}(t)$ is
\begin{equation} \label{app:genQSM2}
\mathcal{O}_{J_r,L}(t)=\mathcal{O}_{J_r,L}\left[\alpha^r(t),\alpha^\varphi(t)\right]=\sum_{\Delta j_r,\,\Delta l}\,\mathcal{O}^{J_r,L}_{\Delta j_r,\Delta l}\exp\left\{-i\Delta j_r\alpha^r(t)-i\Delta l\alpha^\varphi(t)\right\}\,.
\end{equation}
Here the Fourier coefficients $\mathcal{O}^{J_r,L}_{\Delta j_r,\Delta l}$ are related as usual to $\mathcal{O}_{J_r,L}\left[\alpha^r,\alpha^\varphi\right]$ as
\begin{equation} \label{eq:genQSM2tt}
\mathcal{O}^{J_r,L}_{\Delta j_r,\Delta l}=\int\,\frac{d\alpha^r}{2\pi}\,\int\,\frac{d\alpha^\varphi}{2\pi}\,\mathcal{O}_{J_r,L}\left[\alpha^r,\alpha^\varphi\right]\,\exps{i\Delta j_r\alpha^r+i\Delta l\alpha^\varphi}\,.
\end{equation}
Just like the case of $r(t)$ and $\chi(t)$, finding the Fourier coefficients of $\mathcal{O}_{J_r,L}(t)$ usually requires a full solution of the classical EOM of the system. Such a solution, and the subsequent double-inverse Fourier transform \eqref{eq:genQSM2tt} are generally unknown analytically, and one has to resort to numerical integration. The main result of the paper is the explicit analytic solution for these Fourier coefficients as the classical limit of quantum matrix elements. This solution is possible whenever the quantum eigenstates of the system are known (exactly or perturbatively), as is the case for the hydrogen atom, and even for motion in the background of a Schwarzschild black hole.

\section{Quantum-to-Classical Relations, a First Glance}\label{sec:QtoC}

As a first glance of the compelling relations between classical motion and its quantum counterpart, consider the Hamiltonian \eqref{eq:Ham}, now taken as a \textit{quantum Hamiltonian}. Importantly, the system has the same symmetries both classically and quantum mechanically\footnote{In other words, time translation and rotational invariance are not anomalous.}, and so for every classical action variable there is a corresponding operator commuting with the Hamiltonian. The eigenvalues of these operators are the quantum numbers of the system. For a spherically symmetric Hamiltonian, one conventionally works in an eigenbasis $\left|j_r,l,m\right\rangle$ labeled by the radial quantum number $j_r$, the angular momentum quantum number $l$, and the ``magnetic'' quantum number $m$, so that
\begin{eqnarray} \label{eq:eig}
H\left|j_r,l,m\right\rangle&=&E_{j_r,l}\left|j_r,l,m\right\rangle\nonumber\\[5pt]
L^2\left|j_r,l,m\right\rangle&=&\hbar^2l(l+1)\left|j_r,l,m\right\rangle\nonumber\\[5pt]
L_z\left|j_r,l,m\right\rangle&=&\hbar m\left|j_r,l,m\right\rangle\,.
\end{eqnarray}
For the special case of the hydrogen atom, the spectrum is degenerate and we conventionally label energies by the \textit{principal} quantum number $n=j_r+l+1$. The quantum numbers $j_r,\,l,\,m$ correspond to the classical conserved quantities $J_r$, $L\equiv|\vec{L}|$ and $M\equiv L_z$. To make this correspondence more explicit, we write
\begin{equation} \label{eq:corrs}
j_r\rightarrow \hbar^{-1}J_r~~,~~l\rightarrow \hbar^{-1}L~~,~~m\rightarrow \hbar^{-1}M\equiv\hbar^{-1}L_z\,.
\end{equation}
The meaning is simple; whenever we have a quantum expression and we want to take its classical limit, we should first 
replace $j_r,\,l,\,m$ in that expression by their classical counterparts in \eqref{eq:corrs}, and then take the $\hbar\rightarrow 0$ limit. In practice, we will always specialize to motion in the XY plane, for which $L_z=L$. Below, when we take the classical limit, we would often use the notation $v_{J_r,L}$ to denote the classical limit of a variable $v_{j_r,l}$; this means taking
$v_{j_r,l}$ and making the substitution $(j_r,l,m)=\hbar^{-1}(J_r,L,M)$. We will also use the subscript $J_r,\,L$ to emphasize the dependence of general classical quantities on $J_r$ and $L$. For example, we write $\Upsilon^i$ as $\Upsilon^i_{J_r,L}$ to emphasize their $J_r$ and $L$ dependence. For the hydrogen atom, which exhibits a degeneracy, we will replace the radial quantum number $j_r$ by the principal quantum number $n$, and relate $n$ to the classical $N$ by $n=\hbar^{-1}N$.

As a first non-trivial application of the correspondence principle, let us derive the classical fundamental frequency from its quantum eigenstates. The derivation here is valid for generic radial potentials, not just the simple Coulomb one. We now prove the following quantum-to-classical identities
\begin{eqnarray}\label{eq:masterfreq}
\Upsilon^r_{J_r,L}&=&\lim_{\hbar\rightarrow 0}\frac{E_{j_r,l}-E_{j_r-\Delta j_r,l}}{\hbar\Delta j_r}\nonumber\\[5pt]
\Upsilon^\varphi_{J_r,L}&=&\lim_{\hbar\rightarrow 0}\frac{E_{j_r,l}-E_{j_r,l-\Delta l}}{\hbar\Delta l}\,.
\end{eqnarray}
Here we wrote $\Upsilon^i_{J_r,L}$ to remind the reader that they depend on the classical action variables of the system. The expression on the RH side also depends on these classical quantities, since, before taking the $\hbar\rightarrow 0$ limit, one substitutes $(j_r,l,m)=\hbar^{-1}(J_r,L,M)$, in accordance with the general prescription \eqref{eq:corrs} linking quantum numbers and classical conserved quantities. According to the prescription \eqref{eq:corrs} used in \eqref{eq:masterfreq}, as $\hbar\rightarrow 0$,  $j_r,l\rightarrow \infty$, exactly as we expect from a classical limit. Their products $J_r=\hbar j_r,\,
L=\hbar l
,\,M=\hbar m$ are finite, dimensionful quantities characterizing the state of the classical system. All the while, $\Delta j_r,\,\Delta l$ and $\Delta m$ remain finite integers. The relation \eqref{eq:masterfreq} may seem surprising, but it is nothing but the semiclassical quantization condition for energies in the WKB approximation \cite{Ghatak}, which becomes exact in the $\hbar\rightarrow 0$ limit. To see this, consider the leading WKB expression for a radial wavefunction $R_{j_r,l}(r)$ (in any spherically symmetric potential),
\begin{eqnarray}\label{eq:hwkbf}
R^{{\rm{WKB}}}_{j_r,l}(r)\propto\,\sin\left[\hbar^{-1}S^r_{j_r,l}(r)+\frac{\pi}{4}\right]\,,
\end{eqnarray}
where $S^r_{j_r,l}(r)$ in \eqref{eq:hwkbf} is the radial Hamilton-Jacobi function (with a $+$ sign up-front),
\begin{eqnarray}\label{eq:HJF}
S^r_{j_r,l}(r)&=&\int_{r_{min}}^r\,dr'\,\sqrt{U^r_{j_r,l}(r')}\nonumber\\[5pt]
U^r_{j_r,l}(r)&\equiv&2\mu\,\left(E_{j_r,l}-V(r)-\frac{\hbar^2 l(l+1)}{2\mu r^2}\right)\,.
\end{eqnarray}
The WKB quantization condition then comes from the requirement that 
\begin{eqnarray}\label{eq:quantcond}
J_r(E=E_{j_r,l})\equiv \frac{1}{\pi}S^r_{j_r,l}(r_{max})= \hbar\,\left(j_r+\frac{1}{2}\right)\,,
\end{eqnarray}
which comes from sowing the oscillating WKB wavefunction in the allowed zone $r\in[r_{min},r_{max}]$ to decaying WKB wavefunctions at the turning points. The quantization condition \eqref{eq:quantcond} is the key to proving \eqref{eq:masterfreq}. Taking the difference between \eqref{eq:quantcond} at $j_r$ and at $j_r-\Delta j_r$, we have
\begin{eqnarray}\label{eq:quantcond2}
\lim_{\hbar\rightarrow 0}\frac{J_r(E=E_{j_r,l})-J_r(E=E_{j_r-\Delta j_r,l})}{\hbar \Delta j_r}=1\,,
\end{eqnarray}
while on the other hand 
\begin{eqnarray}
\lim_{\hbar\rightarrow 0}\frac{J_r(E=E_{j_r,l})-J_r(E=E_{j_r-\Delta j_r,l})}{\hbar \Delta j_r}=\frac{\partial J_r(E_{J_r,L})}{\partial E_{J_r,L}}\,\lim_{\hbar\rightarrow 0}\frac{E_{j_r,l}-E_{j_r-\Delta j_r,l}}{\hbar \Delta j_r}\,.
\end{eqnarray}
 Identifying $1/\Upsilon^r_{J_r,L}=\partial J_r/\partial E_{J_r,L}$ according to \eqref{eq:Upcalc}, we have
\begin{eqnarray}\label{eq:quantcond3}
\lim_{\hbar\rightarrow 0}\frac{1}{ \Upsilon^r_{J_r,L}}\frac{E_{j_r,l}-E_{j_r-\Delta j_r,l}}{\hbar \Delta j_r}=1\,,
\end{eqnarray}
from which the first equation of \eqref{eq:masterfreq} follows. As we stressed before, we always substitute $(j_r,l,m)=\hbar^{-1}(J_r,L,M)$ before taking the $\hbar\rightarrow 0$ limit. One can also explicitly check \eqref{eq:quantcond3} for the special case of a Coulomb potential $V(r)=-K/r$, for which $E_{j_r,l}=E_n=-\frac{mK^2}{2\hbar^2n^2}$ are the discrete energy eigenvalues of the hydrogen atom, with $n=j_r+l+1$.
Similarly, from \eqref{eq:quantcond}, we have
\begin{eqnarray}\label{eq:quantcond4}
\lim_{\hbar\rightarrow 0}\,\frac{S^r_{j_r,l}(r_{max})-S^r_{j_r,l-\Delta l}(r_{max})}{\hbar{\Delta l}}\,=0\,,
\end{eqnarray}
so that
\begin{eqnarray}\label{eq:quantcond5}
\lim_{\hbar\rightarrow 0}\,\left\{\frac{\partial J_r}{\partial E_{J_r,L}}\frac{E_{j_r,l}-E_{j_r,l-\Delta l}}{\hbar \Delta l}-\int^{r_{max}}_{r_{min}}\,dr'\,\frac{L}{ r'^2 \sqrt{U^r_{J_r,L}(r')}}\right\}=0\,.
\end{eqnarray}
Simplifying this expression using \eqref{eq:Upcalc2}, we have
\begin{eqnarray}\label{eq:quantcond6}
\lim_{\hbar\rightarrow 0}\,\left\{\frac{1}{\Upsilon^r_{J_r,L}}\frac{E_{j_r,l}-E_{j_r,l-\Delta l}}{\hbar \Delta l}-\frac{\Upsilon^\varphi_{J_r,L}}{\Upsilon^r_{J_r,L}}\right\}=0\,,
\end{eqnarray}
from which is the second equation in \eqref{eq:masterfreq} follows. 

\section{The Quantum Spectral Method For Spherically Symmetric Potentials}\label{sec:QSM}

In this section we present the main idea of this paper: the calculation of the Fourier coefficients $\mathcal{O}^{J_r,L}_{\Delta j_r,\Delta l}$ for any classical observable $\mathcal{O}_{J_r,L}$ as the $\hbar\rightarrow 0$ limit of the \textit{quantum matrix elements} associated with $\mathcal{O}$, taken as a \textit{quantum operator} between eigenstates whose eigenvalues are associated with the classical $J_r,\,L$. We present our method for spherically symmetric potentials to reduce complexity, though it is equally applicable to integrable but not spherically symmetric potentials, like motion in the background of a Kerr black hole. This is because all we need in order to apply our method effectively is a maximal set of conserved quantities / operators.

Without further ado, let us present the ``master equation'' of our paper:
\begin{eqnarray}\label{eq:master}
\mathcal{O}^{J_r,L}_{\Delta j_r,\Delta l}\,=\,\lim_{\hbar\rightarrow 0}\,\sum_{\Delta m}\,\left\langle j_r-\Delta j_r,l-\Delta l,l-\Delta m\right|\,\mathcal{O}\,\left|j_r,l,l\right\rangle\,.
\end{eqnarray}
Here, $\mathcal{O}$ is regarded as an operator, and $(j_r,l)=\hbar^{-1}(J_r,L)$. Here we set the magnetic quantum number $m$ on the RHS to be $m=l$. This is because we are focusing, without loss of generality, on motion in the XY plane, for which $L_z=L$. As we can see, the $\Delta j_r$ difference between the two eigenstates that appear in \eqref{eq:master} coincides with the Fourier integer "harmony" of the radial motion, and similarly for $\Delta l$ and the azimuthal Fourier decomposition.
Due to spherical symmetry, the $\Delta m$ does not have an associated fundamental frequency, and it is summed over in \eqref{eq:master}.
The QSM provides an analytic map between quantum matrix elements and classical observables, which holds non-perturbatively in the interaction strength of the system. For the place of the QSM within the landscape of quantum-to-classical methods, see Appendix~\ref{app:quantumtoclassicalladscape}.

In this section we provide a technical yet straightforward proof of \eqref{eq:master}, again using the WKB method which becomes exact in the $\hbar\rightarrow 0$ limit. Finally, while our proof focuses on an energy conserving, non-radiating system, in Section~\ref{sec:ESF} we are able to apply the QSM to a quasi-periodic radiating system as part of PA perturbation theory, using the method of \textit{osculating orbits}.
The proof that we outline here for the QSM master equation \eqref{eq:master} holds for any spherically symmetric potential. We begin with the quantum wavefunction for the eigenstate $\left|j_r,l,m\right\rangle$, namely
\begin{eqnarray}\label{eq:PsiWKBp}
\Psi_{j_r,l,m}(r)=(-1)^l\,R_{j_r,l}(r)\,Y_{l}^{m}(\theta,\varphi)\,,
\end{eqnarray}
where $R_{j_r,l}(r)$ is the radial wavefunction and $Y_{l}^{m}(\theta,\varphi)$ is a spherical harmonic. As we are interested in the classical limit, we can take the $\mathcal{O}(\hbar^0)$ WKB expression for $R_{j_r,l}$ to prove the QSM. Our wavefunction \eqref{eq:PsiWKBp} has a $(-1)^l$ pre-factor with respect to the conventionally normalized wavefunction of, say, the hydrogen atom. This choice makes the master equation \eqref{eq:master} more aesthetic -- without this factor, the latter would have included a factor of $(-1)^{\Delta l}$ up-front.
\subsection{Radial WKB Wavefunction and Integration}\label{sec:Iint}
The radial wavefunction $R_{j_r,l}(r)$ is given, up to next-to-leading order in $\hbar$, by its WKB form
\begin{eqnarray}\label{eq:RWKBp}
R^{\rm{WKB}}_{j_r,l}(r)=\sqrt{\frac{\mu}{T^r_{j_r,l}}}\,\frac{2}{r\,[U^r_{j_r,l}(r)]^{1/4}}\,\sin\left(\frac{1}{\hbar}S^r_{j_r,l}(r)+\frac{\pi}{4}\right)\,,
\end{eqnarray}
where the radial action $S^r_{j_r,l}(r)$ and the effective polynomial $U^r_{j_r,l}(r)$ were given in \eqref{eq:HJF}. Importantly, the classical limit of the normalization constant $T^r_{j_r,l}$ is the classical radial period $T^r_{J_r,L}$, as can be seen from\footnote{Technically, the integration region should be $(0,\infty)$. However, the classically forbidden region, namely outside ($r_{\rm{min}}
,r_{\rm{max}}$), is exponentially decaying and its contribution vanishes in the classical limit.}
\begin{eqnarray}\label{eq:nrm}
\lim_{\hbar\rightarrow 0}\,\int_{r_{min}}^{r_{max}}\,r^2\,R^{\rm{WKB}*}_{j_r,l}(r)\,R^{\rm{WKB}}_{j_r,l}(r)\,dr=\,\frac{2}{T^r_{j_r,l}}\,\int_{r_{min}}^{r_{max}}\,\frac{\mu}{\sqrt{U^r_{j_r,l}(r)}}\,dr=1\,,
\end{eqnarray}
by virtue of \eqref{eq:Upcalc}. Here the $\sin$ factor became $1/2$ in the $\hbar\rightarrow 0$ limit -- this would not be the case for matrix elements between $j_r$ and $j'_r\neq j_r$. From now on we drop the WKB label on $R_{j_r,l}(r)$.
To evaluate matrix elements in the classical limit, we need the radial integral
\begin{eqnarray}\label{eq:snpr}
&&I^{J_r,L}_{\Delta j_r,\Delta l}[f]\equiv\lim_{\hbar\rightarrow 0} \int_{r_{min}}^{{r_{max}}} dr\,r^2\,f(r)\,R^*_{j_r-\Delta j_r,l-\Delta l}R_{j_r,l}(r)\,.
\end{eqnarray}
As usual, we take $(j_r,l)=\hbar^{-1}(J_r,L)$ before taking the $\hbar\rightarrow 0$ limit. From \eqref{eq:RWKBp} we have,
\begin{eqnarray}\label{eq:sn}
&&I^{J_r,L}_{\Delta j_r,\Delta l}[f]=\lim_{\hbar\rightarrow 0} \frac{4\mu}{T^r_{j_r,l}}\,\int_{r_{min}}^{{r_{max}}} dr\,\frac{f(r)}{\sqrt{U^r_{j_r,l}(r)}}\,\sin\left(\frac{1}{\hbar}S^r_{j_r,l}(r)+\frac{\pi}{4}\right)\sin\left(\frac{1}{\hbar}S^r_{j_r-\Delta j_r,l-\Delta l}(r)+\frac{\pi}{4}\right)\,.\nonumber\\
\end{eqnarray}
The above integral looks much better when we change integration variables from $r$ to $\alpha_r$ using \eqref{eq:EOMral} (with a $+$ sign since $0\leq\alpha^r<\pi$),
\begin{eqnarray}\label{eq:sn70}
&&I^{J_r,L}_{\Delta j_r,\Delta l}[f]=\lim_{\hbar\rightarrow 0} 4\int_{0}^{\pi} \frac{d\alpha^r}{2\pi}\,f(\alpha^r)\,\sin\left(\frac{1}{\hbar}S^r_{j_r,l}(\alpha^r)+\frac{\pi}{4}\right)\sin\left(\frac{1}{\hbar}S^r_{j_r-\Delta j_r,l-\Delta l}(\alpha^r)+\frac{\pi}{4}\right)\,.\nonumber\\
\end{eqnarray}
In the above equation we used a slight abuse of notation in writing $S^r_{j_r,l}(\alpha^r)$ and $f(\alpha^r)$ instead of the more exact $S^r_{j_r,l}(r(\alpha^r))$ and $f[r(\alpha^r)]$. Because $r(\pi-\alpha^r)=r(\pi+\alpha^r)$ for $0\leq \alpha^r\leq\pi$, we can double the integration region for $\alpha^r$ while multiplying the result by a $1/2$,
\begin{eqnarray}\label{eq:sn77}
&&I^{J_r,L}_{\Delta j_r,\Delta l}[f]=2\lim_{\hbar\rightarrow 0} \int_{0}^{2\pi} \frac{d\alpha^r}{2\pi}\,f(\alpha^r)\,\sin\left(\frac{1}{\hbar}S^r_{j_r,l}(\alpha^r)+\frac{\pi}{4}\right)\sin\left(\frac{1}{\hbar}S^r_{j_r-\Delta j_r,l-\Delta l}(\alpha^r)+\frac{\pi}{4}\right)\,.\nonumber\\
\end{eqnarray}
Using a basic trigonometric identity we then have
\begin{eqnarray}\label{eq:sn71}
&&I^{J_r,L}_{\Delta j_r,\Delta l}[f]=\nonumber\\[5pt]
&&\lim_{\hbar\rightarrow 0}\int_{0}^{2\pi} \frac{d\alpha^r}{2\pi}\,f(\alpha^r)\,\left[\cos\left(\frac{S^r_{j_r,l}(\alpha^r)-S^r_{j_r-\Delta j_r,l-\Delta l}(\alpha^r)}{\hbar}\right)+\sin\left(\frac{S^r_{j_r,l}(\alpha^r)+S^r_{j_r-\Delta j_r,l}(\alpha^r)}{\hbar}\right)\right]\,.\nonumber\\
\end{eqnarray}
In the second term $S^r_{j_r,l}(r)+S^r_{j_r-\Delta j_r,l-\Delta l}(r)=\mathcal{O}(\hbar^0)$ and so this term gives a vanishing oscillatory contribution, so that
\begin{eqnarray}\label{eq:sn72}
&&I^{J_r,L}_{\Delta j_r,\Delta l}[f]=\lim_{\hbar\rightarrow 0} \int_{0}^{2\pi} \frac{d\alpha^r}{2\pi}\,f(\alpha^r)\,\cos\left(\frac{S^r_{j_r,l}(\alpha^r)-S^r_{j_r-\Delta j_r,l-\Delta l}(\alpha^r)}{\hbar}\right)\,.
\end{eqnarray}
On the other hand we know that
\begin{eqnarray}\label{eq:sn2}
&&\lim_{\hbar\rightarrow 0}\frac{S^r_{j_r,l}(\alpha^r)-S^r_{j_r-\Delta j_r,l-\Delta l}(\alpha^r)}{\hbar}=\lim_{\hbar\rightarrow 0}\left[\frac{S^r_{j_r,l}(\alpha^r)-S^r_{j_r-\Delta j_r,l}(\alpha^r)}{\hbar}+\frac{S^r_{j_r-\Delta j_r,l}(\alpha^r)-S^r_{j_r-\Delta j_r,l-\Delta l}(\alpha^r)}{\hbar}\right]=\nonumber\\[5pt]
&&\lim_{\hbar\rightarrow 0}\left[\Delta j_r\frac{\partial S^r_{J_r,L}(\alpha^r)}{\partial E_{J_r,L}}\frac{E_{j_r,l}-E_{j_r-\Delta j_r,l}}{\hbar \Delta j_r}+\Delta l\left(\frac{\partial S^r_{J_r,L}(\alpha^r)}{\partial E_{J_r,L}}\frac{E_{j_r-\Delta j_r,l}-E_{j_r-\Delta j_r,l-\Delta l}}{\hbar \Delta l}-\frac{\partial S^r_{J_r,L}(\alpha^r)}{\partial L}\right)\right]=\nonumber\\[5pt]
&&\Delta j_r\frac{\partial S^r_{J_r,L}(\alpha^r)}{\partial E_{J_r,L}}\Upsilon^r_{Jr, L}+\Delta l\left(\frac{\partial S^r_{J_r,L}(\alpha^r)}{\partial E_{J_r,L}}\Upsilon^\varphi_{Jr,L}-\frac{\partial S^r_{J_r,L}(\alpha^r)}{\partial L}\right)\,.
\end{eqnarray}
To get the last line of \eqref{eq:sn2}, we made use of \eqref{eq:masterfreq}. All-in-all, we have
\begin{eqnarray}\label{eq:sn73}
&&I^{J_r,L}_{\Delta j_r,\Delta l}[f]=\int_{0}^{2\pi} \frac{d\alpha^r}{2\pi}\,f(\alpha^r)\,\cos\left\{\Delta j_r\frac{\partial S^r_{J_r,L}(\alpha^r)}{\partial E_{J_r,L}}\Upsilon^r_{Jr,L}+\Delta l\left(\frac{\partial S^r_{J_r,L}(\alpha^r)}{\partial E_{J_r,L}}\Upsilon^\varphi_{Jr,L}-\frac{\partial S^r_{J_r,L}(\alpha^r)}{\partial L}\right)\right\}\,.\nonumber\\
\end{eqnarray}
We now use \eqref{eq:Ader6} from Appendix~\ref{app:WKB} that allows us to relate $I^{J_r,L}_{\Delta j_r,\Delta l}[f]$ to an integral over $\alpha^\varphi$ as
\begin{eqnarray}\label{eq:sn731}
&&I^{J_rL}_{\Delta j_r,\Delta l}[f]=\int_{0}^{2\pi} \frac{d\alpha^r}{2\pi}\,\int_{0}^{2\pi} \frac{d\alpha^\varphi}{2\pi}\,f(\alpha^r)\,\cos\left\{\Delta j_r \alpha^r+\Delta l\,\left[\alpha^\varphi-\varphi(\alpha^r,\alpha^\varphi)\right]\right\}\,.
\end{eqnarray}
Finally, we use the fact that $r(2\pi-\alpha_r)=r(\alpha_r)$ while $\varphi(2\pi-\alpha_r,2\pi-\alpha_\varphi)={2\pi}-\varphi(\alpha_r,\alpha_\varphi)$. This allows us to write \eqref{eq:sn731} as
\begin{eqnarray}\label{eq:sn732}
&&I^{J_r,L}_{\Delta j_r,\Delta l}[f]=\int_{0}^{2\pi} \frac{d\alpha^r}{2\pi}\,\int_{0}^{2\pi} \frac{d\alpha^\varphi}{2\pi}\,f(\alpha^r)\,e^{-i\Delta l\varphi(\alpha^r,\alpha^\varphi)}\,\exps{i\Delta j_r \alpha^r+i\Delta l\,\alpha^\varphi}\,,
\end{eqnarray}
which already looks very close to a (double) inverse Fourier transform in $\alpha^r$ and $\alpha^\varphi$.
\subsection{From Matrix Elements to Fourier Coefficients}
Now consider the matrix element for an operator $\mathcal{O}(r,\theta,\varphi)$, leaving the generalization to momentum dependent operators to Appendix~\ref{app:momWKB}. First, let's decompose $\mathcal{O}(r,\theta,\varphi)$ in spherical harmonics as
\begin{eqnarray}
\mathcal{O}(r,\theta,\varphi)=\sum_{l_\gamma=0}^\infty\,\sum_{m_\gamma=-l_\gamma}^{l_\gamma}\,\mathcal{O}_{l_\gamma,m_\gamma}(r)\,Y_{l_\gamma}^{m_\gamma}\left(\theta,\varphi\right)\,.
\end{eqnarray}
Now we can take the classical limit of the matrix element of $\mathcal{O}$. As we are considering, without loss of generality, motion in the XY plane with $L_z=L$ and so, accordingly, our initial $m$ is equal to $l$. The matrix element is
\begin{eqnarray}\label{eq:ME}
&&\lim_{\hbar\rightarrow 0}\,\sum_{\Delta m}\,\left\langle j_r-\Delta j_r,l-\Delta l,l-\Delta m\right|\,\mathcal{O}\,\left|j_r,l,l\right\rangle=\nonumber\\[5pt]
&&\lim_{\hbar\rightarrow 0}\,\sum_{\Delta m}\,\sum_{l_\gamma,m_\gamma}\,(-1)^{\Delta l}\,I^{J_r,L}_{\Delta j_r,\Delta l}\left[\mathcal{O}_{l_\gamma,m_\gamma}(r)\right]\,\left\langle l',m'\right|Y_{l_\gamma}^{m_\gamma}\left(\theta,\varphi\right)\left|l,l\right\rangle\,.
\end{eqnarray}
Here the $(-1)^{\Delta l}$ factor came from the overall $(-1)^l$ phase in the definition of our eigenfunctions \eqref{eq:PsiWKBp}. We remind the reader that in the equation above $\theta,\varphi$ are operators whose expectation value is taken between the bra and the ket. The classical limit of the angular part is computed in \eqref{eq:ang1} of Appendix~\ref{app:clas}, which reads
\begin{eqnarray}
\lim_{\hbar\rightarrow 0}\left\langle l-\Delta l,l-\Delta m\right|Y_{l_\gamma}^{m_\gamma}\left(\theta,\varphi\right)\left|l,l\right\rangle =\delta_{\Delta l,\Delta m}\delta_{-\Delta l,m_\gamma}\,(-1)^{m_\gamma}\,Y_{l_\gamma}^{m_{\gamma}}\left(\pi/2,0\right)\,.
\end{eqnarray}
On the other hand $I^{J_r,L}_{\Delta j_r,\Delta l}\left[\mathcal{O}_{l_\gamma,m_\gamma}(r)\right]$ is computed from \eqref{eq:sn732}. We get
\begin{eqnarray}\label{eq:ME2}
&&\lim_{\hbar\rightarrow 0}\,\sum_{\Delta m}\,\left\langle j_r-\Delta j_r,l-\Delta l,l-\Delta m\right|\,\mathcal{O}\,\left|j_r,l,l\right\rangle=\int_{0}^{2\pi} \frac{d\alpha^r}{2\pi}\,\int_{0}^{2\pi} \frac{d\alpha^\varphi}{2\pi}\nonumber\\[5pt]
&&\sum_{l_{\gamma},m_\gamma}\,\left\{\mathcal{O}^{J_r,L}_{l_\gamma,m_{\gamma}}[r(\alpha^r)]\,Y_{l_\gamma}^{m_\gamma}\left(\pi/2,\varphi(\alpha^r,\alpha^\varphi)\right)\right\}\,\exps{i\Delta j_r \alpha^r+i\Delta l\,\alpha^\varphi}\,.
\end{eqnarray}
Here we dropped the $\delta_{-\Delta l,m_\gamma}$ factor, which is redundant since the $\alpha^\varphi$ integral vanishes for $m_\gamma\neq -\Delta l$. Resumming the $l_\gamma,\,m_\gamma$ to get back $\mathcal{O}_{J_r,L}$, we have
\begin{eqnarray}
&&\lim_{\hbar\rightarrow 0}\,\sum_{\Delta m}\,\left\langle j_r-\Delta j_r,l-\Delta l,l-\Delta m\right|\,\mathcal{O}\,\left|j_r,l,l\right\rangle=\nonumber\\[5pt]
&&\int_{0}^{2\pi} \frac{d\alpha^r}{2\pi}\,\int_{0}^{2\pi} \frac{d\alpha^\varphi}{2\pi}\mathcal{O}_{J_r,L}[r(\alpha^r),\pi/2,\varphi(\alpha^r,\alpha^\varphi)]\,\exps{i\Delta j_r \alpha^r+i\Delta l\,\alpha^\varphi}
=\mathcal{O}^{J_r,L}_{\Delta j_r,\Delta l}\,.\nonumber\\
\end{eqnarray}
The last line is simply the definition of the classical Fourier coefficients in \eqref{eq:genQSM2tt}.
This completes the proof of the QSM master equation \eqref{eq:master}.
\section{Keplerian Motion in a Coulomb Potential: A System With A Single Fundamental Frequency}\label{sec:Coul}
In the first three applications of the QSM presented in this paper, we choose to focus on a special case of \eqref{eq:Ham}, namely the case of a Coulomb potential
\begin{eqnarray}\label{eq:Coulpot}
V(r)=-\frac{K}{r}\,,
\end{eqnarray}
where $K=\frac{Z q^2}{4\pi}$ (we take $\varepsilon_{0}=1$), $\mu$ is the reduced mass, $Z$ is the atomic number and $q$ is the electron Coulomb charge. An explicit evaluation of \eqref{eq:aa1} for the Coulomb potential \eqref{eq:Coulpot} reveals that
\begin{eqnarray}\label{eq:defN}
N\equiv J_r+J_\theta+J_\varphi=J_r+L=\sqrt{-\frac{\mu K^2}{2E}}\,.
\end{eqnarray}
The variable $N$ has a natural interpretation as the classical limit of the principle quantum number of the hydrogen atom, via $n=\hbar^{-1}N$. Furthermore, the Coulomb case is special in the sense that the motion is doubly degenerate so that $H(J_r,J_\theta,J\varphi)=H(J_r+J_\theta+J_\varphi)$ and $\Upsilon^\theta_{N,L}=\Upsilon^\varphi_{N,L}=\Upsilon^r_{N,L}\equiv\Upsilon_{N,L}$. In the Coulomb case we will always use the more appropriate label ${N,L}$ instead of $J_r,\,L$. In other words, in the Coulomb case, the motion involves a single angle variable $\alpha$ and a single fundamental frequency $\Upsilon_{N,L}$.  The fundamental frequency and time period are then
\begin{eqnarray}\label{eq:Ups}
\Upsilon_{N, L}=\frac{\mu K^2}{N^3}=\frac{2}{K}\sqrt{-\frac{2E^3}{\mu }}~~,~~T_{N, L}=2\pi\Upsilon^{-1}_{N,L}=\pi K\sqrt{-\frac{\mu}{2E^3}}\,.
\end{eqnarray}
and $\alpha=\Upsilon_{N, L} t$. Since the motion is singly periodic, we can express any time-dependent observable $\mathcal{O}_{N,L}(t)=\mathcal{O}_{N,L}[\alpha(t)]$ in a single Fourier series. In this case it is more convenient to label observables in terms of $N$ and $L$ via \eqref{eq:defN}. The Fourier representation of $\mathcal{O}_{N,L}(t)$ is then
\begin{equation} \label{eq:genQSM2of2}
\mathcal{O}_{N,L}\left(t\right)=\mathcal{O}_{N,L}\left[\alpha(t)\right]=\sum_{\Delta n}\,\mathcal{O}^{N,L}_{\Delta n}\exps{-i\Delta n\alpha(t)}\,,
\end{equation}
where $\Delta n$ is an integer (the index of the $\Delta n_{th}$ Fourier harmony), and
\begin{equation} \label{eq:genQSM2tof}
\mathcal{O}^{N,L}_{\Delta n}=\int_0^{2\pi}\,\frac{d\alpha}{2\pi}\,\mathcal{O}_{N,L}[\alpha]\exps{i\Delta n\alpha}\,.
\end{equation}
The extra degeneracy of the Coulomb Hamiltonian allows for an explicit analytical solution for $r_{N,L}(\alpha),\,\varphi_{N,L}(\alpha)$. This solution is simply \textit{Keplerian motion}. For bound orbits with $E<0$, the motion is parametrized by the ellipse \cite{Goldstein2011}
\begin{equation} \label{eq:Kepell}
r_{N,L}(\alpha)=\frac{p}{1+e\cos[\varphi_{N,L}(\alpha)]}\,,
\end{equation}
where the semi-latus-rectum $p$ and the eccentricity $e$ are related to $E,\,L$ by
\begin{equation} \label{eq:pe}
p=\frac{L^2}{\mu K}~~~,~~e=\sqrt{1-\frac{L^2}{N^2}}~~\rightarrow~~N=\sqrt{\frac{\mu K p}{1-e^2}}\,.
\end{equation}
Similarly to any other observable of the system, the radius $r_{N,L}(\alpha)$ can be expressed in a single Fourier series as
\begin{equation} \label{eq:rfourt}
r_{N,L}\left(\alpha\right)=\sum_{\Delta n=-\infty}^{\infty}\,r^{N,L}_{\Delta n}\exps{-i\Delta n\alpha}\,,
\end{equation}
where the coefficients are related to $r[\alpha]$ via
\begin{eqnarray}\label{eq:rco}
r^{N,L}_{\Delta n}=\int_0^{2\pi}\,\frac{d\alpha}{2\pi}\,r_{N,L}(\alpha)\exps{i\Delta n\alpha}\,.
\end{eqnarray}
We put the labels $N,\,L$ on $r_{N,L}$ and its Fourier coefficients to emphasize that they depend on the conserved action variables of the system. 
The expansion \eqref{eq:rfourt} is more commonly rearranged as
\begin{equation} \label{eq:rfourt1}
r_{N,L}\left(\alpha\right)=\sum_{\Delta n=0}^{\infty}\,\widetilde{r}^{N,L}_{\Delta n}\cos\left[{\Delta n\alpha}\right]\,,
\end{equation}
where $\widetilde{r}^{N,L}_{\Delta n=0}=r^{N,L}_{\Delta n=0}$ and $\widetilde{r}^{N,L}_{\Delta n}=2r^{N,L}_{\Delta n}$ otherwise. An explicit solution of the EOM \eqref{eq:EOMrvarphi} with the Coulomb potential \eqref{eq:Coulpot} yields \cite{Brouwer1961}
\begin{eqnarray} \label{ClassLimRadialtext}
&&\widetilde{r}^{N,L}_{\Delta n}=\frac{p}{1-e^2}\begin{cases}1+\frac{\,e^2}{2}& \Delta n=0\\-\frac{2e}{\Delta n^2}\frac{d}{de}J_{\Delta n}(e\Delta n)&\Delta n> 0\end{cases}\,.
\end{eqnarray}
$\varphi(\alpha)$, on the other hand, is commonly expressed using the eccentric anomaly $\beta(\alpha)$ as
\begin{eqnarray}\label{eq:betadef}
\beta(\alpha) \equiv 2\arctan\left\{\sqrt{\frac{1-e}{1+e}}\tan[\varphi(\alpha)/2]\right\}~~~,~~~
\varphi(\alpha)=2\arctan\left\{\sqrt{\frac{1+e}{1-e}}\tan\left[\beta(\alpha)/2\right]\right\}\,.\nonumber\\
\end{eqnarray}
The Fourier series for $\beta(\alpha)$ is then \cite{Brouwer1961}
\begin{eqnarray}\label{eq:balpha}
\beta(\alpha)&=&\alpha+2\sum_{\Delta n=1}^\infty\,\frac{1}{\Delta n
}J_{\Delta n}(\Delta n \,e)\,\sin\left(\Delta n \alpha\right)\,.
\end{eqnarray}
In the parlance of the Kepler problem and celestial motion, one usually refers to the angle $\alpha$ as the \textit{mean anomaly}.
\section{First Application of the QSM: Time-Dependent Keplerian Motion}\label{sec:Kep}
In this section we demonstrate the QSM by applying it to bound Keplerian motion in the potential \eqref{eq:Coulpot}. In particular, we will use it to obtain the Fourier coefficients $\widetilde{r}^{N,L}_{\Delta n}$ as the classical limit of matrix elements of the \textit{quantum operator} $r$.

The attentive reader will not be surprised that the Hamiltonian \eqref{eq:Ham} with the Coulomb potential \eqref{eq:Coulpot} is also the Hamiltonian for the hydrogen atom, whose quantum eigenstates are conventionally labeled as $\left|n,l,m\right\rangle$ so that
\begin{eqnarray} \label{eq:eig1}
H\left|n,l,m\right\rangle&=&E_{n}\left|n,l,m\right\rangle\nonumber\\[5pt]
L^2\left|n,l,m\right\rangle&=&
\hbar^2l(l+1)\left|n,l,m\right\rangle\nonumber\\[5pt]
L_z\left|n,l,m\right\rangle&=&\hbar m\left|n,l,m\right\rangle\,.
\end{eqnarray}
Here $n$ is the \textit{principal quantum number}, and it is equal to $j_r+l+1$. Due to the degeneracy of the Coulomb potential, the (negative) energy eigenvalues for bound states depend on $j_r$ and $l$ only through their sum $n$, namely 
\begin{eqnarray} \label{eq:hydE}
E_{n}=-\frac{\mu K^2}{2\hbar^2n^2}\,.
\end{eqnarray}
These eigenenergies are related to the degenerate fundamental frequency $\Upsilon_{N,L}$ via the degenerate version of \eqref{eq:masterfreq}, 
\begin{eqnarray}\label{eq:masterfreqdeg}
\Upsilon_{N,L}&=&\lim_{\hbar\rightarrow 0}\frac{E_{n}-E_{n-\Delta n}}{\hbar\Delta n}\,.
\end{eqnarray}
Finally, the hydrogenic wavefunctions are famously given by
\begin{eqnarray}\label{eq:hydwf}
\Psi_{n,l,m}(r)=(-1)^l\,R_{n,l}(r)\,Y_{l}^{m}(\theta,\varphi)\,,
\end{eqnarray}
where
\begin{eqnarray} \label{eq:HydrogenKummer}
&&R_{n,l}\left(r\right)=\nonumber\\
&&\frac{1}{\left(2l+1\right)!}\sqrt{\frac{\left(n+l\right)!}{\left(n-l-1\right)!2n}}\left(\frac{2Z}{na_{0}}\right)^{l+3/2}\exp\left(-\frac{Zr}{na_{0}}\right)r^{l}\, \,_{1}F_{1}\left(-n+l+1;2l+2;\frac{2Zr}{na_{0}}\right).\nonumber\\
\end{eqnarray}
Here $a_{0}=\frac{4\pi \hbar^{2}}{\mu q^{2}}=\frac{\hbar^2 Z}{K\mu}$ is the Bohr radius, and $Z$ is the atomic number. Here we included again a $(-1)^l$ phase factor with respect to the conventional hydrogen atom definition.

The adaptation of our master equation \eqref{eq:master} to the Coulomb / hydrogen problem is
\begin{eqnarray}\label{eq:masterhyd}
\mathcal{O}^{N,L}_{\Delta n}\,=\,\lim_{\hbar\rightarrow 0}\,\sum_{\Delta l,\,\Delta m}\,\left\langle n-\Delta n,l-\Delta l,l-\Delta m\right|\,\mathcal{O}\,\left|n,l,l\right\rangle\,.
\end{eqnarray}
From here on, whenever we deal with this degenerate case we opt to use the principal quantum number $n$ rather than the radial one $j_r$, to conform with the usual notation of the quantum hydrogen atom. The principal quantum number is related to the action variable $N$ defined in \eqref{eq:defN} via
\begin{eqnarray}\label{eq:Nrel}
n=\hbar^{-1}N\,.
\end{eqnarray}
We will use \eqref{eq:Nrel} whenever we consider classical limits, \textit{before} taking $\hbar\rightarrow 0$. From \eqref{eq:masterhyd}, all we have to do to obtain\footnote{Remember that $\widetilde{r}^{N,L}_{\Delta n=0}=r^{N,L}_{\Delta n=0}$ and $\widetilde{r}^{N,L}_{\Delta n}=2r^{N,L}_{\Delta n}$ otherwise.} $r^{N,L}_{\Delta n}$ is compute
\begin{eqnarray}\label{eq:masterhyd1}
r^{N,L}_{\Delta n}\,&=&\,\lim_{\hbar\rightarrow 0}\,\sum_{\Delta l,\,\Delta m}\,\,\left\langle n-\Delta n,l-\Delta l,l-\Delta m\right|\,r\,\left|n,l,l\right\rangle\nonumber\\[5pt]
&=&\,\lim_{\hbar\rightarrow 0}\,\left\langle n-\Delta n,l,l\right|\,r\,\left|n,l,l\right\rangle\,.
\end{eqnarray}
In the last line only $\Delta l=\Delta m=0$ contributes, because $r$ is a \textit{scalar operator}. To compute \eqref{eq:masterhyd1} we have to first (a) calculate the full quantum matrix element; and (b) set $(n,l)=\hbar^{-1}(N,L)$ and take the $\hbar\rightarrow 0$ limit. Note that we \textit{cannot} use WKB to compute
\eqref{eq:masterhyd1}, since this will bring us back to the classical inverse Fourier transform we are trying to compute. Specifically for $r(\alpha)$ of the Kepler/Coulomb problem, the result is known to be \eqref{ClassLimRadialtext} and so our QSM calculation of $r^{N,L}_{\Delta n}$ can be seen as a \textit{validation} of our method. In the next section we will use the QSM to compute Fourier coefficients that are not previously known analytically.

The explicit form of the matrix element in \eqref{eq:masterhyd1} is 
\begin{eqnarray}\label{eq:exp}
\left\langle n-\Delta n,l,l\right|\,r\,\left|n,l,l\right\rangle=\left\langle n-\Delta n,l\right|\,r\,\left|n,l\right\rangle\,&=&\int_0^{\infty}\,dr\,r^3\,R^*_{n-\Delta n,l}(r)R_{n,l}(r)\,.
\end{eqnarray}
It is evaluated using Gordon's integral \cite{Gordon1929,Matsumoto1991},
\begin{equation} \label{eq:IntIdenKummer}
\int_{0}^{\infty}e^{-s r}r^{\rho-1}\,_{1}F_{1}\left(a;b;pr\right)\,_{1}F_{1}\left(c;d;qr\right)\,dr =s^{-\rho}\,\Gamma\left(\rho\right)F_{2}\left(\rho,a,c,b,d;\frac{p}{s},\frac{q}{s}\right)\,.
\end{equation}
Here $F_2$ is the second Appell Function \cite{bateman1953higher}. In our present case \eqref{eq:IntIdenKummer} is applied with
\begin{eqnarray} \label{eq:IntIdenKummerspec}
&&a=-n+\Delta n+l+1~~,~~c=-n+l+1~~,~~b=d=2l+2~~,~~\rho=2l+4~~,\nonumber\\[5pt]&&~~s=\frac{Z}{a_0}\left(\frac{1}{n}+\frac{1}{n-\Delta n}\right)~~,~~p=\frac{2Z}{n a_0}~~,~~q=\frac{2Z}{(n-\Delta n) a_0}\,.
\end{eqnarray}
In this case the $F_2$ degenerates to a sum of regularized Gauss hypergeometric functions ${}_2\widetilde{F}_1$, and the matrix element becomes
\begin{eqnarray}~\label{eq:quantkep}
&&\left\langle n',l\right| r \left| n ,l\right\rangle=-\frac{p\hbar^{2}}{L^2}\frac{2^{2l+2}\left(nn'\right)^{l+2}}{\left(n+n'\right)^{2l+4}}\left(\frac{n-n'}{n+n'}\right)^{n-n'-2}\,\sqrt{\frac{\left(n+l\right)!\left(n-l-1\right)!}{\left(n'+l\right)!\left(n'-l-1\right)!}}\times\nonumber\\[5pt]
&&\,\left[\,{}_{2}\widetilde{F}_{1}\left(l-n'+1;n+l;n-n';\left(\frac{n-n'}{n+n'}\right)^2\right)-\right.\nonumber\\[5pt]
&&~~~~~~~\left.{(n-l)(n+l+1)}\left(\frac{n-n'}{n+n'}\right)^{2}~{}_{2}\widetilde{F}_{1}\left(l-n'+1;n+l+2;n-n'+2;\left(\frac{n-n'}{n+n'}\right)^2\right)\right]\,,\nonumber\\
\end{eqnarray}
where $n'=n-\Delta n$, and we used the definition \eqref{eq:pe} of $p$. In the special case where $n=n'$, we have
\begin{eqnarray}\label{eq:nteqn}
&&\left\langle n,l\right| r \left| n ,l\right\rangle=\frac{p\hbar^{2}}{L^2}\frac{3n^2-l(l+1)}{2}\,.
\end{eqnarray}
\subsection{Classical Limit}
We now set $(n,l)=\hbar^{-1}(N,L)$ in \eqref{eq:quantkep} and \eqref{eq:nteqn}, and subsequently take the $\hbar\rightarrow 0$ limit. To do that, we need to learn how to take the $\hbar\rightarrow 0$ limit of the ${}_2\widetilde{F}_1$ in \eqref{eq:quantkep}. Specifically, the ${}_2\widetilde{F}_1$ in \eqref{eq:quantkep} are of the form
\begin{eqnarray}\label{eq:Ftilform}
{}_2\widetilde{F}_1\left(-\hbar^{-1}A,\hbar^{-1}B ,c,\hbar^2 \delta\right)\,,
\end{eqnarray}
where
\begin{eqnarray}\label{eq:Ftilformpar}
A=N-L~~,~~B=N+L~~,~~\delta=\frac{\Delta n^2}{4N^2}\,,
\end{eqnarray}
and $c=\Delta n$ ($c=\Delta n+2$) for the first (second) hypergeometric function in \eqref{eq:quantkep}. 
The definition of the Gauss hypergeometric function is
\begin{eqnarray}\label{eq:Gausshyp}
{}_2\widetilde{F}_1\left(-\hbar^{-1}A,\hbar^{-1}B ,c,\hbar^2 \delta\right)=\frac{1}{\Gamma(c)}\,\lim_{\hbar\rightarrow 0}\,\sum_l\,\frac{(-A\hbar^{-1})_l(B\hbar^{-1})_l}{l!\,(c)_l}(\hbar^2 \delta)^l\,,
\end{eqnarray}
where $(a)_n\equiv \Gamma(a+n)/\Gamma(a)$ is the Pochhammer symbol, whose classical limit is
\begin{eqnarray}\label{eq:poclimt}
\lim_{\hbar\rightarrow 0}(\hbar^{-1} P)_n=\lim_{\hbar\rightarrow 0}(\hbar^{-1} P)^n~~,~~{\rm\, for \,any \,}P\,.
\end{eqnarray}
Using this limit in \eqref{eq:Gausshyp}, we have
\begin{eqnarray}\label{eq:fd}
&&\lim_{\hbar\rightarrow 0}\,{}_2\widetilde{F}_1\left(-\hbar^{-1}A,\hbar^{-1}B ,c,\hbar^2 \delta\right)=\frac{1}{\Gamma(c)}\,\sum_l\,\frac{\left(-A B \delta\right)^l}{l!\,(c)_l}={\left(\frac{2}{\zeta}\right)}^{c-1}J_{c-1}(\zeta)\,,\nonumber\\
\end{eqnarray}
where $J$ is the Bessel function of the first kind, and $\zeta=2\sqrt{AB\delta}=e\Delta n$, with the eccentricity $e$ defined in \eqref{eq:pe}. Substituting this in the $\hbar\rightarrow 0$ limit of \eqref{eq:quantkep}, we get
\begin{eqnarray}~\label{eq:quantkepcl2}
\lim_{\hbar\rightarrow 0}\left\langle n',l\right| r \left| n ,l\right\rangle&=&-\frac{e}{2}\frac{p}{1-e^2}\frac{1}{\Delta n}\,\left[J_{\Delta n-1}(e\Delta n)-J_{\Delta n+1}(e\Delta n)\right]\nonumber\\[5pt]
&=&-\frac{p\,e}{1-e^2}\frac{1}{\Delta n^2}\,\frac{d}{de}J_{\Delta n}(e\Delta n)
\end{eqnarray}
For $n'= n$ we can take the classical limit of \eqref{eq:nteqn} directly and get
\begin{eqnarray}~\label{eq:resquantclskepeq}
\lim_{\hbar\rightarrow 0}\left\langle n,l\right| r \left| n ,l\right\rangle=\frac{p}{1-e^2}\left(1+\frac{e^2}{2}\right)\,.
\end{eqnarray}

Substituting these in \eqref{eq:masterhyd1}, we see that we reproduce the classical Fourier coefficients \eqref{ClassLimRadialtext} exactly (see footnote 4).

\section{All-order EM Radiation with the QSM}\label{sec:EM}
In our treatment of the Kepler problem, we were fortunate enough to have the known Fourier-series solution to compare with. In this section we present a first application of the QSM to a classical problem whose all-order analytical solution, we believe, is not previously known. In fact, the conventional treatment involves a reduction to a series of radial integrals that require numerical evaluation.
The problem we address in this section is the calculation of the retarded EM field $A^\mu_{\rm ret}$ generated by a classical \textit{non-relativistic}\footnote{The relativistic generalization is straightforward, as can be seen from the relativistic treatment of the Schwarzschild case in Section~\ref{sec:Schwarz}.} electron moving along a (quasi-) Keplerian orbit. By quasi-Keplerian, we mean that the trajectory is given by \eqref{eq:rfourt1}, but allowing for a slow change in $N(t)$, $L(t)$ and $\dot{\alpha}(t)$ over time. When $\dot{\alpha}$ is constant, the motion is purely Keplerian. In this section we use the generic symbols $N,\,L$ and suppress their possible slow time dependence. To apply the results of this section for quasi-periodic motion at the adiabatic (0PA) level, simply substitute $N=N(t),\,L=L(t)$ and $\dot{\alpha}=\Upsilon_{N(t),L(t)}$ in the expressions of this section. The EM current $J^\mu$ of the quasi-Keplerian electron is
\begin{eqnarray}\label{eq:curre}
J^\mu(t',x')=q\,v^\mu_{N,L}(t')\,\delta^{(3)}\left[\vec{x}'-\vec{r}_{N,L}(t')\right]\,,
\end{eqnarray} 
where $v^\mu_{N,L}(t')=\left(1,\partial_{t'}\vec{r}_{N,L}(t')\right)$ and $q$ is the electric charge of the electron. Here $\vec{r}_{N,L}(t')$ is the time dependent position of a source in a \textit{quasi}-Keplerian orbit. That is to say, $\vec{r}_{N,L}(t')=\vec{r}_{N,L}[\alpha'(t')]$ and the $\alpha'$ dependence is the one from \eqref{eq:rfourt1} with $\alpha\rightarrow\alpha'$. 

The retarded EM field $A^\mu_{\rm ret}$ generated by the electron is then obtained by integrating $J^\mu$ together with the retarded Green's function $G_{\rm ret}^{\mu\nu}$,
\begin{eqnarray}\label{eq:fret}
A^{\mu}_{\rm ret}(t,\vec{x})&=&\int\,d^4x'\,G_{\rm ret}^{\mu\nu}(t,\vec{x};t',\vec{x}')\,J_\nu(t',x')\,.
\end{eqnarray}
The retarded Green's function for the EM field is famously given by \cite{Jackson1998}
\begin{eqnarray}~\label{eq:GreenretPS}
G_{\rm ret}^{\mu\nu}(t,\vec{x};t',\vec{x}')=\eta^{\mu\nu}\frac{\Theta(t-t')}{4\pi R}\,\delta(t-t'-R)\,,
\end{eqnarray}
where $R=|\vec{x}-\vec{x}'|$, and $\eta^{\mu\nu}$ is the Minkowski metric\footnote{In Section~\ref{sec:Schwarz} we work in a Schwarzschild metric with a mostly-plus signature.} (in the $(+,-,-,-)$ signature). This is the Green's function used to derive the Li\'{e}nard-Wiechert retarded potential in electromagnetism. For our purposes it is convenient to use the Fourier representation of the delta function
\begin{eqnarray}~\label{eq:GreenretFour}
G_{\rm ret}^{\mu\nu}(t,\vec{x};t',\vec{x}')=\eta^{\mu\nu}\frac{\Theta(t-t')}{2\pi}\,\int_{-\infty}^{\infty}\,d\omega \,e^{-i\omega(t-t')}\,\frac{e^{i\omega R}}{4\pi R}\,.
\end{eqnarray}
We now expand $e^{i\omega R}/(4\pi R)$ in multipoles and get
\begin{eqnarray}~\label{eq:GreenretMulti2}
G_{\rm ret}^{\mu\nu}(t,\vec{x};t',\vec{x}')&=&\eta^{\mu\nu}\frac{\Theta(t-t')}{2\pi}\,\int_{-\infty}^{\infty}\,d\omega \,e^{-i\omega(t-t')}\,\times\nonumber\\[5pt]
&&\left\{i\omega\,\sum_{l_{\gamma}=0}^\infty\,j_{l_{\gamma}}(\omega r_<)\,h^{(1)}_{l_{\gamma}}(\omega r_>)\,\sum_{m_{\gamma}=-l_{\gamma}}^{l_{\gamma}}\,Y_{l_{\gamma}}^{m_{\gamma}}(\theta',\varphi')Y_{l_{\gamma}}^{m_{\gamma}*}(\theta,\varphi)\right\}\,,~~~~~~~~~
\end{eqnarray}
where $\Theta$ is the Heaviside function, $Y_{l_\gamma}^{m_\gamma}$ is a spherical harmonic, $j_{l_\gamma}$ is a spherical Bessel function, and $h^{(1)}_{l_\gamma}$ is a spherical Hankel function of the first kind. Finally, $\{r_<,r_>\}=\{{\rm min}(r,r'),{\rm max}(r,r')\}$. From here on we will focus on the case where $r$ is larger than $r'$, so that $r_>=r,\,r_<=r'$.
Substituting the expressions \eqref{eq:curre} for the current and  \eqref{eq:GreenretMulti2} for the Green's function in \eqref{eq:fret}, and integrating over $\vec{x}'$ using the delta function in \eqref{eq:curre}, we get
\begin{eqnarray}~\label{eq:GreenretMulti1}
A^\mu_{\rm ret}(t,\vec{x})&=&\frac{q}{\mu}\,\sum_{l_{\gamma}=0}^\infty\,\sum_{m_{\gamma}=-l_{\gamma}}^{l_{\gamma}}\,Y_{l_{\gamma}}^{m_{\gamma}{*}}(\theta,\varphi)\,\int\,dt'\,\frac{\Theta(t-t')}{2\pi}\,\int_{-\infty}^{\infty}\,d\omega \,e^{-i\omega(t-t')}\,i\omega\,h^{(1)}_{l_{\gamma}}(\omega r)\,\mathcal{M}^{\mu,N,L}_{l_{\gamma},m_{\gamma}}(t')\nonumber\\[5pt]
\mathcal{M}^{\mu,N,L}_{l_{\gamma},m_{\gamma}}(t')&\equiv &j_{l}(\omega r_{N,L}(t'))\,Y_{l_{\gamma}}^{m_{\gamma}}(\hat{r}_{N,L}(t'))\,p^\mu_{N,L}(t')\,.
\end{eqnarray}
The time dependent observable $\mathcal{M}^{\mu,N,L}_{l_{\gamma},m_{\gamma}}(t')$ depends periodically on $\alpha'$,
\begin{equation} \label{eq:genQSM2of}
\mathcal{M}^{\mu,N,L}_{l_{\gamma},m_{\gamma}}\left(t'\right)=\mathcal{M}^{\mu,N,L}_{l_{\gamma},m_{\gamma}}\left[\alpha'(t')\right]=\sum_{\Delta n}\,\mathcal{M}^{\mu,N,L}_{l_{\gamma},m_{\gamma},\Delta n}\exps{-i\Delta n\alpha'(t')}\,.
\end{equation}
Substituting this time dependence in \eqref{eq:GreenretMulti1}, we get
\begin{eqnarray}~\label{eq:GreenretMulti3}
A^\mu_{\rm ret}(t,\vec{x})&=&\frac{q}{\mu}\,\sum_{\Delta n}\,\sum_{l_{\gamma}=0}^\infty\,\sum_{m_{\gamma}=-l_{\gamma}}^{l_{\gamma}}\,Y_{l_{\gamma}}^{m_{\gamma}{*}}(\theta,\varphi)\,\int\,dt'\,\frac{\Theta(t-t')}{2\pi}\,\nonumber\\[5pt]
&&\int_{-\infty}^{\infty}\,d\omega\,i\omega\,h^{(1)}_{l_{\gamma}}(\omega r)\,\exps{-i\Delta n\alpha'(t')-i\omega(t-t')}\,\mathcal{M}^{\mu,N,L}_{l_{\gamma},m_{\gamma},\Delta n}\,.
\end{eqnarray}
Equation~\eqref{eq:GreenretMulti3} is generic and exact, and we shall use it in the next section to calculate the waveform corresponding to an adiabatic inspiral, in the far-field approximation. For the rest of this section, we will focus on the special case of exact Keplerian motion, in which case $N$, $L$ and $\Upsilon_{NL}$ are truly constant, and $\alpha'=\Upsilon_{NL} t'$. In this case we can explicitly carry the $t'$ and $\omega$ integrals, obtaining
\begin{eqnarray}\label{eq:greencoh3}
&&A^\mu_{\rm ret}=\frac{iq}{\mu}\,\sum_{l_{\gamma}=0}^\infty\sum_{m_{\gamma}=-l_{\gamma}}^{l_\gamma}\,\sum_{\Delta n}\,e^{-i\Delta n\,\alpha}\,\omega_{\Delta n}\,h^{(1)}_{l_{\gamma}}(\omega_{\Delta n} r)\,Y_{l_\gamma}^{m_{\gamma}{*}}(\theta,\varphi)\, \mathcal{M}^{\mu,N,L}_{l_{\gamma},m_{\gamma},\Delta n}\,,
\end{eqnarray}
where $\alpha=\Upsilon_{NL} t$ and $ \omega_{\Delta n}\equiv \Upsilon_{NL} \Delta n$. We see that the retarded field emitted by a Keplerian electron inherits its periodic dependence on $\alpha=\Upsilon_{NL} t$.
\subsection{Using the QSM to get the Fourier Coefficients}
Here we summarize the computation of the classical Fourier coefficients in \eqref{eq:genQSM2of} with the QSM. As a special case of the degenerate version \eqref{eq:masterhyd} of the master equation, we have
\begin{eqnarray}\label{eq:masterhydrad}
&&\mathcal{M}^{\mu,N,L}_{l_{\gamma},m_{\gamma},\Delta n}\,=\,\lim_{\hbar\rightarrow 0}\,\sum_{\Delta l,\,\Delta m}\,{\mathcal{M}^{\mu,N,L}_{l_{\gamma},m_{\gamma},\mathbf{\Delta}}(\omega)},\nonumber\\&&{\mathcal{M}^{\mu,N,L}_{l_{\gamma},m_{\gamma},\mathbf{\Delta}}(\omega)\equiv\left\langle n{'},l',m'\right|\,j_{l}(\omega r)\,Y_{l_{\gamma}}^{m_{\gamma}}(\hat{r})\,p^\mu\,\left|n,l,l\right\rangle\,,}
\end{eqnarray}
where $r,\hat{r}$ and $p^i=-i\hbar\partial^{i}$ are now read as quantum operators (on the other hand, $p^0=\mu$). Additionally, {$\mathbf{\Delta}\equiv(\Delta n,\Delta l,\Delta m)$, and $(n',l',m')=(n,l,{l})-\mathbf{\Delta}$.}
Explicitly, 
\begin{eqnarray}\label{eq:Amumatel0}
&&{(-1)^{\Delta l}}\mathcal{M}^{0,N,L}_{l_{\gamma},m_{\gamma},
{\mathbf{\Delta}}}=\mu\,\underbrace{\left\{\lim_{\hbar\rightarrow 0}\left\langle n', l'\right|\,j_{l_{\gamma}}(\omega r)\left| n, l\right\rangle\right\}}_{{\rm Eq.}\,\text{\eqref{eq:rad1}}}~\underbrace{\left\{\lim_{\hbar\rightarrow 0}\left\langle l' , m'\right| Y_{l_\gamma}^{m_\gamma}(\theta,\varphi)\left| l ,l \right\rangle\right\}}_{{\rm Eq.}\,\text{\eqref{eq:ang1}}}\,.\nonumber\\
\end{eqnarray}
Here the $(-1)^{\Delta l}$ factor came from the overall $(-1)^l$ phase in the definition of our eigenfunctions \eqref{eq:hydwf}. On the other hand, we have
\begin{eqnarray}\label{eq:Amumatelvec}
&&{(-1)^{\Delta l}}\overrightarrow{\mathcal{M}}^{N,L}_{l_{\gamma},m_{\gamma},{\mathbf{\Delta}}}=\nonumber\\[5pt]
&&-i\underbrace{\left\{\lim_{\hbar\rightarrow 0}\left\langle n', l'\right|\,r^{-1}j_{l_{\gamma}}(\omega r)\left| n, l\right\rangle\right\}}_{{\rm Eq.}\,\text{\eqref{eq:rad2}}}~\underbrace{\left\{\lim_{\hbar\rightarrow 0}\,\hbar \sum_{q=-1}^1\,\left\langle l' , m'\right| Y_{l_\gamma}^{m_\gamma}(\theta,\varphi)(\overrightarrow{\nabla}_{\Omega})_q\left| l ,l \right\rangle\,\vec{\varepsilon}_q\,\right\}}_{{\rm Eq.}\,\text{\eqref{eq:ang3}}}\nonumber\\[5pt]
&&-i\underbrace{\left\{\lim_{\hbar\rightarrow 0}\hbar\left\langle n', l'\right|\,j_{l_{\gamma}}(\omega r)\,\partial_r\left| n, l\right\rangle\right\}}_{{\rm Eq.}\,\text{\eqref{eq:rad3}}}~\underbrace{\left\{\lim_{\hbar\rightarrow 0}\sum_{q=-1}^1\,\left\langle l' , m'\right| Y_{l_\gamma}^{m_\gamma}(\theta,\varphi)(\hat{r})_q\left| l ,l \right\rangle\,\vec{\varepsilon}_q\,\right\}}_{{\rm Eq.}\,\text{\eqref{eq:ang2}}}\,,\nonumber\\
\end{eqnarray}
where $\vec{\varepsilon}_0=\hat{z},\,\vec{\varepsilon}_{\pm}=\tfrac{1}{\sqrt{2}}\left(\mp\hat{x}+i\hat{y}\right)$ and $(\vec{v})_q=\vec{v}\cdot\vec{\varepsilon}_{q}^{\,*}$. The calculation itself is rather technical, and we relegate it to our Appendix~\ref{app:clas}. In the above equations, we added tags to indicate the relevant equations in the appendix where the corresponding classical matrix elements are calculated. 
\subsection{Results}
\vspace{10pt}
\begin{figure}[ht!]
\begin{center}
\includegraphics[width=0.75\linewidth]{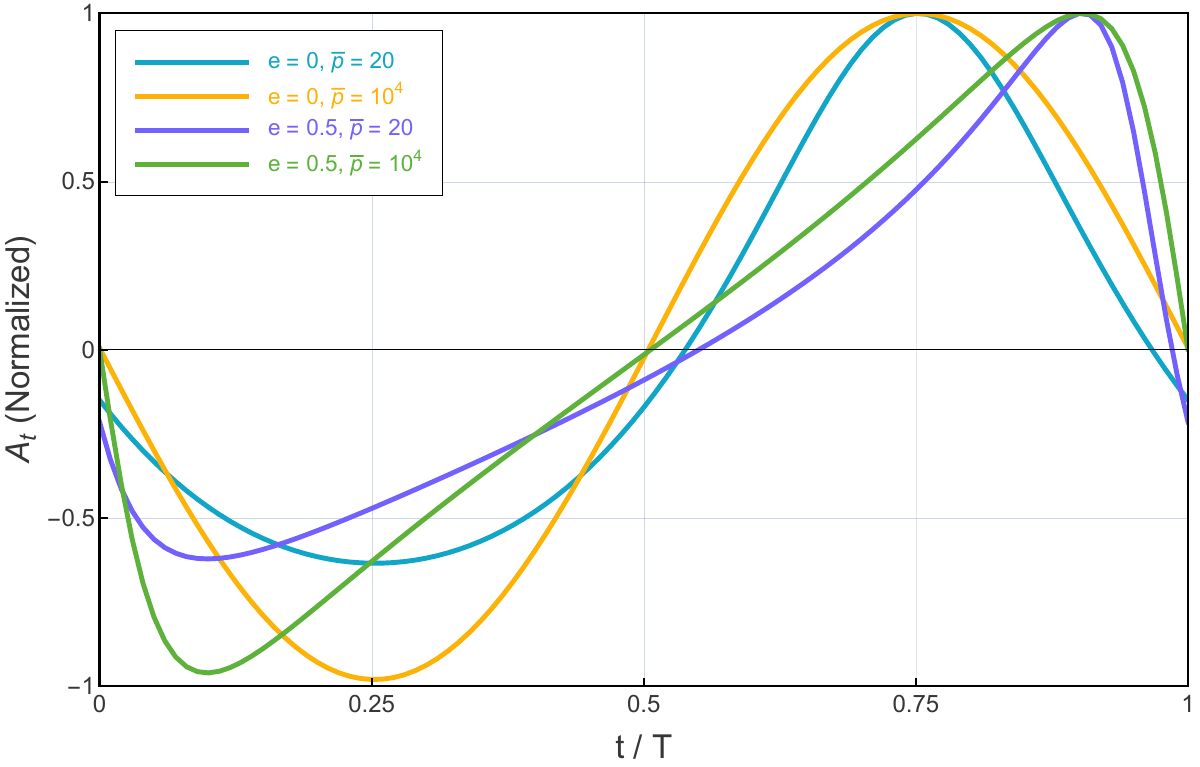}
\caption{Keplerian waveforms. This figure shows the  waveform --$A_t$-- radiated over one period by an electron undergoing Keplerian motion in four cases of the orbital parameters $(e,\Bar{p})$: (i) $(0,20)$; (ii) $(0,10^4)$; (iii) $(0.5,20)$; and (iv) $(0.5,10^4)$, where $\Bar{p}=p\mu/K$. The observation point is on the x-axis, far away from the electron's orbit. For each case, the horizontal and vertical axes are normalized by the orbital period $T=2\pi/\Upsilon$ and the maximum of the waveform, respectively.}\label{fig:KWFs}
\end{center}
\end{figure}

\begin{figure}[ht!]
     \centering
     \begin{subfigure}{0.46\textwidth}
         \centering
         \includegraphics[width=\textwidth]{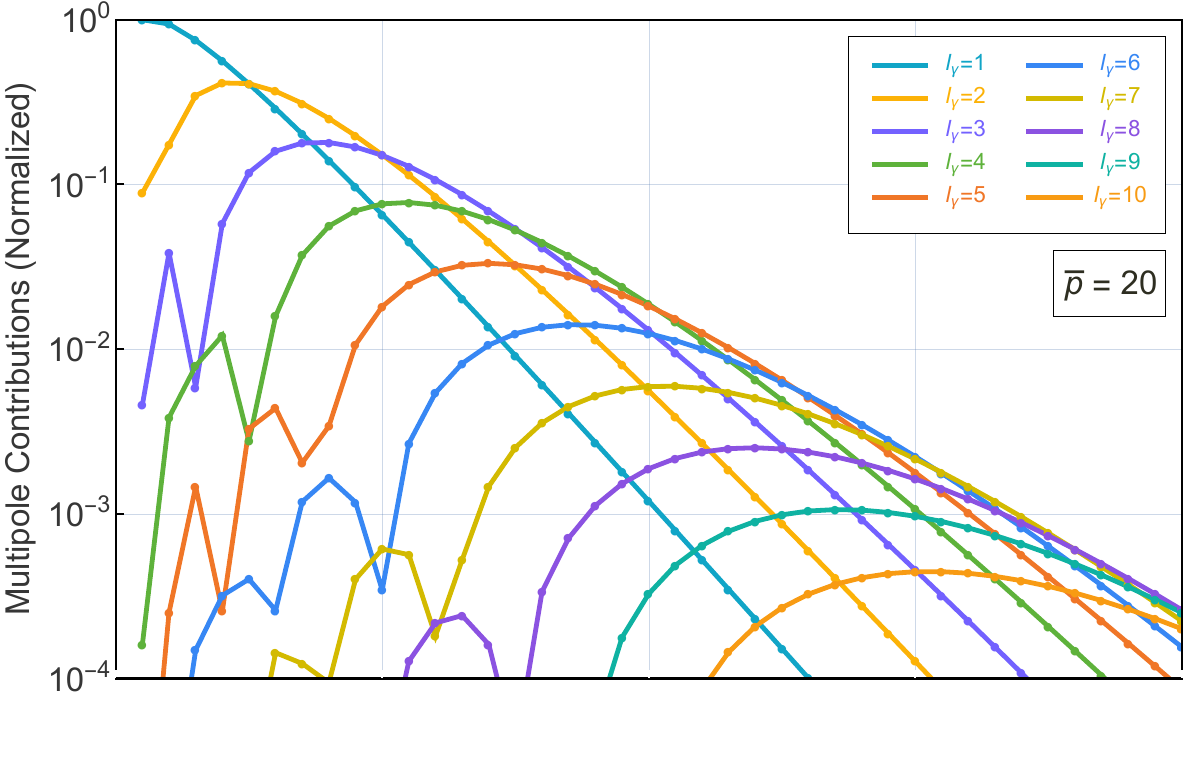}   \end{subfigure}
     \hspace{25pt}
     \begin{subfigure}{0.455\textwidth}
         \centering  \includegraphics[width=\textwidth]{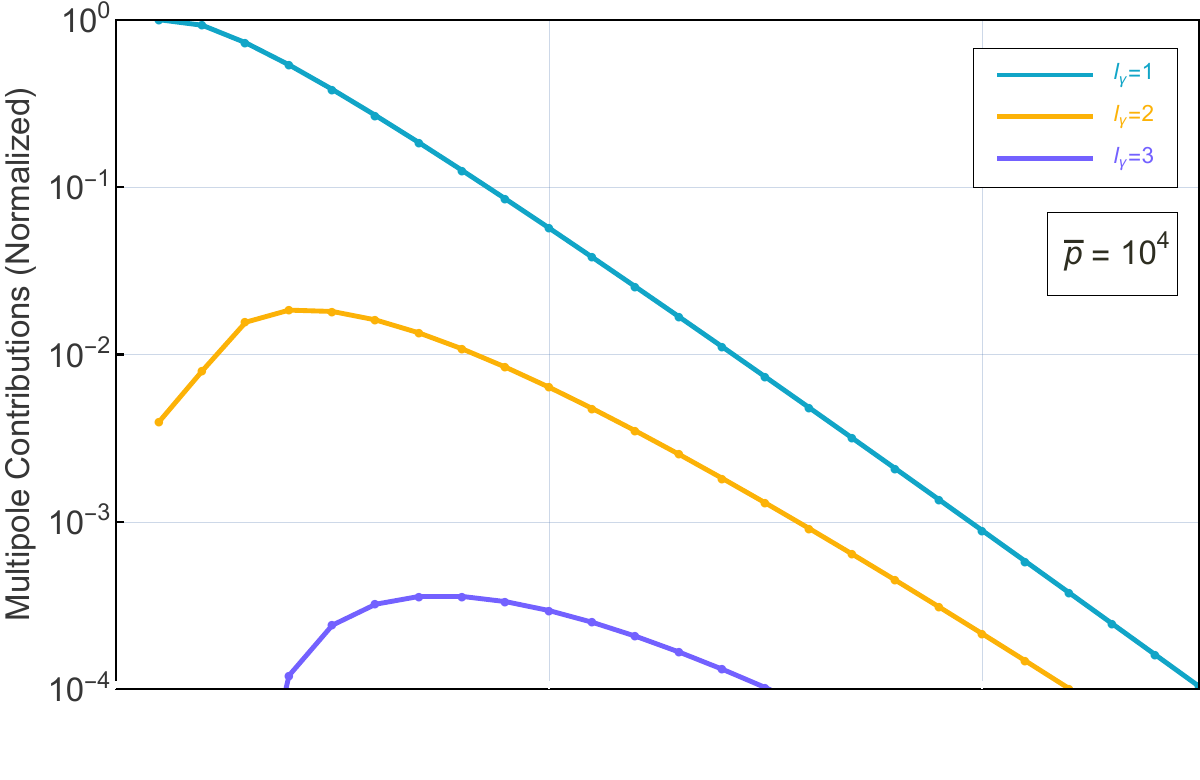}
     \end{subfigure}\\
     \vspace*{-15pt}
        \hspace{-2pt}  \begin{subfigure}{0.453\textwidth}
         \centering \includegraphics[width=\textwidth]{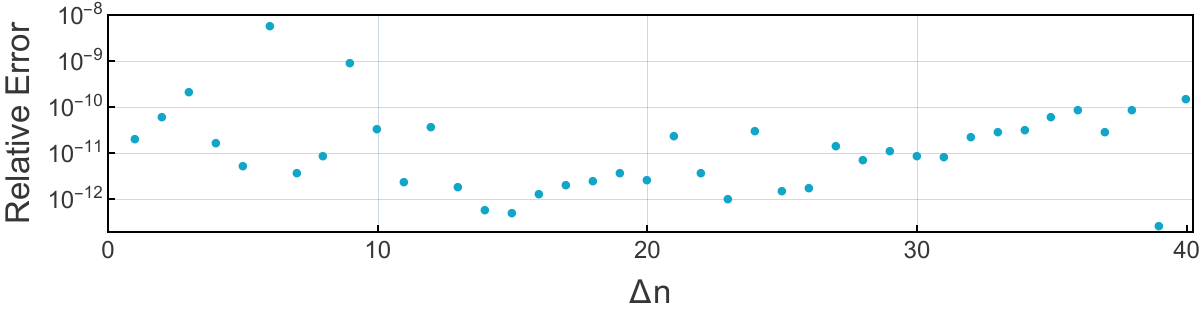}
         \caption{Multipole contributions and relative error for $(e,\Bar{p})=(0.5,20)$. We show the first 10 leading multipole contributions,  ignoring the monopole $(l_\gamma=0)$ contribution.}
         \label{fig:error20}
     \end{subfigure}
     \hspace{29pt}
     \begin{subfigure}{0.452\textwidth}
         \centering \includegraphics[width=\textwidth]{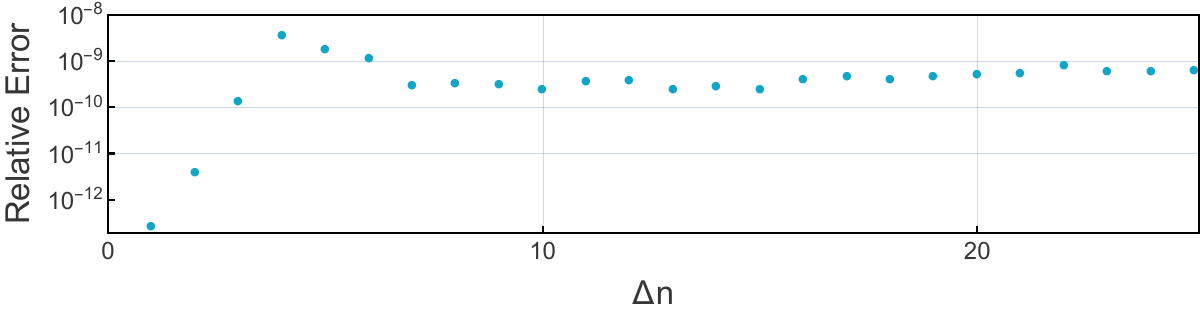}
         \caption{Multipole contributions and relative error for $(e,\Bar{p})=(0.5,10^4)$. We show the first 3 leading multipole contributions,  ignoring the monopole $(l_\gamma=0)$ contribution.}
         \label{fig:error10k}
     \end{subfigure}
        \caption{Multipole contributions and equality with the classical integrals. This figure shows the contributions of the dominant multipoles to the Fourier coefficients of $A_t$ in \eqref{eq:greencoh3}, where we exclude $h^{(1)}_{l_{\gamma}} $ for aesthetic purposes. We take two sets of orbital elements $(e,\Bar{p})$, where $\Bar{p}=p\mu/K$. The vertical axis is normalized by the maximum in each panel. The relative error of the Fourier coefficients with respect to the (numerical) classical  integrals is shown in the bottom inserts.  }\label{fig:f}        
\end{figure}

In Fig. \ref{fig:KWFs}, we present waveforms in $A_t$ corresponding to various Keplerian orbits. Specifically, we consider the four combinations of circular ($e=0$) versus elliptic ($e=0.5$), and "fast" ($p\mu/K=20$) versus "slow" ($p\mu/K=10^4$) orbits. Note that the asymmetry between the maximum and minimum of the waveforms is due to the observation point being on the x-axis. The qualitative features of the waveforms can be understood as follows. For circular orbits, it can be shown that the Fourier coefficients of \eqref{eq:greencoh3} vanish unless $\Delta n= m_{\gamma}$ can be satisfied, whereas the Fourier coefficients of elliptic curves are generally non-vanishing. Moreover, unlike "fast" orbits, "slow" orbits are well-described using the dipole-approximation ($l_\gamma=1$), as can be seen in Fig. \ref{fig:f}. Hence, "slow" circular orbits have sinusoidal waveforms, in contrast to more involved waveforms in the general case. Finally, the broadening (narrowing) of the minimum (maximum) of the waveforms is due to the Doppler effect. 

The multipole contributions to the Fourier coefficients of \eqref{eq:greencoh3}, for "fast" and "slow" elliptic orbits, are presented in Fig. \ref{fig:f}. As mentioned above, in the "fast" orbit case (Fig. \ref{fig:error20}), the Fourier coefficients receive substantial contributions from higher multipoles, whereas, in the "slow" orbit case (Fig. \ref{fig:error10k}), the dipole is by far the dominant contribution.

As can be seen from the relative error plots in Fig. \ref{fig:f}, the Fourier coefficients of  \eqref{eq:greencoh3} are indeed equal to the standard coefficients obtained from classical electromagnetism, expressed using (numerical) Bessel integrals.

\vspace{15pt}
\newpage
\section{EM Self-Force with the QSM}\label{sec:ESF}

The natural application of the QSM is in the calculation of the self-force on an inspiralling body -- a crucial ingredient for the determination of its trajectory and associated waveform. In this section, we calculate the self-force for the case of an inspiralling classical electron, along with its adiabatic trajectory and associated radiation (waveform). The calculation is done using the method of
\textit{osculating orbits} \cite{Brouwer1961}. This is a parameterization of the true trajectory of an inspiralling object as a slow transition between different Keplerian orbits, each one momentarily tangent (''osculating") to the true trajectory. 
Practically, this means that in addition to the mean anomaly $\alpha(t)$, we need to keep track of the time-dependent orbital elements $(e(t),\,p(t))$ (equivalently $E(t)$ and $L(t)$) of the osculating orbit to describe the inspiralling trajectory. Importantly, the method of osculating orbits is not an approximation but rather a parameterization of the full trajectory. The osculation method becomes particularly useful when the self-force is relatively weak compared with the external force (Coulomb force here), i.e., as long as the energy and angular momentum loss rates are slow, i.e. $\frac{T_{N,L}\dot{E}}{E}\sim \frac{T_{N,L}\dot{L}}{L}\ll 1 $, where $T_{N,L}$ is the rotation period. This allows for the perturbative treatment of the self-force, giving rise to the PA expansion \cite{Hinderer2008}. In this work, we limit ourselves to the leading order in the PA expansion, also called the adiabatic (0PA) order. At this order, the dissipation rates of $\dot{E}$ and $\dot{L}$ are calculated for the osculating Keplerian orbit, without back-reaction.

Putting things more concretely, consider the EOM for a classical electron in a central potential, with an additional small self-force $\vec{F}$:
\begin{eqnarray}\label{eq:EOMF}
\mu\ddot{\vec{r}}_{\rm{insp}}+K\frac{\vec{r}_{\rm{insp}}}{r^3_{\rm{insp}}}=\varepsilon \vec{F}\,.
\end{eqnarray}
Here, $\vec{r}_{\rm{insp}}$ denotes the position of the inspiralling electron, and the $\varepsilon$ in front of the force $\vec{F}$ is a power-counting parameter. Expanding in it gives us PA perturbation theory. The self-force itself may also depend on $\varepsilon$, both explicitly and through its dependence on the trajectory of the inspiralling electron. We can solve \eqref{eq:EOMF} order-by-order by evaluating $\vec{F}$, and its partial derivatives on the osculating orbit parametrized by the orbital elements $(e(t),\,p(t))$ and the mean anomaly $\alpha(t)$. Explicitly, the trajectory is given by \eqref{eq:rfourt1}-\eqref{eq:balpha}, with time-dependent orbital elements\footnote{Equivalently $E(t)$ and $L(t)$.} $(e(t),p(t))$ and $\alpha(t)$ which slowly deviates from $\Upsilon_{N,L} t$. The time dependence of $e(t),\,p(t)$, and $\alpha(t)$ is derived from the osculation (tangentiality) conditions between the osculating and true orbit, namely:
\begin{eqnarray}\label{eq:osc}
\vec{r}_{N(t),L(t)}[\alpha(t)]&=&\vec{r}_{\rm{insp}}(t)\nonumber\\[5pt]
\frac{d}{dt}\left\{{\vec{r}}_{N(t),L(t)}[\alpha(t)]\right\}&=&\dot{\vec{r}}_{\rm{insp}}(t)\,,
\end{eqnarray}
where $\vec{r}_{\rm{insp}}$ satisfies the full EOM \eqref{eq:EOMF}, which in turn depends on the self-force $\vec{F}$. These relations greatly simplify in our case of interest, in which $\vec{F}$ is the force due to \textit{electromagnetic radiation-reaction}. This force is purely dissipative, and so it is enough to know $dE/dt$ and $dL/dt$ (alongside $\alpha(t)$) to determine the complete time evolution of the system.
The EM self-force $\vec{F}$ is given by the Lorentz force induced by the radiated EM field, it is clear that the energy and angular momentum loss rates are given by
\begin{eqnarray}\label{eq:dEdL}
\frac{dE}{dt}&=&q\,\vec{v}_{\rm{insp}}\cdot \vec{E}^{\rm reg}\nonumber\\[5pt]
\frac{d\vec{L}}{dt}&=&q\,\vec{r}_{\rm{insp}}\,\times\,\left[\vec{E}^{\rm reg}+\vec{v}_{\rm{insp}}\times \vec{B}^{\rm reg}\right]\,,
\end{eqnarray}
where $\vec{v}_{\rm{insp}}$ is the particle's velocity, $E^{\rm reg}_i=\partial_{[0}A^{\rm reg}_{i]}$, and $B^{\rm reg}_i=-\frac{1}{2}\epsilon_{ijk}\partial^{[j}A^{\rm reg;k]}$, evaluated at $(e(t),p(t),\alpha(t))$. The label ''reg" on $A^{\rm reg}_\mu$ means the part of the EM field which is regular at the position of the inspiralling electron -- we comment on this more in the next section. 
Equations \eqref{eq:dEdL} are supplemented by an equation for $d\alpha/dt$, which is derived from the osculation condition \eqref{eq:osc}. Together, these three equations constitute a full set of differential equations that determine the time evolution of the inspiral. However, the right-hand side of \eqref{eq:dEdL} is a highly nonlinear function of $\alpha$, which in practice makes the system challenging for numerical integration. Reference \cite{VanDeMeent2018} addressed this problem by performing a clever change of variables called a near-identity transformation (NIT). At 0PA and under a particular choice of the NIT\footnote{This is the choice made in section 2.6.4 of \cite{VanDeMeent2018} and section 6.2.5 of \cite{Pound2021}}, the $\alpha$ dependence is eliminated from the right-hand side, and the evolution equations reduce to
\begin{eqnarray}\label{eq:dEdLavg}
\frac{dE}{dt}&=&q\,\left\langle\vec{v}_{\rm{insp}}\cdot \vec{E}^{\rm reg}\right\rangle_{\alpha}\nonumber\\[5pt]
\frac{d\vec{L}}{dt}&=&q\,\left\langle\vec{r}_{\rm{insp}}\,\times\,\left[\vec{E}^{\rm reg}+\vec{v}_{\rm{insp}}\times \vec{B}^{\rm reg}\right]\right\rangle_{\alpha}\,,
\end{eqnarray}
where $\left\langle\mathcal{O}(\alpha)\right\rangle_{\alpha}\equiv\frac{1}{2\pi}\int^{2\pi}_0\,\mathcal{O}\,d\alpha$.
Equations \eqref{eq:dEdLavg} are supplemented by an equation for $d\alpha/dt$, which at 0PA is given by
\begin{eqnarray}\label{eq:eqal}
\frac{d\alpha}{dt}=\Upsilon_{N(t),L(t)}=\frac{\mu K^2}{N^3(t)}\,,
\end{eqnarray}
which is the slowly-varying version of \eqref{eq:Ups}. $N(t),\,L(t)$ and $E(t)$ are all related to the orbital elements $p(t),\,e(t)$ via the slowly-varying version of \eqref{eq:pe}, namely
\begin{equation} \label{eq:pe1}
p(t)=\frac{L^2(t)}{K\mu}~~~,~~e(t)=\sqrt{1-\frac{L^2(t)}{N^2(t)}}~~\rightarrow~~N(t)=\sqrt{\frac{\mu K p(t)}{1-e^2(t)}}\,.
\end{equation}
Together, the three equations \eqref{eq:dEdLavg}-\eqref{eq:eqal} constitute a full set of differential equations that determine the time evolution of the inspiral. 

\subsection{Adiabatic radiation as the classical limit of spontaneous emission}
To calculate the right-hand side of \eqref{eq:dEdLavg}, we need the regulated field $A^\mu_{\rm reg}$ generated by an electron in an osculating Keplerian orbit, and evaluated on the very same orbit. By Dirac's famous prescription \cite{Dirac1938} (see \cite{Detweiler:2002mi} for a curved space generalization), this regulated field is given by $A^\mu_{\rm reg}=\frac{1}{2}\left(A^\mu_{\rm ret}-A^\nu_{\rm adv}\right)$, where $A_{\rm adv}$ is the \textit{advanced} gauge field that is obtained using the advanced Green's function. The regular field $A^\mu_{\rm reg}$ has to be evaluated \textit{at the momentary position of the particle}, which at 0PA is simply the osculating orbit itself, without any back-reaction. In fact \cite{Dirac1938}, all we have to do to get the $A^\mu_{\rm reg}$ is start from $A^\mu_{\rm ret}(r)$ and take $h^{(1)}_{l_\gamma}\to j_{l_\gamma}$. All-in-all we have
\begin{eqnarray}\label{eq:Areg}
&&A^\mu_{\rm reg}(t,r)=\frac{iq}{\mu}\,\sum_{l_{\gamma}=0}^\infty\sum_{m_{\gamma}=-l_{\gamma}}^{l_\gamma}\,\sum_{
{\mathbf{\Delta}}}\,e^{-i {\Delta n\, \alpha}}\,\omega_{\Delta n}\,j_{l_{\gamma}}(\omega_{\Delta n} r)\,Y_{l_\gamma}^{m_\gamma*}(\hat{r})\, {}\mathcal{M}^{\mu,N,L}_{l_{\gamma},m_{\gamma},{\mathbf{\Delta}}}{(\omega_{\Delta n})}\,.\nonumber\\
\end{eqnarray}
Here we used~\eqref{eq:masterhydrad}, and we remind that $\mathbf{\Delta}=(\Delta n,\Delta l,\Delta m)$. The resulting electric field is then
\begin{eqnarray}\label{eq:Ereg}
&&\vec{E}_{\rm reg}(t,r)=\partial^t\vec{A}_{\rm reg}(t,r)-\vec{\nabla}A^0_{\rm reg}(t,r)=\frac{q}{\mu}\cdot\sum_{l_{\gamma}=0}^\infty\sum_{m_{\gamma}=-l_{\gamma}}^{l_\gamma}\,\sum_{{\mathbf{\Delta}}}\,e^{-i {\Delta n\, \alpha}}\,\omega_{\Delta n}^{{2}} {}\nonumber\\[5pt]
&&\left[{-}j_{l_{\gamma}}(\omega_{\Delta n} r)\,Y_{l_\gamma}^{m_\gamma{*}}(\hat{r})\vec{\mathcal{M}}^{N,L}_{l_{\gamma},m_{\gamma},{\mathbf{\Delta}}}{+}\frac{i}{\omega_{\Delta n}}\vec{\nabla}\left[j_{l_{\gamma}}(\omega_{\Delta n} r)\,Y_{l_\gamma}^{m_\gamma{*}}(\hat{r})\right]\mathcal{M}^{0,N,L}_{l_{\gamma},m_{\gamma},{\mathbf{\Delta}}}\right]\,.\nonumber\\
\end{eqnarray}
where we used $\dot{\alpha}=\Upsilon_{N,L}$, which is correct for this calculation at 0PA, as well as $\omega_{\Delta n}=\Upsilon_{N,L}\Delta n$. We remind the reader that at 0PA, whenever we write $N,L$ we mean the slowly time dependent $N(t)$ and $L(t)$, with $N(t)$ related to $E(t)$ by \eqref{eq:pe1}. Plugging \eqref{eq:Ereg} in \eqref{eq:dEdLavg}, we get at 0PA
\begin{eqnarray}\label{eq:dE2}
&&\frac{dE}{dt}=\frac{q^2}{ \mu^2}\,\,\lim_{\hbar\rightarrow 0}\sum_{{\mathbf{\Delta}}}\,{}\omega_{\Delta n}^{{2}}\,\int_{0}^{2\pi}\,\frac{d\alpha}{2\pi}\,e^{-i\Delta n {\,\alpha}}\,\sum_{l_{\gamma}=0}^\infty\sum_{m_{\gamma}=-l_{\gamma}}^{l_\gamma}\,\left(\mathcal{O}^{N,L*}_{l_\gamma,m_\gamma}(\alpha)\right)^{\mu}\left(\mathcal{M}^{N,L}_{l_\gamma,m_\gamma,{\mathbf{\Delta}}}\right)_\mu\,.~~~~
\end{eqnarray}
In the last expression we separated the $\mu$ index for readability. Here the variable $\mathcal{O}^{\mu,N,L}_{l_\gamma,m_\gamma}(\alpha)$ is defined by
\begin{eqnarray}
\mathcal{O}^{0,N,L}_{l_\gamma,m_\gamma}(\alpha)&=&-\frac{i}{\omega_{\Delta n}}\vec{\nabla}\left[j_{l_{\gamma}}(\omega_{\Delta n} r_{N,L}(\alpha))\,Y_{l_\gamma}^{m_\gamma{*}}(\hat{r}_{N,L}(\alpha))\right]\cdot \vec{p}_{N,L}(\alpha)\nonumber\\[5pt]
\vec{\mathcal{O}}^{N,L}_{l_\gamma,m_\gamma}(\alpha)&=&\vec{\mathcal{M}}^{N,L}_{l_\gamma,m_\gamma}(\alpha)
\,.
\end{eqnarray}
Similarly to any other operator, $\mathcal{O}^{\mu,N,L}_{l_\gamma,m_\gamma}(\alpha)$ is periodic in $\alpha$ and we can Fourier expand it as
\begin{equation} \label{eq:genQSM2of1}
\mathcal{O}^{\mu,N,L}_{l_{\gamma},m_{\gamma}}\left(\alpha\right)=\sum_{\Delta \widetilde{n}}\,\mathcal{O}^{\mu,N,L}_{l_{\gamma},m_{\gamma},\Delta \widetilde{n}}\exps{-i\Delta \widetilde{n}\alpha}\,.
\end{equation}
{From the QSM, we know that
\begin{eqnarray}\label{eq:QSMO}
&&\mathcal{O}^{\mu,N,L}_{l_{\gamma},m_{\gamma},\Delta \tilde{n}}\,=\,\lim_{\hbar\rightarrow 0}\,\sum_{\Delta \tilde{l},\,\Delta \tilde{m}}\,\mathcal{O}^{\mu,N,L}_{l_{\gamma},m_{\gamma},\tilde{\mathbf{\Delta}}},\nonumber\\&&\mathcal{O}^{\mu,N,L}_{l_{\gamma},m_{\gamma},\tilde{\mathbf{\Delta}}}\equiv\left\langle \tilde{n}{'},\tilde{l}',\tilde{m}'\right|\,\mathcal{O}^{\mu}_{l_{\gamma},m_{\gamma}}\,\left|n,l,l\right\rangle\,,
\end{eqnarray}
where $\tilde{\mathbf{\Delta}}\equiv(\Delta \tilde{n},\Delta \tilde{l},\Delta \tilde{m})$ and $(\tilde{n}{'},\tilde{l}',\tilde{m}')=(n,l,l)-\tilde{\mathbf{\Delta}}$}.
Substituting this expansion into \eqref{eq:dE2} and integrating over $\alpha$, we get a $2\pi\delta_{\Delta n,\Delta \tilde{n}}$ factor. {Moreover, as can be seen from Appendix~\ref{0PA:SE}, the matrix element~\eqref{eq:QSMO} has the same structure as $\mathcal{M}^{\mu,N,L}_{l_{\gamma},m_{\gamma},\mathbf{\Delta}}$ so that in the classical limit, there is no need to sum over $\Delta\tilde{l},\Delta\tilde{m}$ and we can just set them to be $\Delta l,\Delta m$. Overall, this implies that we can set $\tilde{\mathbf{\Delta}}=\mathbf{\Delta}$} and so
\begin{eqnarray}\label{eq:dE2t}
\frac{dE}{dt}&=&\lim_{\hbar\rightarrow 0}\,\frac{q^2}{\mu^2}\,\sum_{{\mathbf{\Delta}}}\,\omega_{\Delta n}^{{2}}\,\left\{\sum_{l_{\gamma}=0}^\infty\sum_{m_{\gamma}=-l_{\gamma}}^{l_\gamma}\,\left(\mathcal{O}^{N,L*}_{l_\gamma,m_\gamma,{\mathbf{\Delta}}}\right)^{\mu}\left(\mathcal{M}^{N,L}_{l_\gamma,m_\gamma,{\mathbf{\Delta}}}\right)_\mu\right\}\nonumber\\[5pt]
&=&-\lim_{\hbar\rightarrow 0}\,\sum_{{\Delta n>0,\Delta l,\Delta m}}\,(E_n-E_{n'})\,\Gamma_{s.e.}\,,
\end{eqnarray}
where {$n'=n-\Delta n$ and we used $\lim_{\hbar\to 0} \frac{E_n-E_{n'}}{\hbar}=\omega_{\Delta n}$. We also defined}
\begin{eqnarray}\label{eq:dE2tG}
\Gamma_{s.e.}\equiv-\frac{{2}q^2\omega_{\Delta n}}{\hbar\mu^2}\sum_{l_{\gamma}=0}^\infty\sum_{m_{\gamma}=-l_{\gamma}}^{l_\gamma}\,\left(\mathcal{O}^{N,L*}_{l_\gamma,m_\gamma,{\mathbf{\Delta}}}\right)^{\mu}\left(\mathcal{M}^{N,L}_{l_\gamma,m_\gamma,{\mathbf{\Delta}}}\right)_\mu
\,.
\end{eqnarray}
In Appendix~\ref{0PA:SE}, we prove two key results regarding this expression. The first result is
\begin{eqnarray}\label{eq:dE2tGt}
\Gamma_{s.e.}=-\frac{{2}q^2\omega_{\Delta n}}{\hbar\mu^2}\sum_{l_{\gamma}=0}^\infty\sum_{m_{\gamma}=-l_{\gamma}}^{l_\gamma}\,\left(\mathcal{M}^{N,L*}_{l_\gamma,m_\gamma,{\mathbf{\Delta}}}\right)^{\mu}\left(\mathcal{M}^{N,L}_{l_\gamma,m_\gamma,{\mathbf{\Delta}}}\right)_\mu+\mathcal{O}(\hbar^0)
\,.
\end{eqnarray}
The second result is
\begin{eqnarray} \label{SE3main}
\Gamma_{s.e.}=\frac{q^{2}\left(E_{n}-E_{n'}\right)}{8\pi^2\hbar^2\mu^{2}}\int d^{2}\Omega_{k}\,\left.\left\{ \sum_{\sigma}\left|\left\langle n',l',m'\right|e^{-i\vec{k}\cdot\vec{x}}\left(\vec{\varepsilon}^{~\sigma}_{\vec{k}}\cdot\vec{p}\right)\left|n,l,m\right\rangle \right|^{2}\right\} \right|_{k=\left(E_{n}-E_{n'}\right)/\hbar}\,.
\end{eqnarray}
where $\vec{\varepsilon}^{~\sigma}_{\vec{k}}$ are the usual EM transverse polarization vectors, $\vec{x}$ is the position operator, {$\vec{k}$ is the wave-vector of the emitted photon, and $k=\left|\vec{k}\right|$}. This is simply the expression for the rate for spontaneous emission in the quantum hydrogen atom, as calculated via Fermi's golden rule.
We thus see that the 0PA energy loss rate from a classical electron is simply the $\hbar\rightarrow0$ limit of the expression for the energy loss rate due to \textit{spontaneous emission} in the quantum hydrogen atom.

Though we will not present this here explicitly, one can start from the second equation in \eqref{eq:dEdLavg} and follow similar steps to arrive at 
\begin{eqnarray} \label{SE1main}
\frac{dL}{dt}&=&-\lim_{\hbar\rightarrow 0}\,\sum_{\Delta n>0{,\Delta l,\Delta m}}\,\hbar\,\left(l-l'\right)\,\Gamma_{s.e.}\,,
\end{eqnarray}
or, in other words, the rate of 0PA \textit{angular momentum loss} from a classical electron is the $\hbar\rightarrow0$ limit of the angular momentum loss due to quantum spontaneous emission.
\subsection{0PA Trajectory and the Emitted EM Waveform}
\begin{figure}[ht!]
\begin{center}
\includegraphics[width=0.6\linewidth]{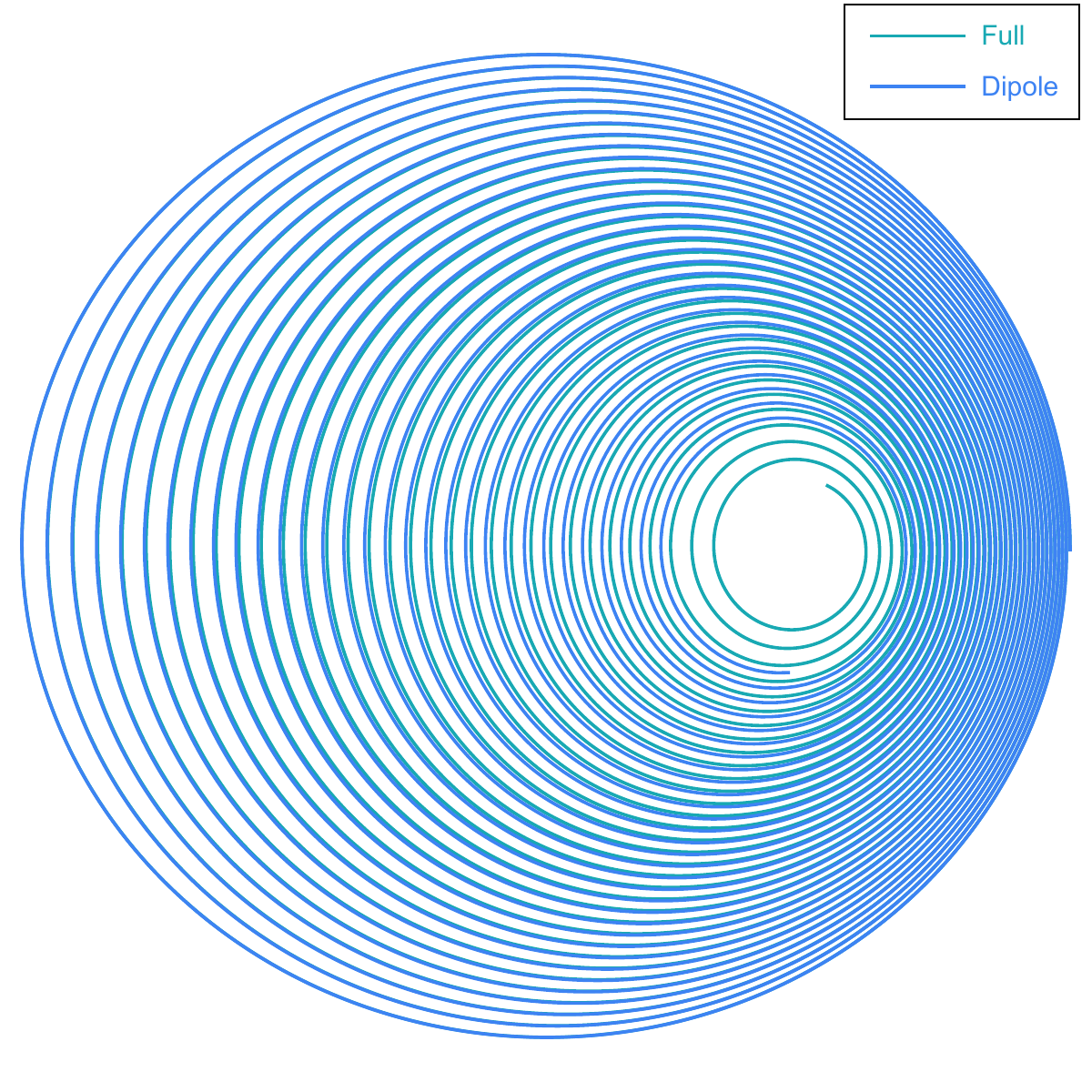}
\caption{Full versus dipole inspiral. This figure shows the adiabatic (0PA) trajectory (green). The dipole trajectory (blue), obtained by retaining only the dipole-order contributions in \eqref{eq:dE2tGt}, is also shown. The initial values of the orbital parameters are $(e,\Bar{p})=(0.5,300)$. Moreover, in this figure alone, we took a particularly small value of $Z$ to make the inspiral faster and the evolution more discernible. Both trajectories start together from the same position, and run for the same duration. As seen in this figure, the full trajectory spirals deeper than the dipole one.}\label{fig:InspTraj}
\end{center}
\end{figure}
To obtain the full 0PA trajectory, we can now integrate~\eqref{eq:dE2t}, \eqref{SE1main} and \eqref{eq:eqal} numerically to extract $(e(t),p(t),\alpha(t))$ -- see also \eqref{eq:pe1} for the relations between $e(t),\,p(t),\,E(t),\,L(t)$ and $N(t)$. The distance from the center and azimuthal angle are then obtained by plugging $(e(t),p(t),\alpha(t))$ in \eqref{eq:rfourt1} and \eqref{eq:balpha}, which are still valid at 0PA order.

\begin{figure}[ht!]
     \centering
     \begin{subfigure}[T]{0.35\textwidth}
         \centering
         \includegraphics[width=0.98\textwidth]{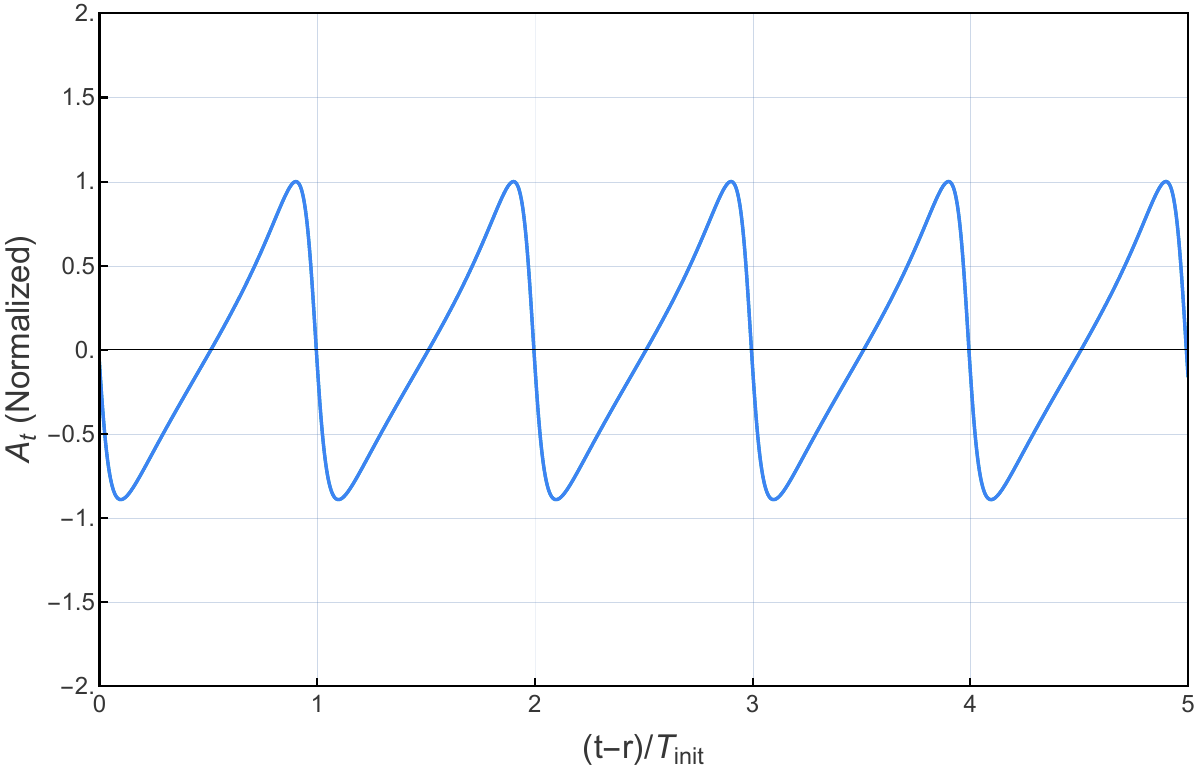}         
         \caption{The (normalized) waveform produced during the first 5 periods in the inspiral. The full (green) waveform does not appear because it coincides with the dipole one in this duration.}\label{fig:inspWF1}
     \end{subfigure}
     \hspace{20pt}
     \begin{subfigure}[T]{0.35\textwidth}
        \centering 
        \includegraphics[width=\textwidth]{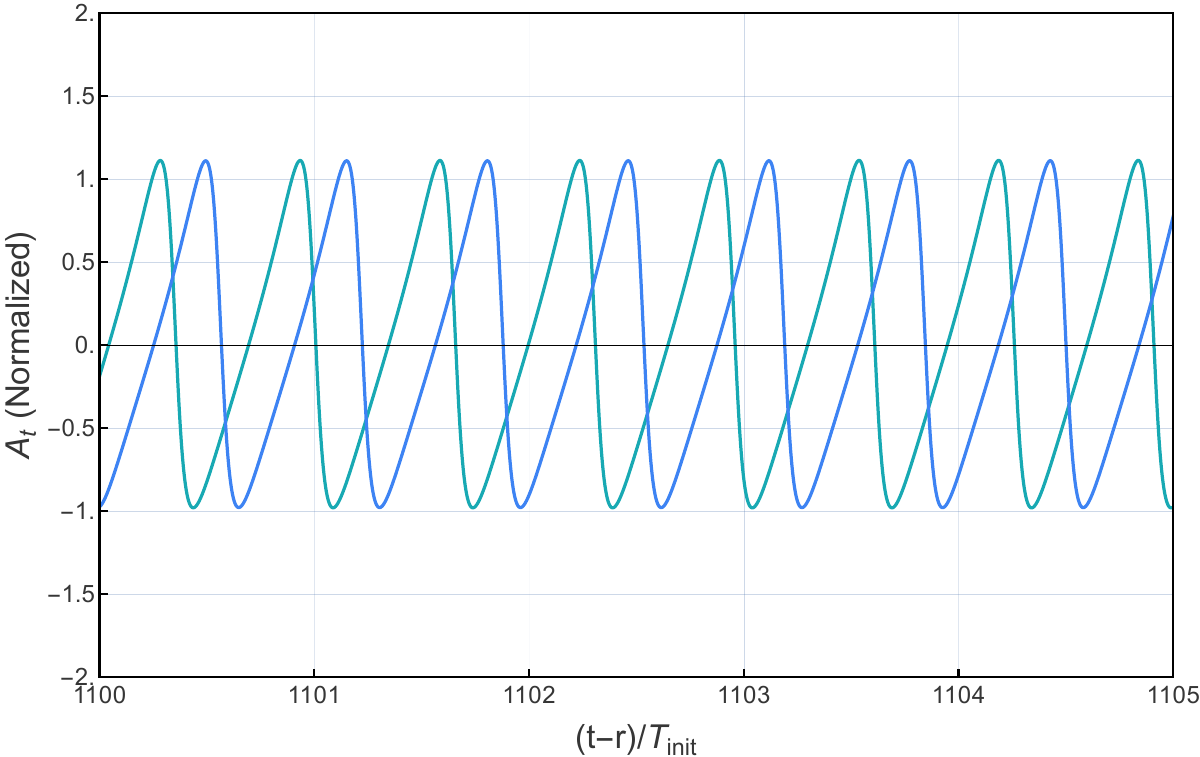}
         \caption{The (normalized) waveform produced from $t=1100\, 
 T_{\rm{init}}$ to $t=1105\, T_{\rm{init}}$.}\label{fig:inspWF2}
     \end{subfigure}
     \vspace{30pt}
     \begin{subfigure}{\textwidth}
         \centering
         \includegraphics[width=0.7\textwidth]{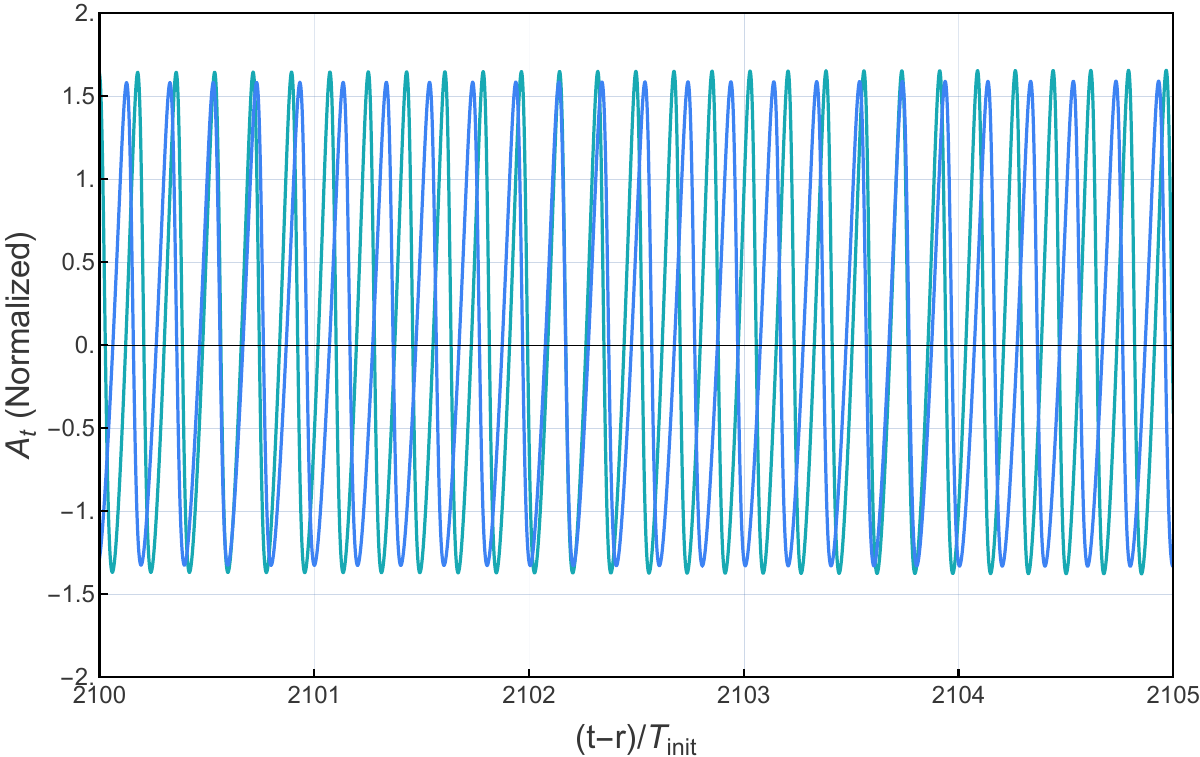}    
         \caption{The (normalized) waveform produced from $t=2100\, 
 T_{\rm{init}}$ to $t=2105\, T_{\rm{init}}$.}\label{fig:inspWF3}
     \end{subfigure}
        \caption{Full versus dipole waveform. This figure shows (in green) the adiabatic (0PA) waveform -- $A_{t}$ -- radiated by the electron for $Z=4\pi$. The dipole waveform (blue), obtained by retaining only the dipole-order contributions in \eqref{eq:dE2tGt}, is also shown. The observation point is on the x-axis, far away from the electron's inspiral. 
         The initial orbital parameters of the electron's trajectory are given by $(e_0,\Bar{p}_0)=(0.5,300)$.
         The horizontal axis denotes the retarded time, $t-r$, normalized by the period of the initial Keplerian orbit, $T_{\rm{init}}$. The vertical axis is normalized by the maximum of the initial Keplerian waveform.}\label{fig:three graphs}
\end{figure}

Once the inspiralling trajectory is known, the associated waveform can be extracted using the expression~\eqref{eq:GreenretMulti3}. This time, we cannot use the simplified expression~\eqref{eq:greencoh3}, which is only valid for a Keplerian source. In practice,~\eqref{eq:GreenretMulti3} is cumbersome to evaluate, and so, as is the case for gravitational inspirals, one usually resorts to a far-field expansion for waveform generation. Starting from~\eqref{eq:GreenretMulti3}, we take the far-field limit where $h^{(1)}_{l_\gamma}(\omega r)$ attains its asymptotic value $(-i)^{l_\gamma+1} e^{i\omega r}/(\omega r)$. Furthermore, in the far-field regime, all time and distance scales are hierarchically larger than the inspiral period, and so we can use the stationary phase (saddle point) approximation for the exponential in~\eqref{eq:GreenretMulti3}. The stationary phase condition over $
\omega$ and $t'$ gives $t'=t-r\equiv t_{\rm{ret}}$  and $\omega=\dot{\alpha}'\equiv\Delta n \Upsilon_{N,L}(t')$.
Finally, we have 
\begin{eqnarray}~\label{eq:GreenretMulti2t}
A^\mu_{\rm ret}(t,\vec{x})&=&\frac{q}{\mu r}\,\sum_{{\mathbf{\Delta}}}\,\sum_{l_{\gamma}=0}^\infty\,\sum_{m_{\gamma}=-l_{\gamma}}^{l_{\gamma}}\,e^{-i\Delta n\, \alpha(t_{\rm{ret}})}\,(-i)^{l_{\gamma}+1}\,Y_{l_{\gamma}}^{m_{\gamma}*}(\theta,\varphi)\,\mathcal{M}^{\mu,N,L}_{l_{\gamma},m_{\gamma},{\mathbf{\Delta}}}{(\Delta n \Upsilon_{N,L}(t_{\rm{ret}}))}\,.\nonumber\\
\end{eqnarray}
This far-field approximation to the emitted EM field is easy to evaluate. A sample inspiralling trajectory is presented in Fig.~\ref{fig:InspTraj} (with a particularly small value of $Z$ to make the inspiral faster and the evolution more discernible), and a typical waveform is shown in Fig. \ref{fig:three graphs}. The corresponding dipole-order inspiral and waveforms, obtained by retaining only the dipole ($l_\gamma=1$) contributions in \eqref{eq:dE2tGt}, are also shown. At the beginning of the inspiral, the dipole-approximation is valid, and as such the trajectories (Fig.~\ref{fig:InspTraj}) and waveforms (Fig.~\ref{fig:inspWF1}) match. As the inspiral continues, the dipole-approximation breaks down, and a phase mismatch develops (Fig.~\ref{fig:inspWF2}). Finally, at later times, a frequency and amplitude mismatch also develops (Fig.~\ref{fig:inspWF3}). This outlines the importance of higher multipoles for the accurate description of inspiral dynamics.
\newpage
\newpage
\section{The QSM for Relativistic Motion in Schwarzschild}\label{sec:Schwarz}
So far, our application of the QSM was strictly in the context of \textit{non-relativistic} motion in flat space, under the influence of a spherically symmetric potential. This, though, was a mere convenience, since all we need for the QSM is a field equation whose WKB solutions involve the Hamilton-Jacobi action of the classical system. Remarkably, this line-of-thought works seamlessly for relativistic motion as well, so long as we consider a Klein-Gordon Hamiltonian instead of the non-relativistic \eqref{eq:Ham}. In this section we do not \textit{apply} the QSM for relativistic motion in a Schwarzschild background (we leave this for future work). Instead, we show how to generalize the WKB proof of the QSM in Section~\ref{sec:QSM} to the Schwarzschild case. We begin by reviewing the classical Hamilton-Jacobi theory for Schwarzschild, as developed in \cite{Carter:1968ks,Fujita2009,VanDeMeent2018}.

\subsection{Hamilton-Jacobi for Schwarzschild}
The seminal work \cite{Carter:1968ks} showed that classical point particle motion in curved space can be obtained from the  Hamiltonian\footnote{In \cite{Carter:1968ks}, time is normalized to be dimensionless. Here we keep time in its natural units, and so we include factors of $1/\mu$ in the Hamiltonian.}
\begin{eqnarray}\label{eq:hamsch}
H=\frac{1}{2\mu}g^{\mu\nu}p_\mu p_\nu\,,
\end{eqnarray}
where the $p_\mu$ are the relativistic conjugate momenta{, $\mu$ is the particle's mass,} and $g^{\mu\nu}$ is the metric. In our case, $g^{\mu\nu}$ is the Schwarzschild metric given by
\begin{eqnarray}
g_{\mu\nu}=\colmatf{-\frac{\Delta(r)}{r^2}&0&0&0\\0&\frac{r^2}{\Delta(r)}&0&0\\0&0&r^2&0\\0&0&0&r^2\sin\theta}\,,
\end{eqnarray}
where $\Delta(r)=r\,(r-2GM)$ and $\mu,\nu\in\{t,r,\theta,\varphi\}$. The conjugate momenta are related to derivatives of the coordinates $x^\mu=(t,r,\theta,\varphi)$ with respect to proper time $\tau$, 
\begin{eqnarray}\label{eq:pssch}
p_{\mu}=\mu g_{\mu\nu}\dot{x}^{\nu}\,.
\end{eqnarray}
The Hamiltonian \eqref{eq:hamsch} is clearly conserved, and, in fact, for a relativistic particle it equals $-\mu/2$ where $\mu$ is the rest mass of the particle. Furthermore, it allows us to write down the relativistic version of the Hamilton-Jacobi equation,
\begin{eqnarray}\label{eq:SHJSch}
-\mu^2=g^{\mu\nu}\frac{\partial S_{HJ}}{\partial x^\mu}\frac{\partial S_{HJ}}{\partial x^\nu}\,,
\end{eqnarray}
where
\begin{eqnarray}
S_{HJ}(t,r,\theta,\varphi)=-Et+S_r(r)+S_\theta(\theta)+L_z\varphi\,,
\end{eqnarray}
is the relativistic Hamilton-Jacobi action. Here we already used the fact that the Hamilton-Jacobi equation is separable in Schwarzschild, and also that $L_z$ is conserved by the cyclicity of $\varphi$. $S_\theta(\theta)$ is the same as in the flat-space case \eqref{eq:radialangular}, while the radial action is
\begin{eqnarray}
S^r(r)&=&\pm\int_{r_{\rm min}}^r\,\frac{\sqrt{U^r(r')}}{\Delta(r')}\,dr'\nonumber\\[5pt]
U^r(r)&=&r^4\left[E^2-\frac{\Delta(r)}{r^2}\left(\frac{L^2}{r^2}+\mu^2\right)\right]\,.
\end{eqnarray}
Once again, the $+$ sign should be used to compute observables in the $r_{\rm min}\rightarrow r_{\rm max}$ half of each radial cycle, while the $-$ sign should be used in the $r_{\rm max}\rightarrow r_{\rm min}$ half of each cycle. For $E^2<\mu^2$, we have bound geodesics librating between $r_{\rm min}$ and $r_{\rm max}$, two consecutive real roots of $\Delta^2(r)U^r(r)$. The EOM derived from the Hamilton-Jacobi action are
\begin{eqnarray}
p_i=\frac{\partial S_{HJ}}{\partial q_i}~~,~~i\in\{t,r,\varphi\}\,,
\end{eqnarray}
or, using \eqref{eq:pssch},
\begin{eqnarray}
\dot{t}=\frac{E r^2}{\mu\Delta(r)}~~,~~\dot{r}=\pm\frac{1}{\mu r^2}\sqrt{U^r(r)}~~,~~\dot{\varphi}=\frac{L}{\mu r^2}\,,
\end{eqnarray}
where the dot is a derivative with respect to proper time $\tau$. Similarly to the flat-space case, We can define the action variables as usual as 
\begin{eqnarray}
J_i\equiv\frac{1}{2\pi}\oint\,dq_i\,p_i~~~,~~~i\in\{r,\varphi\}\,.
\end{eqnarray}
By spherical symmetry, $J_\varphi=L$ (without loss of generality the motion is in the XY plane and $L=L_z$). Furthermore
\begin{eqnarray}\label{eq:aa}
J_r=\frac{1}{\pi}\,S^r(r_{\rm max})\,,
\end{eqnarray}
where the $+$ sign is taken for $S^r(r_{\rm max})$. For the sake of defining fundamental frequencies, we follow the common practice of defining ``Mino" time \cite{Mino:2003yg} $\lambda$, which in the Schwarzschild case is just\footnote{Conventionally Mino time is defined so that $d\lambda=r^{-2}d\tau$ but then $\lambda$ does not have units of time. To remedy this, we rescaled the conventional definition by $(L^2/\mu^2)$ so that $d\lambda=(L^2/\mu^2)r^{-2}d\tau$ and $\lambda$ has units of time. Since $d\varphi=(L/\mu)r^{-2}d\tau$, we can simply define the Mino time to be $\lambda\equiv(L/\mu)\varphi$.} $\lambda=\varphi L/\mu$. The radial and azimuthal fundamental frequencies (with respect to Mino time) are then \cite{Fujita2009}
\begin{eqnarray}\label{eq:fundrsch}
\Upsilon^r_{Jr,L}&=&\frac{2\pi}{T^r_{Jr,L}}~~,~~\Upsilon^\varphi_{Jr,L}=\frac{\varphi}{\lambda}=\frac{\mu}{L}\,,
\end{eqnarray}
and the radial time period (in Mino time) is
\begin{eqnarray}
T^r_{Jr,L}=2\int_{\lambda(r_{\rm min})}^{\lambda(r_{\rm max})}\,d\lambda=\frac{2L^2}{\mu^2}\int_{\tau(r_{\rm min})}^{\tau(r_{\rm max})}\,\frac{1}{r^2(\tau)}\,d\tau=\frac{2L^2}{\mu^2}\int_{r_{\rm min}}^{r_{\rm max}}\,\frac{\mu}{\sqrt{U^r(r)}}dr\,.
\end{eqnarray}
On the other hand, the \textit{dimensionless} $\Upsilon^t_{Jr,L}$ is given by
\begin{eqnarray}\label{eq:fundtsch}
\Upsilon^t_{Jr,L}&=&\Upsilon^r_{Jr,L}~\frac{t(r_{\rm max})-t(r_{\rm min})}{\pi}\,,
\end{eqnarray}
where
\begin{eqnarray}\label{eq:fundtsch2}
t(r_{\rm max})-t(r_{\rm min})=\int_{t(r_{\rm min})}^{t(r_{\rm max})}\,dt=\int_{r_{\rm min}}^{r_{\rm max}}\,\frac{Er^4}{\Delta(r)\sqrt{U^r(r)}}dr\,.
\end{eqnarray}
It is common and useful in the literature to define \text{Boyer-Lindquist} frequencies as 
\begin{eqnarray}\label{eq:BLfreq}
\Omega^i_{J_r,L}\equiv \frac{\Upsilon^i_{J_r,L}}{\Upsilon^t_{J_r,L}}~~,~~i\in\{r,\varphi\}\,.
\end{eqnarray}
The latter are the temporal frequencies as measured by a distant observer \cite{Fujita2009,Pound2021}. Correspondingly, we can define the Boyer-Lindquist angle variables
\begin{eqnarray}\label{eq:angsch}
\tilde{\alpha}^i=\Omega^i_{Jr,L}t~~,~~i\in\{r,\varphi\}\,.
\end{eqnarray}
Though these are not true angle variables since they don't grow linearly with proper time, they are nevertheless the natural angle variables for the double-Fourier expansion of time-dependent observers, namely
\begin{equation} \label{app:genQSM2sch}
\mathcal{O}_{J_r,L}(t)=\mathcal{O}_{J_r,L}\left[\tilde{\alpha}^r(t),\tilde{\alpha}^\varphi(t)\right]=\sum_{\Delta j_r,\,\Delta l}\,\mathcal{O}^{J_r,L}_{\Delta j_r,\Delta l}\exp\left\{-i\Delta j_r\tilde{\alpha}^r(t)-i\Delta l\tilde{\alpha}^\varphi(t)\right\}\,.
\end{equation}
The Fourier coefficients $\mathcal{O}^{J_r,L}_{\Delta j_r,\Delta l}$ are related as usual to $\mathcal{O}_{J_r,L}\left[\tilde{\alpha}^r,\tilde{\alpha}^\varphi\right]$ via
\begin{equation} \label{eq:genQSM2ttsch}
\mathcal{O}^{J_r,L}_{\Delta j_r,\Delta l}=\int\,\frac{d\tilde{\alpha}^r}{2\pi}\,\int\,\frac{d\tilde{\alpha}^\varphi}{2\pi}\,\mathcal{O}_{J_r,L}\left[\tilde{\alpha}^r,\tilde{\alpha}^\varphi\right]\,\exps{i\Delta j_r\tilde{\alpha}^r+i\Delta l\tilde{\alpha}^\varphi}\,.
\end{equation}
\subsection{WKB Proof of the QSM for Schwarzschild}
The essential ingredient of the WKB proof in Section~\ref{sec:QSM} was the relation between the WKB eigenfunctions \eqref{eq:RWKBp} of the Hamiltonian \eqref{eq:Ham} and the Hamilton-Jacobi radial function $S^r_{j_r,l}(r)$. A similar relation persists for the Hamiltonian \eqref{eq:hamsch}. This time, we are interested in $H$ acting not on an eigenfunction but rather on a \textit{field} $\Phi$. The constraint $H=-\mu/2$ then gives
\begin{eqnarray}\label{eq:KG1}
H\Phi&=&-\frac{\mu}{2}\Phi\,.
\end{eqnarray}
Importantly, $\Phi$ here does not have the interpretation of a wavefunction, but rather of a field. This does not matter for us, since all we need is for the WKB solution for $\Phi$ to be related to $S^r(r)$. Quantum mechanically, we interpret $p_\mu$ as $-i\hbar\partial_\mu$, and so \eqref{eq:KG1} gives us the massive \textit{Klein-Gordon} equation in a Schwarzschild background,
\begin{eqnarray}\label{eq:KGPhi}
\left(g^{\mu\nu}\partial_\mu \partial_\nu-\frac{\mu^2}{\hbar^2}\right) \Phi=0\,.
\end{eqnarray}
Separation of variables in a spherically symmetric problem naturally gives us
\begin{eqnarray}\label{eq:phsch}
\Phi_{j_r,l,m}=(-1)^{l}\,e^{i \hbar^{-1} E_{j_r,l} t}\,R_{j_r,l}(r)\,Y_{l}^m(\theta,\varphi)\,.
\end{eqnarray}
The attentive reader may be worried that the solution to the Klein-Gordon equation could be unstable in the presence of the black hole horizon. Furthermore, one could worry that there may not be stable solutions. We are happy to dispel both of these worries for the following reasons:
\begin{enumerate}
\item For us, the horizon is always behind an angular momentum barrier. The effect of tunnelling into the horizon always vanishes in the classical limit that we are interested in. For this reason, all of our bound states are strictly stable in the $\hbar\rightarrow 0$ limit.
\item Contrary to the case of \textit{massless} excitations of Schwarzschild/Kerr, which only have a bound orbit at the light ring, our Klein-Gordon is \textit{massive}. The mass leads to a finite potential well that gives rise to honest bound states.
\end{enumerate}

The leading WKB solution for $R_{j_r,l}$ in \eqref{eq:phsch} is
\begin{eqnarray}
R_{j_r,l}(r)=\sqrt{\frac{2E_{j_r,l}\,\Omega^r_{j_r,l}}{\pi}}\,\frac{1}{[U^r_{j_r,l}(r)]^{1/4}}\,\sin\left(\frac{1}{\hbar}S^r_{j_r,l}(r)+\frac{\pi}{4}\right)\,,
\end{eqnarray}
which is similar to \eqref{eq:RWKBp}, this time with
\begin{eqnarray}\label{eq:Sschw}
S^r_{j_r,l}(r)&=&\int_{r_{\rm min}}^r\,\frac{\sqrt{U^r_{j_r,l}(r')}}{\Delta(r')}\,dr'\nonumber\\[5pt]
U^r_{j_r,l}(r)&=&r^4\left[E^2_{j_r,l}-\frac{\Delta(r)}{r^2}\left(\frac{\hbar^2 l(l+1)}{r^2}+\mu^2\right)\right]\,.
\end{eqnarray}
Repeating the arguments of Section~\ref{sec:QSM}, we have
\begin{eqnarray}
\lim_{\hbar\rightarrow 0}\,\int_{r_{\rm min}}^{r_{\rm max}}\,dr\,\frac{r^4}{\Delta(r)}\,R^*_{j_r,l}(r)R_{j_r,l}(r)\,=\,\lim_{\hbar\rightarrow 0}\,\Omega^r_{j_r,l}\,\frac{1}{\pi}\int_{r_{\rm min}}^{r_{\rm max}}\,dr\,\,\frac{E_{j_r,l}r^4}{\Delta(r)\sqrt{U^r_{j_r,l}(r)}}=1\,.
\end{eqnarray}
Here the $r^4/\Delta(r)=g_{rr}g_{\theta\theta}g_{\varphi\varphi}$ is the (non-invariant) 3D volume element in Schwarzschild (remember that $\theta=\pi/2$). Hence, indeed, $R_{j_r,l}$ is correctly normalized.
The analog of \eqref{eq:masterfreq} for the Schwarzschild case is
\begin{eqnarray}\label{eq:masterfreqsch}
\Omega^r_{J_r,L}&=&\lim_{\hbar\rightarrow 0}\frac{E_{j_r,l}-E_{j_r-\Delta j_r,l}}{\hbar\Delta j_r}\nonumber\\[5pt]
\Omega^\varphi_{J_r,L}&=&\,\lim_{\hbar\rightarrow 0}\frac{E_{j_r,l}-E_{j_r,l-\Delta l}}{\hbar\Delta l}\,.
\end{eqnarray}
Eq.~\eqref{eq:masterfreqsch} is proven by a straightforward generalization of Section~\ref{sec:QtoC}, which we omit here for brevity. Finally, we can adapt the WKB proof of Section~\ref{sec:QSM} to the present case. Defining as usual
\begin{eqnarray}
\mathcal{O}(r,\theta,\varphi)=\sum_{l_\gamma=0}^\infty\,\sum_{m_\gamma=-l_\gamma}^{l_\gamma}\,\mathcal{O}_{l_\gamma,m_\gamma}(r)\,Y_{l_\gamma}^{m_\gamma}\left(\theta,\varphi\right)\,,
\end{eqnarray}
the relevant classical matrix element is
\begin{eqnarray}\label{eq:MEsch}
&&\,\lim_{\hbar\rightarrow 0}\,\sum_{\Delta m}\,\left\langle j_r-\Delta j_r,l-\Delta l,l-\Delta m\right|\,\mathcal{O}\,\left|j_r,l,l\right\rangle=\lim_{\hbar\rightarrow 0}\,\sum_{\Delta m}\,\sum_{l_\gamma,m_\gamma}\delta_{\Delta l,\Delta m}\delta_{-\Delta l,m_\gamma}\nonumber\\[5pt]
&&\,\frac{\Omega^r_{{{j_r,l}}}}{\pi}\int_{r_{\rm min}}^{r_{\rm max}}\,\frac{E_{{{j_r,l}}}r^4}{\Delta(r)\sqrt{U^r_{{{j_r,l}}}(r)}}\,\mathcal{O}_{l_\gamma,m_\gamma}(r)\,Y_{l_\gamma}^{m_{\gamma}}\left(\pi/2,0\right)\,\cos\left(\frac{S^r_{j_r,l}(r)-S^r_{j_r-\Delta j_r,l-\Delta l}(r)}{\hbar}\right)dr\,.\nonumber\\
\end{eqnarray}
Repeating the steps in Section~\ref{sec:Iint}, this time for $S^r_{j_r,l}$ given by \eqref{eq:Sschw}, we have
\begin{eqnarray}
&&\,\lim_{\hbar\rightarrow 0}\,\sum_{\Delta m}\,\left\langle j_r-\Delta j_r,l-\Delta l,l-\Delta m\right|\,\mathcal{O}\,\left|j_r,l,l\right\rangle=\nonumber\\[5pt]
&&\int_{0}^{2\pi} \frac{d\tilde{\alpha}^r}{2\pi}\,\int_{0}^{2\pi} \frac{d\tilde{\alpha}^\varphi}{2\pi}\mathcal{O}_{J_r,L}[r(\tilde{\alpha}^r),\pi/2,\varphi(\tilde{\alpha}^r,\tilde{\alpha}^\varphi)]\,\exps{i\Delta j_r \tilde{\alpha}^r+i\Delta l\,\tilde{\alpha}^\varphi}
=\mathcal{O}^{J_r,L}_{\Delta j_r,\Delta l}\,,\nonumber\\
\end{eqnarray}
c.f. the definition \eqref{eq:genQSM2ttsch} of the Fourier coefficients in the double Fourier expansion in Boyer-Lindquist angles. This proves the QSM for motion in a Schwarzschild background.
\section{Future Applications of the QSM}\label{sec:future}
The main future application of the QSM would be to calculate the GSF analytically. A standard practice in the field when introducing new computational methods is to first demonstrate them in the simpler case of a \textit{scalar} self-force in a black-hole background \cite{Barack2023,Bini:2024icd,Long:2024ltn}. This is particularly useful when the previewed method is not inherently related to GR, as is the case here. By a scalar self-force we mean that the inspiralling body is charged under a new, massless scalar. As the body moves in the BH background, it emits a scalar field, much like the emission of the EM field in Section~\ref{sec:EM}. In turn, the emitted scalar generates radiation-reaction on the inspiralling mass, the scalar, curved space analog of \eqref{eq:dEdL}. In this section, we will not explicitly calculate the scalar self-force -- we leave that for future work. We will, however, outline the required steps to calculate it and by analogy 1st order GSF in Schwarzschild. 

The Green's function for a massless scalar in Schwarzschild is the solution to the inhomogeneous Klein-Gordon equation
\begin{eqnarray}
g^{\mu\nu}\partial_\mu \partial_\nu G^\phi_{\rm ret}(x,x')=\frac{1}{\sqrt{g}}\delta^{(4)}(x-x')\,,
\end{eqnarray}
where $g$ is the Schwarzschild metric. It is given by \cite{Bini2022a}
\begin{eqnarray}\label{eq:GFschphi}
&&G^{\phi}_{\rm ret}(x,x')=\frac{\Theta(t-t')}{2\pi}\,\int_{-\infty}^{\infty}\,d\omega \,e^{-i\omega (t-t')}\,\left\{i\omega\,\sum_{l_\phi=0}^{\infty}\frac{1}{W}\,R^{\rm{in}}_{l_\phi}(\omega r_<)R^{\rm{up}}_{l_\phi}(\omega r_>)\right.\nonumber\\[5pt]
&&\left.\,\sum_{m_\phi=-l_{\phi}}^{l_{\phi}}\,Y^{m_\phi*}_{l_\phi}(\theta',\varphi')Y_{l_\phi}^{m_\phi}(\theta,\varphi)\right\}\,,
\end{eqnarray}
where $\{r_<,r_>\}=\{{\rm min}(r,r'),{\rm max}(r,r')\}$, and we will focus on the case where $r'<r$. Here $R^{\rm{in}}_{l_\phi }$ ($R^{\rm{up}}_{l_\phi}$) are the solutions of the Schwarzschild-background Klein-Gordon equation with ''in" ("up")-type boundary conditions, and $W$ is their (constant) Wronskian
\begin{eqnarray}\label{eq:Wronskian}
W=\Delta\,\left(R^{\rm{in}}_{l_\phi }\partial_rR^{\rm{up}}_{l_\phi }-R^{\rm{up}}_{l_\phi }\partial_rR^{\rm{in}}_{l_\phi }\right)\,.
\end{eqnarray}
We will not provide explicit expressions for $R^{\rm{in}}_{ l_\phi }$ and $R^{\rm{up}}_{l_\phi }$, but merely comment that they are given by confluent Heun functions \cite{Bini2022a}, and can be evaluated as a sum of hypergeometric functions \cite{MST}. This sum truncates at finite orders in the PM expansion. 
Using the Green's function \eqref{eq:GFschphi}, we can express the retarded scalar field emitted by a body on a Schwarzschild geodesic as
\begin{eqnarray}\label{eq:fieldphi}
&&\phi_{\rm ret}(x)=\int d^4x'\,G^{\phi}_{\rm ret}(t,\vec{x};t',\vec{x}{'})\,\delta^{(3)}(\vec{x}'-\vec{x}_{J_r,L}(t'))\,,
\end{eqnarray}
where $\vec{x}_{J_r,L}(\tau)$ is the position of the body on the massive Schwarzschild geodesic parametrized by the conserved action variables $J_r$ and $L$, as a function of \textit{coordinate} time. In analogy with Section~\ref{sec:EM}, we need to calculate a matrix element associated with the source on the Schwarzschild geodesic, namely
\begin{eqnarray}\label{eq:MEsch1}
{\mathcal{M}^{J_r,L}_{l_{\phi},m_{\phi}}(t')}=\,R^{\rm{in}}_{l_\phi}(\omega r_{J_r,L}(t'))\,Y^{m_\phi{*}}_{l_\phi}(\hat{r}_{J_r,L}(t'))\,,
\end{eqnarray}
which is a particular case of an observable of the kind of $\mathcal{O}_{J_r,L}(t')$ defined in \eqref{app:genQSM2sch}. Consequently, to apply the QSM in this case, we need to compute
\begin{eqnarray}\label{eq:MEsch2}
&&\,\lim_{\hbar\rightarrow 0}\,\sum_{\Delta m}\,\left\langle j_r-\Delta j_r,l-\Delta l,l-\Delta m\right|\,{\mathcal{M}_{l_{\phi},m_{\phi}}}\,\left|j_r,l,l\right\rangle\,,
\end{eqnarray}
where the $\left|j_r,l,l\right\rangle$ are the full quantum solutions of \eqref{eq:KGPhi}, and ${\mathcal{M}_{l_{\phi},m_{\phi}}}$ is the operator version of~\eqref{eq:MEsch1} . As in the EM case, a WKB approximation does not help when \textit{applying} the QSM, only in \textit{proving} it. So we really need to fully compute \eqref{eq:MEsch2}. While the angular part of this integral is trivial, the radial part is essentially a triple integral over Heun functions:
\begin{eqnarray}\label{eq:Isch}
I=\int_{r_{sch}}^\infty dr\,\frac{r^4}{\Delta}\,R^{\rm{in}}_{l_\phi}(\omega r)R^{\rm{bound}*}_{j_r-\Delta j_r,l-\Delta l}(r)R^{\rm{bound}}_{j_r,l}(r)\,dr\,,
\end{eqnarray}
where $R^{\rm{bound}}_{j_rl}(r)$ are bound state radial functions for the massive Klein-Gordon equation \eqref{eq:KGPhi}, which are also Heun functions. Carrying this integral to all orders in $G$ seems like a daunting task, and indeed it would require a special function identity that we currently do not know. Nevertheless, the MST method \cite{MST} allows us to represent each one of the functions in this integral as a finite sum over hydrogenic radial functions, at any given PM order. Essentially, the problem would then reduce to a finite PM sum over hydrogen matrix elements of the form \eqref{eq:masterhyd}, and there's no particular complication in taking it to as high-order in $G$ as we see fit. We leave this explicit calculation for future work.

Once $\phi_{\rm ret}(x)$ is computed, we would be able to extract its regular part in the $x\rightarrow x_{\rm geodesic}$ limit. This can done using the conventional mode-sum \cite{Barack:1999wf} or puncture methods (see e.g. \cite{Pound2021}), and is orthogonal to our QSM calculation. Finally, we reiterate that the QSM can be used in a very similar manner in the full 1st order GSF case, albeit with slightly more elaborate expressions. Nevertheless, the heart of the computation in the full GSF case would also be an integral similar to \eqref{eq:Isch}. The analytical result for 1st order GSF naturally fits into a 0PA calculation of a gravitational inspiral, the gravity analog of Section~\ref{sec:ESF}.

Finally, note that the reward in performing this integral is quite significant -- as it can be PM expanded and compared to the impressive multi-loop calculations of the PM program (see, e.g. \cite{Damour:2016abl,Bern:2019crd,Bern:2019nnu,Damour:2019lcq,Bern2022,Bini2022,Bern2022a,Dlapa2023a,Kaelin2023,Jakobsen2023,Damour2023,Dlapa2024,Driesse2024}), as was recently done successfully for the unbound case in \cite{Bini:2024icd,Long:2024ltn}.

\section{Conclusions}\label{sec:conc}

In this paper we presented the Quantum Spectral Method (QSM) for the calculation of classical Fourier coefficients from their corresponding quantum matrix elements. The QSM allows for the exact analytical calculation of classical quantities that so far have only been calculated numerically. We also proved the QSM for general motion is a spherically symmetric potential, using the WKB method.

To demonstrate the power of the QSM, we applied it to the case of a classical electron moving in a $1/r$ potential. By taking the classical limit of hydrogen-like atom matrix elements, we were able to analytically compute to all multipole orders: (a) time dependent Keplerian motion, (b) the EM field radiated by a classical electron in a (quasi)-Keplerian orbit, and (c) an adiabatic EM inspiral and its associated waveform. Lastly, we uncovered a direct connection between the quantum expressions for spontaneous emission, and the orbit-averaged expressions for the classical energy and angular momentum loss \eqref{eq:dE2t} and \eqref{SE1main}.

As the classical limit of quantum bound states, the QSM is non-perturbative in the coupling constant, and so it is naturally suitable for self-force/PA calculations. This is in contrast to other quantum-to-classical approaches which start from scattering amplitudes, and are thus inherently perturbative in the coupling constant. The perturbativity of these quantum-to-classical approaches makes them well suited for PM \cite{Bjerrum-Bohr:2014zsa,Bjerrum-Bohr:2014zsa,Arkani-Hamed:2017jhn,Guevara:2017csg,Cachazo:2017jef,Bjerrum-Bohr:2018xdl,Cheung:2018wkq,Arkani-Hamed:2019ymq,Cristofoli:2019neg,Chung:2019duq,Bern:2019crd,Bern:2019nnu,Bjerrum-Bohr:2019kec,Cheung:2020gyp,Cheung2020,Bern:2020buy,Bern:2020gjj,Bjerrum-Bohr:2021din,Bjerrum-Bohr:2021vuf,Herrmann2021,Herrmann2021a,Chen2022,Bern:2021dqo,Bern2022a,Bern2022,Bern2023a,Bern2023,Cristofoli:2019neg,Bini2018a,Damour:2019lcq,Damour:2020tta,Bini2022,Damour2023} and PN \cite{Blanchet:2013haa,Foffa:2012rn,Foffa:2019hrb} calculations, but not for the PA approach \cite{Barack:2018yvs,Pound2021}, which is perturbative in the mass ratio, but not in the coupling. It would be very interesting to compare between our approach and the perturbative quantum-to-classical approach, in the spirit of \cite{Barack2023}. For example, the results of this work could be compared to the perturbative EM calculations in \cite{Manohar2022,Bern2023}, and the future gravitational application of the QSM should be compared to a resummation of self-force EFT \cite{Cheung2023,Kosmopoulos2023}.

In addition to our explicit application of the QSM to analytically compute EM self-force, we also proved the QSM for a body moving in a background Schwarzschild metric. We then sketched the future application of the QSM to calculate the retarded scalar field radiated by a body with scalar charge in Schwarzschild -- a standard toy model that captures many of the essential elements of GSF. We showed that the calculation in this case comes down to a triple integral over Heun functions, which, using the MST method \cite{MST}, can be reduced to a finite sum over hydrogenic integrals at any finite PM order. In the last part of this paper we considered the Schwarzschild case for simplicity, but it is straightforward to prove the QSM for the case of Kerr BH as well, due to its integrability.

Finally, beyond its computational applications, the QSM provides a new window into the interplay between quantum and classical systems in general. It demonstrates, for the first time, the way the correspondence principle applies quantitatively, exposing the quantum roots of both the conservative and dissipative behavior of quasi-periodic systems. As such, it provides the missing link between adiabatic invariants and quantum mechanics, a century after their concurrent foundation.

\section{Acknowledgements}
OT and MK thank Clifford Cheung and Chia-Hsien Shen for illuminating discussions. We are very grateful to the referee of our paper, who's constructive comments help reshape this paper to be significantly more pedagogical. The work of MK is supported by an ERC STG grant (‘Light-Dark’, grant No. 101040019).
This project has
received funding from the European Research Council
(ERC) under the European Union’s Horizon Europe research and innovation programme (grant agreement No. 101040019). Views and opinions expressed are however those of the author(s) only and do not necessarily reflect those of the European Union. The European Union cannot be held responsible for them. The work of MK is additionally funded in part by the US-Israeli
BSF grant 2016153. OT is additionally supported by
NSF-BSF Physics grant 2022713. 
\appendix
\section{Relation of the QSM to Other Quantum-to-Classical Methods} \label{app:quantumtoclassicalladscape}
\begin{figure}[ht]
\begin{center}
\includegraphics[width=0.75\linewidth]{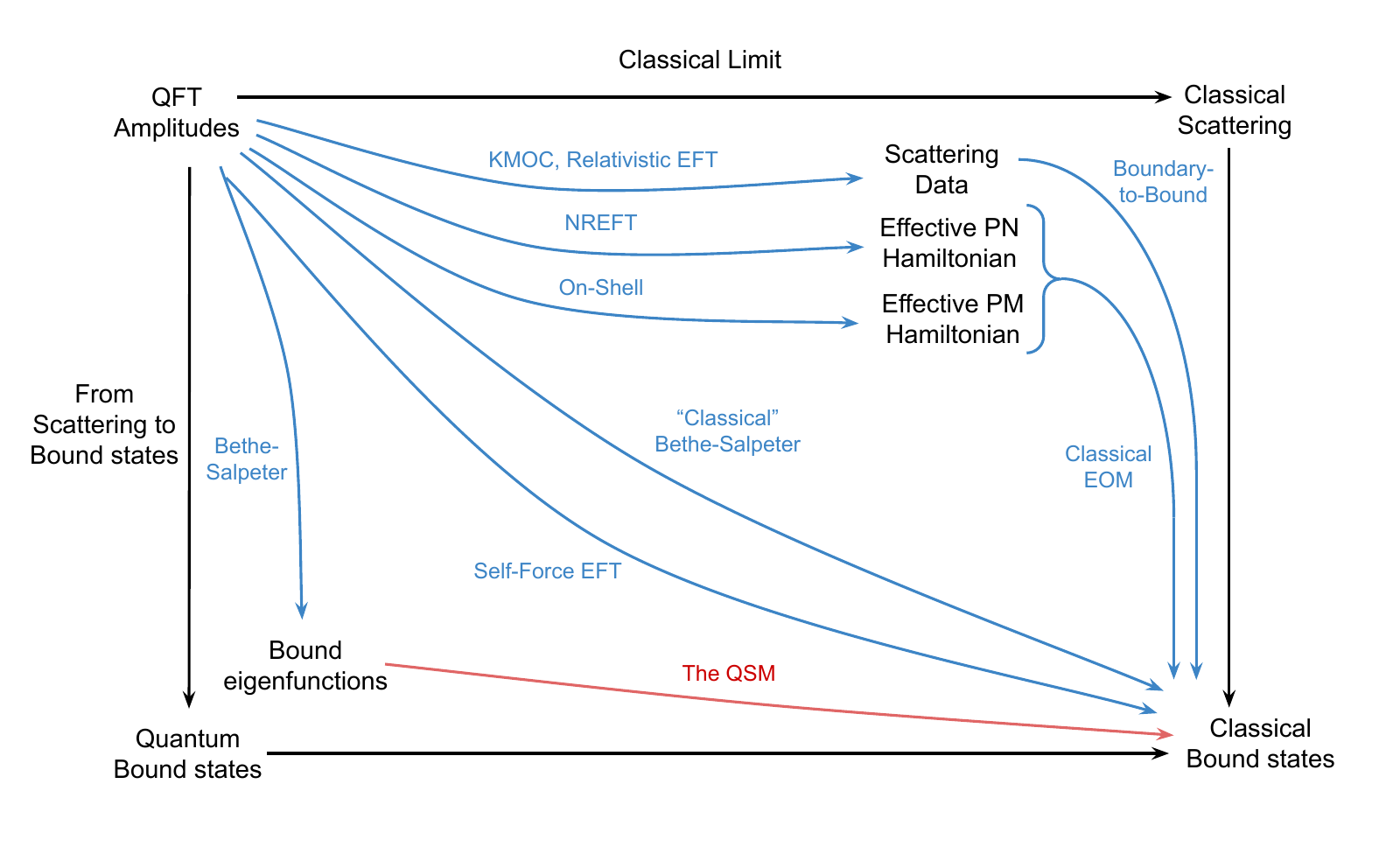}
\caption{The landscape of quantum-to-classical methods. The methods indicated in the plot are KMOC \cite{Kosower:2018adc}, worldline 
\cite{Kalin:2020fhe,Mogull:2020sak,Kalin:2020lmz,Kalin:2020mvi,Dlapa:2021npj,Jakobsen:2021smu,Dlapa2022,Jakobsen2023,Dlapa2023a,Kaelin2023,Dlapa2024,Driesse2024}, NREFT \cite{Goldberger:2004jt,Foffa:2011ub,Porto:2016pyg,Levi:2018nxp,Foffa2017,Foffa:2019yfl,Foffa2019,Foffa2021,Goldberger2022}, On-Shell methods \cite{Bjerrum-Bohr:2014zsa,Bjerrum-Bohr:2014zsa,Arkani-Hamed:2017jhn,Guevara:2017csg,Cachazo:2017jef,Bjerrum-Bohr:2018xdl,Cheung:2018wkq,Arkani-Hamed:2019ymq,Cristofoli:2019neg,Chung:2019duq,Bern:2019crd,Bern:2019nnu,Bjerrum-Bohr:2019kec,Cheung:2020gyp,Cheung2020,Bern:2020buy,Bern:2020gjj,Bjerrum-Bohr:2021din,Bjerrum-Bohr:2021vuf,Herrmann2021,Herrmann2021a,Chen2022,Bern:2021dqo,Bern2022a,Bern2022,Bern2023a,Bern2023}, Boundary-to-Bound \cite{Kalin:2019rwq,Kalin:2019inp,Cho2022a,Gonzo2023}, self-force EFT \cite{Galley2008,Cheung2023,Kosmopoulos2023}, Classical Bethe-Salpeter \cite{Adamo2023}, original Bethe-Salpeter \cite{Salpeter1951}, and the QSM (this work).}\label{fig:landscape}
\end{center}
\end{figure}

The place of the QSM within the landscape of quantum-to-classical methods is illustrated in Fig.~\ref{fig:landscape}.
The grand majority of the current quantum-to-classical methods are based on taking the classical limit of perturbative Quantum Field Theory (QFT) scattering amplitudes, for example in the KMOC method \cite{Kosower:2018adc}, and its recent application to conservative scattering at 5PM \cite{Driesse2024}. In worldline EFT
\cite{Kalin:2020fhe,Mogull:2020sak,Kalin:2020lmz,Kalin:2020mvi,Dlapa:2021npj,Jakobsen:2021smu,Dlapa2022,Jakobsen2023,Dlapa2023a,Kaelin2023,Dlapa2024,Driesse2024}, the classical scattering angle is then translated to bound state data via the Boundary-to-Bound map \cite{Kalin:2019rwq,Kalin:2019inp,Cho2022a,Gonzo2023}. In contrast, the non-relativistic EFT (NREFT) \cite{Goldberger:2004jt,Foffa:2011ub,Porto:2016pyg,Levi:2018nxp,Foffa2017,Foffa:2019yfl,Foffa2019,Foffa2021,Goldberger2022} and On-Shell \cite{Bjerrum-Bohr:2014zsa,Bjerrum-Bohr:2014zsa,Arkani-Hamed:2017jhn,Guevara:2017csg,Cachazo:2017jef,Bjerrum-Bohr:2018xdl,Cheung:2018wkq,Arkani-Hamed:2019ymq,Cristofoli:2019neg,Chung:2019duq,Bern:2019crd,Bern:2019nnu,Bjerrum-Bohr:2019kec,Cheung:2020gyp,Cheung2020,Bern:2020buy,Bern:2020gjj,Bjerrum-Bohr:2021din,Herrmann2021,Herrmann2021a,Chen2022,Bern:2021dqo,Bern2022a,Bern2022,Bern2023a,Bern2023} approaches utilize the classical limit of QFT scattering amplitudes to perform EFT matching to an effective two-body Hamiltonian, which is subsequently used either to calculate inspirals, or as input to an effective one-body Hamiltonian \cite{Albanesi2023,Meent2023,Albertini2022,Albertini2022a,Nagar2022,Albanesi:2021rby,Damgaard:2021rnk,Bini2016,Damour:2016abl,Damour2016a,Damour:2015isa,Damour:2009sm,Damour:2009wj,Buonanno:1998gg}. The Classical Bethe-Salpeter \cite{Adamo2023} approach, on the other hand, aims to go directly from QFT amplitudes to bound state data, via an all-order resummation in the soft limit. Lastly, there has been growing interest in the self-force EFT approach - the EFT for the calculation of two-body gravitational interactions in the extreme mass ratio limit. Within this framework, an effective action for self-forced motion is constructed using a curved-background EFT (e.g., \cite{Galley2008,AdamoAndrea2023}) or by the resummation of PM Feynman diagrams using a flat background EFT \cite{Cheung2023,Kosmopoulos2023}.
Compared to the existing literature, our approach traces an original path; in our work, the classical limit is taken at the level of the quantum bound states, and so is always non-perturbative in the coupling. This makes the QSM particularly suited for the quantum-to-classical map in the context of self-force/PA perturbation theory. The all-order, probe limit quantum-to-classical map was previously explored by one of the current authors in \cite{Kol2022}, in the context of scattering, rather than bound-state motion. The current work extends and formalizes the ideas of \cite{Kol2022}, and applies them to bound states and radiation reaction.

\section{A Useful Classical Identity}\label{app:WKB}
In the WKB proof of the QSM we encounter the expression
\begin{eqnarray}\label{eq:Ader0}
\exp\left\{i\Delta j_r\frac{\partial S^r_{{J_r,L}}(\alpha^r)}{\partial E_{{J_r,L}}} \Upsilon^r_{{J_r,L}}+i\Delta l\,\left[\Upsilon^\varphi\frac{\partial S^r_{{J_r,L}}(\alpha^r)}{\partial E_{{J_r,L}}}-\frac{\partial S^r_{{J_r,L}}(\alpha^r)}{\partial L}\right]\right\}\,.
\end{eqnarray}
We are looking to convert it to the form of an inverse Fourier transform over $\alpha^\varphi$ -- this will help us relate the classical limit of our quantum matrix element to the definition of the Fourier coefficients in $\alpha^r$ and $\alpha^\varphi$.

We begin our discussion by defining the doubly periodic variable
\begin{eqnarray}\label{eq:Adef3}
A_{\Delta j_r,\Delta l}[\alpha^r,\alpha^\varphi]\equiv \exp\left\{i\Delta j_r \alpha^r+i\Delta l\,\left[\alpha^\varphi-\varphi(\alpha^r,\alpha^\varphi)\right]\right\}\,,
\end{eqnarray}
for some integers $\Delta j_r,\,\Delta l$. This variable naturally arises in the WKB proof of the QSM, as we shall see momentarily. We can also consider the time dependent version $A_{\Delta j_r,\Delta l}(t)\equiv A_{\Delta j_r,\Delta l}[\alpha^r(t),\alpha^\varphi(t)]$, remembering that
\begin{eqnarray}\label{eq:Adef1}
&&\alpha^r=\Upsilon^r_{{J_r,L}} t~~,~~\alpha^\varphi=\Upsilon^\varphi_{{J_r,L}} t~~,~~\dot{\varphi}=\frac{L}{{\mu}^2r}\,,
\end{eqnarray}
where the last equation is the EOM \eqref{eq:EOMrvarphi}. For future reference,  we take the time derivative of $A_{\Delta j_r,\Delta l}(t)$, we have
\begin{eqnarray}\label{eq:Adif}
\frac{dA_{\Delta j_r,\Delta l}(t)}{dt} =\left\{{i}\Delta j_r \Upsilon^r_{{J_r,L}}+i\Delta l\,\left[\Upsilon^\varphi_{{J_r,L}}-\frac{L}{{\mu}r^2(t)}\right]\right\}\,A_{\Delta j_r,\Delta l}(t)\,.
\end{eqnarray}
Now for our main derivation. The main variable we are concerned with is the $\alpha^\varphi$ \textit{averaged} value of $A_{\Delta j_r,\Delta l}[\alpha^r,\alpha^\varphi]$, namely
\begin{eqnarray}\label{eq:Adef2}
\overline{A}_{\Delta j_r,\Delta l}[\alpha^r]\equiv\int_0^{2\pi}\frac{d\alpha^\varphi}{2\pi}\,A_{\Delta j_r,\Delta l}[\alpha^r,\alpha^\varphi]\,.
\end{eqnarray}
We will now obtain an explicit expression for $\overline{A}_{\Delta j_r,\Delta l}[\alpha^r]$ from the EOM \eqref{eq:EOMrvarphi}.
Our first step is to differentiate $\overline{A}_{\Delta j_r,\Delta l}[\alpha^r]$ with respect to $\alpha^r$. We get
\begin{eqnarray}\label{eq:Ader}
\Upsilon^r_{{J_r,L}}\frac{d \overline{A}_{\Delta j_r,\Delta l}[\alpha^r]}{d\alpha^r}=\int_0^{2\pi}\frac{d\alpha^\varphi}{2\pi}\,\frac{\partial A_{\Delta j_r,\Delta l}[\alpha^r,\alpha^\varphi]}{\partial \alpha^r}\Upsilon^r_{{J_r,L}}\,,
\end{eqnarray}
Note, however, the identity
\begin{eqnarray}\label{eq:Aderid}
0=\int_0^{2\pi}\frac{d\alpha^\varphi}{2\pi}\,\frac{\partial A_{\Delta j_r,\Delta l}[\alpha^r,\alpha^\varphi]}{\partial \alpha^\varphi}\Upsilon^\varphi_{{J_r,L}}\,,
\end{eqnarray}
which is simply a consequence of the $2\pi$-periodicity of $A_{\Delta j_r,\Delta l}[\alpha^r,\alpha^\varphi]$ in $\alpha^\varphi$. Adding \eqref{eq:Aderid} to \eqref{eq:Ader}, we get
\begin{eqnarray}\label{eq:Ader2}
&&\Upsilon^r_{{J_r,L}}\frac{d \overline{A}_{\Delta j_r,\Delta l}[\alpha^r]}{d\alpha^r}=\int_0^{2\pi}\frac{d\alpha^\varphi}{2\pi}\,\left[\frac{\partial A_{\Delta j_r,\Delta l}[\alpha^r,\alpha^\varphi]}{\partial \alpha^r}\Upsilon^r_{{J_r,L}}+\frac{\partial A_{\Delta j_r,\Delta l}[\alpha^r,\alpha^\varphi]}{\partial \alpha^\varphi}\Upsilon^\varphi_{{J_r,L}}\right]=\nonumber\\[5pt]
&&\int_0^{2\pi}\frac{d\alpha^\varphi}{2\pi}\,\left[\frac{d A_{\Delta j_r,\Delta l}[\alpha^r(t),\alpha^\varphi(t)]}{dt}|_{t=t(\alpha^r,\alpha^\varphi)}\right]\,.
\end{eqnarray}
Using \eqref{eq:Adif}, we now have
\begin{eqnarray}\label{eq:Ader3}
&&\Upsilon^r_{{J_r,L}}\frac{d \overline{A}_{\Delta j_r,\Delta l}[\alpha^r]}{d\alpha^r}=\nonumber\\[5pt]
&&\left\{i\Delta j_r \Upsilon^r_{{J_r,L}}+i\Delta l\,\left[\Upsilon^\varphi_{{J_r,L}}-\frac{L}{{\mu}r^2(\alpha^r)}\right]\right\}\,\int_0^{2\pi}\frac{d\alpha^\varphi}{2\pi}\,\exp\left\{i\Delta j_r \alpha^r+i\Delta l\,\left[\alpha^\varphi-\varphi(\alpha^r,\alpha^\varphi)\right]\right\}=\nonumber\\[5pt]
&&\left\{i\Delta j_r \Upsilon^r_{{J_r,L}}+\Delta l\,\left[\Upsilon^\varphi_{{J_r,L}}-\frac{L}{{\mu}r^2(\alpha^r)}\right]\right\}\,\overline{A}_{\Delta j_r,\Delta l}[\alpha^r]\,.
\end{eqnarray}
We can now change variables to $r$ via the EOM
\begin{eqnarray}\label{eq:rEOM}
\frac{dr}{d\alpha^r}=\frac{1}{\Upsilon^r_{{J_r,L}}}\dot{r}=\frac{1}{\Upsilon^r_{{J_r,L}} {\mu}}\,\sqrt{U^r(r)}\,,
\end{eqnarray}
where we use the $+$ sign for $\dot{r}$, which corresponds to $0\leq \alpha^r<\pi$. Here $U^r(r)=2{\mu}[E_{{J_r,L}}-V(r)-L^2/(2{\mu}r^2)]$. Using the chain rule in \eqref{eq:Ader3} with the help of \eqref{eq:rEOM}, we finally get
\begin{eqnarray}\label{eq:Ader4}
\frac{d \overline{A}_{\Delta j_r,\Delta l}[\alpha^r(r)]}{dr}&=&\left\{i\Delta j_r \Upsilon^r_{{J_r,L}}+i\Delta l\,\left[\Upsilon^\varphi_{{J_r,L}}-\frac{L}{{\mu}r^2}\right]\right\}\,\frac{{\mu}}{\sqrt{U^r(r)}}\overline{A}_{\Delta j_r,\Delta l}[\alpha^r(r)]\,.\nonumber\\
\end{eqnarray}
We can readily integrate \eqref{eq:Ader4} and obtain the explicit expression
\begin{eqnarray}\label{eq:Ader5}
\overline{A}_{\Delta j_r,\Delta l}[\alpha^r]=\exp\left\{i\Delta j_r\frac{\partial S^r_{{J_r,L}}(\alpha^r)}{\partial E_{{J_r,L}}} \Upsilon^r_{{J_r,L}}+i\Delta l\,\left[\Upsilon^\varphi_{{J_r,L}}\frac{\partial S^r_{{J_r,L}}(\alpha^r)}{\partial E_{{J_r,L}}}-\frac{\partial S^r_{{J_r,L}}(\alpha^r)}{\partial L}\right]\right\}\,,\nonumber\\
\end{eqnarray}
where $S^r_{{J_r,L}}(\alpha^r)$ is the radial action
\begin{eqnarray}\label{eq:rac}
S^r_{{J_r,L}}(\alpha^r)=\int_{r_{min}}^{r(\alpha^r)}\,\sqrt{U^r(r')}\,dr'\,,
\end{eqnarray}
again using the $+$ sign in $S^r_{{J_r,L}}$, which corresponds to $0\leq \alpha^r<\pi$. Using the definition of $\overline{A}_{\Delta j_r,\Delta l}[\alpha^r]$, we finally have
\begin{eqnarray}\label{eq:Ader6}
&&\int_0^{2\pi}\frac{d\alpha^\varphi}{2\pi}\,\exp\left\{i\Delta j_r \alpha^r+i\Delta l\,\left[\alpha^\varphi-\varphi(\alpha^r,\alpha^\varphi)\right]\right\}=\nonumber\\[5pt]
&&\exp\left\{i\Delta j_r\frac{\partial S^r_{{J_r,L}}(\alpha^r)}{\partial E_{{J_r,L}}} \Upsilon^r_{{J_r,L}}+i\Delta l\,\left[\Upsilon^\varphi_{{J_r,L}}\frac{\partial S^r_{{J_r,L}}(\alpha^r)}{\partial E_{{J_r,L}}}-\frac{\partial S^r_{{J_r,L}}(\alpha^r)}{\partial L}\right]\right\}\,.
\end{eqnarray}
The RH side of this identity emerges naturally in the WKB proof of the QSM, when we evaluate quantum matrix elements in the WKB limit. The LHS allows us to convert the RHS to the definition of the inverse Fourier transform over $\alpha^\varphi$.

\section{WKB Proof for Momentum-Dependent Operators}\label{app:momWKB}
In our WKB proof we neglected the possibility for operators $\mathcal{O}$ to depend on the conjugate momenta $\vec{p}$. Here we close this gap for operators of the form 
\begin{eqnarray}\label{eq:defovec}
\overrightarrow{\mathcal{O}}(r,\theta,\varphi)=\mathcal{O}(r,\theta,\varphi)\,\overrightarrow{p}\,.
\end{eqnarray}
The generalization to more insertions of $\overrightarrow{p}$ is tedious but straightforward. The double Fourier expansion of $\overrightarrow{\mathcal{O}}_{J_r,L}(r,\theta,\varphi)$ is 
\begin{equation} \label{app:genQSM2app}
\overrightarrow{\mathcal{O}}_{J_r,L}(t)=\overrightarrow{\mathcal{O}}_{J_r,L}\left[\alpha^r(t),\alpha^\varphi(t)\right]=\sum_{\Delta j_r,\,\Delta l}\,\overrightarrow{\mathcal{O}}^{J_r,L}_{\Delta j_r,\Delta l}\exp\left\{-i\Delta j_r\alpha^r(t)-i\Delta l\alpha^\varphi(t)\right\}\,,
\end{equation}
where
\begin{equation} \label{eq:genQSM2ttapp}
\overrightarrow{\mathcal{O}}^{J_r,L}_{\Delta j_r,\Delta l}=\int\,\frac{d\alpha^r}{2\pi}\,\int\,\frac{d\alpha^\varphi}{2\pi}\,\overrightarrow{\mathcal{O}}_{J_r,L}\left[\alpha^r,\alpha^\varphi\right]\,\exps{i\Delta j_r\alpha^r+i\Delta l\alpha^\varphi}\,.
\end{equation}
The QSM relates the Fourier coefficients \eqref{eq:genQSM2ttapp} to the classical limit of quantum matrix elements,
\begin{eqnarray}\label{eq:MEmom}
&&\,\lim_{\hbar\rightarrow 0}\,\sum_{\Delta m}\,\left\langle j_r-\Delta j_r,l-\Delta l,l-\Delta m\right|\,-i\hbar \mathcal{O}\,\overrightarrow{\nabla}\,\left|j_r,l,l\right\rangle\,.
\end{eqnarray}
In the rest of this appendix, we will once again use the WKB method to prove that \eqref{eq:MEmom} gives exactly the Fourier coefficients \eqref{eq:genQSM2ttapp}.
As a first step, we decompose the gradient $\vec{\nabla}$ as \cite{Khersonskii1988}
\begin{eqnarray}\label{eq:nabla}
\overrightarrow{\nabla}=\hat{r}\partial_r+r^{-1}\overrightarrow{\nabla}_{\Omega}\,,
\end{eqnarray}
where $\overrightarrow{\nabla}_{\Omega}=-i\hbar^{-1}\hat{r}\times\overrightarrow{L}$ is the angular part of the gradient. 
We can decompose the {operator} $\mathcal{O}(r,\theta,\varphi)$ in spherical harmonics as usual,
\begin{eqnarray}
\mathcal{O}(r,\theta,\varphi)=\sum_{l_\gamma=0}^\infty\,\sum_{m_\gamma=-l_\gamma}^{l_\gamma}\,\mathcal{O}_{l_\gamma,m_\gamma}(r)\,Y_{l_\gamma}^{m_\gamma}\left(\theta,\varphi\right)\,.
\end{eqnarray}
Substituting this in \eqref{eq:MEmom} and using \eqref{eq:nabla}, we have
\begin{eqnarray}\label{eq:MEapp}
&&\,\lim_{\hbar\rightarrow 0}\,\sum_{\Delta m}\,\left\langle j_r-\Delta j_r,l-\Delta l,l-\Delta m\right|\,-i\hbar\mathcal{O}\overrightarrow{\nabla}\,\left|j_r,l,l\right\rangle=\nonumber\\[5pt]
&&\,\lim_{\hbar\rightarrow 0}\,\sum_{\Delta m}\,\sum_{l_\gamma,m_\gamma}\,{(-1)^{\Delta l}}\,I^{J_r,L}_{\Delta j_r,\Delta l}\left[-i\hbar\mathcal{O}_{l_\gamma,m_\gamma}(r)\partial_r\right]\,\left\langle l',m'\right| Y_{l_\gamma}^{m_\gamma}\left(\theta,\varphi\right)\hat{r}\left|l,l\right\rangle\,+\nonumber\\[5pt]
&&\,\lim_{\hbar\rightarrow 0}\,\sum_{\Delta m}\,\sum_{l_\gamma,m_\gamma}\,{(-1)^{\Delta l}}\,I^{J_r,L}_{\Delta j_r,\Delta l}\left[r^{-1}\mathcal{O}_{l_\gamma,m_\gamma}(r)\right]\,\left\langle l',m'\right|-i\hbar Y_{l_\gamma}^{m_\gamma}\left(\theta,\varphi\right)\overrightarrow{\nabla}_{\Omega}\left|l,l\right\rangle\,.
\end{eqnarray}
\subsection{Radial Derivative}
We first focus on the radial part involving $\partial_r$. Following similar steps to Section~\ref{sec:Iint}, we define
\begin{eqnarray}\label{eq:snprder}
&&iI^{J_r,L}_{\Delta j_r,\Delta l}[-i\hbar f\partial_r]\equiv\lim_{\hbar\rightarrow 0} \int_{r_{min}}^{{r_{max}}} dr\,r^2\,f(r)\,R^*_{j_r-\Delta j_r,l-\Delta l}(r)\hbar\partial_r R_{j_r,l}(r)\,.
\end{eqnarray}
Differentiating the WKB radial wavefunction \eqref{eq:RWKBp} with respect to $r$, we have
\begin{eqnarray}\label{eq:snprder2}
\hbar\partial_r R_{j_r,l}(r)&=&\sqrt{\frac{\mu}{T^r_{j_r,l}}}\,\frac{2}{r\,[U^r_{j_r,l}(r)]^{1/4}}\,\partial_rS^r_{j_r,l}\,\cos\left(\frac{1}{\hbar}S^r_{j_r,l}(r)+\frac{\pi}{4}\right)+\mathcal{O}(\hbar)\nonumber\\[5pt]
&=&\sqrt{\frac{\mu}{T^r_{j_r,l}}}\,\frac{2}{r\,[U^r_{j_r,l}(r)]^{1/4}}\,p_r(r)\,\cos\left(\frac{1}{\hbar}S^r_{j_r,l}(r)+\frac{\pi}{4}\right)+\mathcal{O}(\hbar)\,,
\end{eqnarray}
where the connection to $p_r$ is via the EOM \eqref{eq:EOMrvarphi}. We then have,
\begin{eqnarray}\label{eq:snapp}
&&iI^{J_r,L}_{\Delta j_r,\Delta l}[-i\hbar f\partial_r]=\lim_{\hbar\rightarrow 0} \frac{4\mu}{T^r_{j_r,l}}\,\int_{r_{min}}^{{r_{max}}} dr\,\frac{f(r)\,p_r(r)}{\sqrt{U^r_{j_r,l}(r)}}\,\cos\left(\frac{1}{\hbar}S^r_{j_r,l}(r)+\frac{\pi}{4}\right)\sin\left(\frac{1}{\hbar}S^r_{j_r-\Delta j_r,l-\Delta l}(r)+\frac{\pi}{4}\right)\,.\nonumber\\
\end{eqnarray}
Changing integration variables from $r$ to $\alpha_r$ using \eqref{eq:EOMral} with a $+$ sign, we have
\begin{eqnarray}\label{eq:sn70app}
&&iI^{J_r,L}_{\Delta j_r,\Delta l}[-i\hbar f\partial_r]=\lim_{\hbar\rightarrow 0} 4\int_{0}^{\pi} \frac{d\alpha^r}{2\pi}\,f(\alpha^r)\,p_r(\alpha^r)\,\cos\left(\frac{1}{\hbar}S^r_{j_r,l}(\alpha^r)+\frac{\pi}{4}\right)\sin\left(\frac{1}{\hbar}S^r_{j_r-\Delta j_r,l-\Delta l}(\alpha^r)+\frac{\pi}{4}\right)\,.\nonumber\\
\end{eqnarray}
Using a basic trigonometric identity we then have
\begin{eqnarray}\label{eq:sn71app}
&&iI^{J_r,L}_{\Delta j_r,\Delta l}[-i\hbar f\partial_r]=\nonumber\\[5pt]
&&2\lim_{\hbar\rightarrow 0}\int_{0}^{\pi} \frac{d\alpha^r}{2\pi}\,f(\alpha^r)\,p_r(\alpha^r)\,\left[-\sin\left(\frac{S^r_{j_r,l}(\alpha^r)-S^r_{j_r-\Delta j_r,l-\Delta l}(\alpha^r)}{\hbar}\right)+{\cos}\left(\frac{S^r_{j_r,l}(\alpha^r)+S^r_{j_r-\Delta j_r,l}(\alpha^r)}{\hbar}\right)\right]\,.\nonumber\\
\end{eqnarray}
In the second term $S^r_{j_r,l}(r)+S^r_{j_r-\Delta j_r,l-\Delta l}(r)=\mathcal{O}(\hbar^0)$ and so this term gives a vanishing oscillatory contribution, so that
\begin{eqnarray}\label{eq:sn72app}
&&iI^{J_r,L}_{\Delta j_r,\Delta l}[-i\hbar f\partial_r]=-2\lim_{\hbar\rightarrow 0} \int_{0}^{\pi} \frac{d\alpha^r}{2\pi}\,f(\alpha^r)\,p_r(\alpha^r)\,\sin\left(\frac{S^r_{j_r,l}(\alpha^r)-S^r_{j_r-\Delta j_r,l-\Delta l}(\alpha^r)}{\hbar}\right)\,.\nonumber\\
\end{eqnarray}
Since $r(\pi-\alpha^r)=r(\pi+\alpha^r)$, $p_r(\pi-\alpha^r)=-p_r(-\pi+\alpha^r)$ and $S^r_{j_r,l}(\pi-\alpha^r)=-S^r_{j_r,l}(-\pi+\alpha^r)$ for $0\leq \alpha^r\leq\pi$, we can double the integration region for $\alpha^r$ while multiplying the result by a $1/2$, obtaining
\begin{eqnarray}\label{eq:sn72apprr}
&&iI^{J_r,L}_{\Delta j_r,\Delta l}[-i\hbar f\partial_r]=-\lim_{\hbar\rightarrow 0} \int_{0}^{2\pi} \frac{d\alpha^r}{2\pi}\,f(\alpha^r)\,p_r(\alpha^r)\,\sin\left(\frac{S^r_{j_r,l}(\alpha^r)-S^r_{j_r-\Delta j_r,l-\Delta l}(\alpha^r)}{\hbar}\right)\,.~~~~~~
\end{eqnarray}
Finally, using \eqref{eq:sn2} and \eqref{eq:Ader6}, we now get
\begin{eqnarray}\label{eq:sn731app}
&&iI^{J_rL}_{\Delta j_r,\Delta l}[-i\hbar f\partial_r]=-\int_{0}^{2\pi} \frac{d\alpha^r}{2\pi}\,\int_{0}^{2\pi} \frac{d\alpha^\varphi}{2\pi}\,f(\alpha^r)\,p_r(\alpha^r)\,\sin\left\{\Delta j_r \alpha^r+\Delta l\,\left[\alpha^\varphi-\varphi(\alpha^r,\alpha^\varphi)\right]\right\}\,.\nonumber\\
\end{eqnarray}
Noting also that $r(2\pi-\alpha_r)=r(\alpha_r)$ while $\varphi(2\pi-\alpha_r,2\pi-\alpha_\varphi)={2\pi}-\varphi(\alpha_r,\alpha_\varphi)$ and $p_r(2\pi-\alpha_r)=-p_r(\alpha_r)$. This allows us to write \eqref{eq:sn731app} as 
\begin{eqnarray}\label{eq:sn732app}
&&I^{J_r,L}_{\Delta j_r,\Delta l}[-i\hbar f\partial_r]=\int_{0}^{2\pi} \frac{d\alpha^r}{2\pi}\,\int_{0}^{2\pi} \frac{d\alpha^\varphi}{2\pi}\,f(\alpha^r)\,p_r(\alpha^r)\,e^{-i\Delta l\varphi(\alpha^r,\alpha^\varphi)}\,\exps{i\Delta j_r \alpha^r+i\Delta l\,\alpha^\varphi}\,,\nonumber\\
\end{eqnarray}
which is the momentum-dependent analog of \eqref{eq:sn732}.
\subsection{Angular Part}
Moving on to the angular part, the relevant angular matrix elements for $Y_{l_\gamma}^{m_\gamma}\hat{r}$ and $Y_{l_\gamma}^{m_\gamma}\overrightarrow{\nabla}_{\Omega}$ are given in Appendix~\ref{app:angq}, and their classical limits are presented in Appendix~\ref{app:angc}. Here we present them in a more compact way as
\begin{equation}\label{eq:ang2main}
\begin{gathered} 
\lim_{\hbar\rightarrow 0}\left\langle l-\Delta l,l-\Delta m\right|Y_{l_\gamma}^{m_\gamma}\left(\hat{r}\right)\hat{r}\left|l,l\right\rangle =i\delta_{\Delta l,\Delta m}f^{\Delta l,m_\gamma}_-\,(-1)^{m_{\gamma}}\,Y_{l_\gamma}^{m_{\gamma}}\left(\pi/2,0\right)\,,
\end{gathered} 
\end{equation} 
and
\begin{equation}\label{eq:ang3main} 
\begin{gathered} 
\lim_{\hbar\rightarrow 0}\,\left\langle l-\Delta l,l-\Delta m\right|-i\hbar Y_{l_\gamma}^{m_\gamma}\left(\hat{r}\right)\vec{\nabla}_{\Omega}\left|l,l\right\rangle =L\delta_{\Delta l,\Delta m}f^{\Delta l,m_\gamma}_+\,(-1)^{m_{\gamma}}\,Y_{l_\gamma}^{m_{\gamma}}\left(\pi/2,0\right)\,.
\end{gathered} 
\end{equation} 
Here,
\begin{eqnarray}
f^{\Delta l,m_\gamma}_\pm=\frac{i}{\sqrt{2}}\left(\delta_{-\Delta l,m_\gamma-1}\vec{\varepsilon}_{-}\pm\delta_{-\Delta l,m_\gamma+1}\vec{\varepsilon}_{+}\right)\,,
\end{eqnarray}
and $\vec{\varepsilon}_{\pm}=\tfrac{1}{\sqrt{2}}\left(\mp\hat{x}+i\hat{y}\right)$.
\subsection{Putting it all Together}
Gathering the radial and angular matrix elements, we get 
\begin{eqnarray}\label{eq:ME2app}
&&\,\lim_{\hbar\rightarrow 0}\,\sum_{\Delta m}\,\left\langle j_r-\Delta j_r,l-\Delta l,l-\Delta m\right|\,-i\hbar\mathcal{O}\overrightarrow{\nabla}\,\left|j_r,l,l\right\rangle=\nonumber\\
&&\int_{0}^{2\pi} \frac{d\alpha^r}{2\pi}\,\int_{0}^{2\pi} \frac{d\alpha^\varphi}{2\pi}\,\sum_{l_{\gamma},m_\gamma}\,\left\{\mathcal{O}^{J_r,L}_{l_\gamma,m_{\gamma}}[r(\alpha^r)]\,Y_{l_\gamma}^{m_\gamma}\left(\pi/2,\varphi(\alpha^r,\alpha^\varphi)\right)\right\}\,\exps{i\Delta j_r \alpha^r+i\Delta l\,\alpha^\varphi}\times\nonumber\\[5pt]
&&\left\{p_r(\alpha^r)\,\left[\frac{1}{\sqrt{2}}\left(e^{-i\varphi(\alpha^r,\alpha^\varphi)}\vec{\varepsilon}_{-}-e^{i\varphi(\alpha^r,\alpha^\varphi)}\vec{\varepsilon}_{+}\right)\right]+\frac{L}{r(\alpha^r)}\,\left[-\frac{i}{\sqrt{2}}\left(e^{-i\varphi(\alpha^r,\alpha^\varphi)}\vec{\varepsilon}_{-}+e^{i\varphi(\alpha^r,\alpha^\varphi)}\vec{\varepsilon}_{+}\right)\right]\right\}\,.\nonumber\\
\end{eqnarray}
The factor in the last line is nothing but
\begin{eqnarray}\label{eq:ME2app2}
&&p_r(\alpha^r)\,\hat{r}(\alpha^r,\alpha^\varphi)+p_\varphi(\alpha^r)\,r(\alpha^r)\,\hat{\varphi}(\alpha^r,\alpha^\varphi)=\overrightarrow{p}(\alpha^r,\alpha^\varphi)\,,
\end{eqnarray}
where
\begin{eqnarray}\label{eq:ME2app3}
p_{\varphi}=\frac{L}{r^2}~~,~~\hat{\varphi}=(-\sin\varphi,\cos\varphi,0).
\end{eqnarray}
Remember that the motion is, without loss of generality, limited to the XY plane and so $p_\theta=0$. Finally, resumming the $l_\gamma,\,m_\gamma$ to get back $\mathcal{O}_{J_r,L}$, we have
\begin{eqnarray}
&&\,\lim_{\hbar\rightarrow 0}\,\sum_{\Delta m}\,\left\langle j_r-\Delta j_r,l-\Delta l,l-\Delta m\right|\,-i\hbar\mathcal{O}\overrightarrow{\nabla}\,\left|j_r,l,l\right\rangle=\nonumber\\[5pt]
&&\int_{0}^{2\pi} \frac{d\alpha^r}{2\pi}\,\int_{0}^{2\pi} \frac{d\alpha^\varphi}{2\pi}\mathcal{O}_{J_r,L}[r(\alpha^r),\pi/2,\varphi(\alpha^r,\alpha^\varphi)]\,\overrightarrow{p}(\alpha^r,\alpha^\varphi)\,\exps{i\Delta j_r \alpha^r+i\Delta l\,\alpha^\varphi}
=\overrightarrow{\mathcal{O}}^{J_r,L}_{\Delta j_r,\Delta l}\,.\nonumber\\
\end{eqnarray}
The last line is simply the definition of the classical Fourier coefficients in \eqref{eq:genQSM2ttapp}.
This completes the proof of the QSM master equation \eqref{eq:master} for the momentum-dependent operator $\overrightarrow{\mathcal{O}}_{J_r,L}(r,\theta,\varphi)$.

\section{Identities Related to Spontaneous Emission} \label{0PA:SE}
In this appendix we prove \eqref{eq:dE2tGt}, and show that \eqref{eq:dE2tG} leads to \eqref{SE3main}, or in other words that $\Gamma_{s.e.}$ is indeed the rate for spontaneous emission.
We start by noting two identities relating hydrogenic quantum matrix elements. The first one is
\begin{eqnarray}\label{eq:id1}
&&\sum_{l_{\gamma}=0}^\infty\sum_{m_{\gamma}=-l_{\gamma}}^{l_\gamma}\,\vec{\mathcal{M}}^{N,L*}_{l_\gamma,m_\gamma,{\mathbf{\Delta}}}\cdot\vec{\mathcal{M}}^{N,L}_{l_\gamma,m_\gamma,{\mathbf{\Delta}}}=\sum_{l_{\gamma}=0}^\infty\sum_{m_{\gamma}=-l_{\gamma}}^{l_\gamma}\,\left|\left\langle n',l',m'\right| j_{l_{\gamma}}(\omega_{\Delta n} r)\,Y_{l_\gamma}^{m_\gamma}\,{\vec{p}}\left|n,l,m\right\rangle\right|^2\nonumber\\[5pt]
&&=\frac{1}{(4\pi)^2\,\omega_{\Delta n}^{{2}}}\int\,d^3k\,\delta\left(k-\omega_{\Delta n}\right)\,\left|\left\langle n',l',m'\right| e^{-i\vec{k}\cdot\vec{{x}}}\,{\vec{p}}\left|n,l,m\right\rangle\right|^2\,.
\end{eqnarray}
The equality here is a consequence of the partial wave decomposition of the $e^{-i\vec{k}\cdot\vec{{x}}}$ plane wave, followed by integration over $\vec{k}$. The second one is
\begin{eqnarray}\label{eq:id2}
&&\sum_{l_{\gamma}=0}^\infty\sum_{m_{\gamma}=-l_{\gamma}}^{l_\gamma}\,\mathcal{O}^{0,N,L*}_{l_\gamma,m_\gamma,{\mathbf{\Delta}}} \mathcal{M}^{0,N,L}_{l_\gamma,m_\gamma,{\mathbf{\Delta}}}\nonumber\\[5pt]
&&=\frac{i\mu}{\omega_{\Delta n}}\sum_{l_{\gamma}=0}^\infty\sum_{m_{\gamma}=-l_{\gamma}}^{l_\gamma}\,\left\langle  n',l',m'\right|\vec{\nabla}\left[j_{l_{\gamma}}(\omega_{\Delta n} r)\,Y_{l_\gamma}^{m_\gamma}\right]\cdot\vec{p}\left|n,l,m\right\rangle^*\,\left\langle n',l',m'\right|j_{l_{\gamma}}(\omega_{\Delta n} r)\,Y_{l_\gamma}^{m_\gamma}\left|n,l,m\right\rangle\nonumber\\[5pt]
&&=\frac{i\mu}{(4\pi)^2\omega_{\Delta n}^{{3}}}\int\,d^3k\,\delta\left(k-|\omega_{\Delta n}|\right)\,{\left\langle n',l',m'\right| \vec{\nabla} [e^{-i\vec{k}\cdot\vec{{x}}}]\cdot\vec{p}\left|n,l,m\right\rangle}^*\left\langle n',l',m'\right| e^{-i\vec{k}\cdot\vec{{x}}}\left|n,l,m\right\rangle\nonumber\\[5pt]
&&=\frac{1}{(4\pi)^2\omega_{\Delta n}^{{2}}}\frac{\mu}{\mu+(E_n-E_{n'})/2}\int\,d^3k\,\delta\left(k-\omega_{\Delta n}\right)\,\left|\left\langle n',l',m'\right| e^{-i\vec{k}\cdot\vec{{x}}}\hat{k}\cdot \vec{p
}\left|n,l,m\right\rangle\right|^2\,.
\end{eqnarray}
In passing through~\eqref{eq:id2}, we made use of the identity
\begin{equation}\label{QMiden}
\left\langle n',l',m'\right| e^{-i\vec{k}\cdot\vec{{x}}}\hat{k}\cdot \vec{p
}\left|n,l,m\right\rangle=(\mu+(E_n-E_{n'})/2) \left\langle n',l',m'\right| e^{-i\vec{k}\cdot\vec{{x}}}\left|n,l,m\right\rangle\,.
\end{equation}
A bi-product of \eqref{eq:id2} and \eqref{QMiden} is that
\begin{eqnarray}\label{eq:id3}
&&\sum_{l_{\gamma}=0}^\infty\sum_{m_{\gamma}=-l_{\gamma}}^{l_\gamma}\,\mathcal{O}^{0,N,L*}_{l_\gamma,m_\gamma,\Delta n} \mathcal{M}^{0,N,L}_{l_\gamma,m_\gamma,\Delta n}=\sum_{l_{\gamma}=0}^\infty\sum_{m_{\gamma}=-l_{\gamma}}^{l_\gamma}\,\mathcal{M}^{0,N,L*}_{l_\gamma,m_\gamma,\Delta n} \mathcal{M}^{0,N,L}_{l_\gamma,m_\gamma,\Delta n}+\mathcal{O}(\hbar)\,,
\end{eqnarray}
which proves \eqref{eq:dE2tGt}.
Using the identities \eqref{eq:id1} and \eqref{eq:id2} in \eqref{eq:dE2tG}, we get
\begin{eqnarray}\label{eq:dE2tGtap}
\Gamma_{s.e.}&=&\frac{{2}q^2}{(4\pi)^2\hbar\mu^2 {\omega_{\Delta n}}}\int\,d^3k\,\delta\left(k-\omega_{\Delta n}\right)\times\nonumber\\[5pt]
&&\left\{\left|\left\langle n',l',m'\right| e^{-i\vec{k}\cdot\vec{{x}}}\,{\vec{p}}\left|n,l,m\right\rangle\right|^2-\left|\left\langle n',l',m'\right| e^{-i\vec{k}\cdot\vec{{x}}}\hat{k}\cdot \vec{p
}\left|n,l,m\right\rangle\right|^2\right\}
\,.
\end{eqnarray}
In other words,
\begin{eqnarray} \label{SE3mainapp}
\Gamma_{s.e.}=\frac{q^{2}\left(E_{n}-E_{n'}\right)}{8\pi^2\hbar^2\mu^{2}}\int d^{2}\Omega_{k}\,\left.\left\{ \sum_{\sigma}\left|\left\langle n',l',m'\right|e^{-i\vec{k}\cdot\vec{x}}\left(\vec{\varepsilon}^{~\sigma}_{\vec{k}}\cdot\vec{p}\right)\left|n,l,m\right\rangle \right|^{2}\right\} \right|_{k=\left(E_{n}-E_{n'}\right)/\hbar}+\mathcal{O}(\hbar^0)\,.\nonumber\\
\end{eqnarray}
where $\vec{\varepsilon}^{~\sigma}_{\vec{k}}$ are the usual EM transverse polarization vectors. This proves \eqref{SE3main}.
\section{The Quantum Matrix Elements of Hydrogen-Like Atoms}\label{app:quant}
In this appendix, we give the analytical expressions for the matrix elements used in this work. In particular, since the matrix elements can be separated into spherical and radial parts, we treat them separately, with the spherical elements given first, followed by the radial elements.
\subsection{Angular matrix elements}\label{app:angq}
\subsubsection{Spherical harmonic}
The angular matrix element $\left\langle l',m'\right|Y_{l_\gamma}^{m_\gamma}\left(\hat{r}\right)\left|l,l\right\rangle$ can be evaluated using the Wigner
$3j$ symbols, yielding

\begin{equation} \label{SphericalScalarElem}
\left\langle l',m'\right|Y_{l_\gamma}^{m_\gamma}\left(\hat{r}\right)\left|l,l\right\rangle=\left(-1\right)^{m'}\delta_{m',l+m_\gamma}\sqrt{\frac{\left(2l+1\right)\left(2l_\gamma+1\right)\left(2l'+1\right)}{4\pi}}
\left(\begin{array}{ccc}
l' & l & l_\gamma\\
0 & 0 & 0
\end{array}\right)\left(\begin{array}{ccc}
l' & l & l_\gamma\\
-\left(l+m_\gamma\right) & l & m_\gamma
\end{array}\right).
\end{equation}
\subsubsection{Spherical harmonic times $\hat{r}$}
Vectorial matrix elements are easily computed using a tensor operator basis, namely, an arbitrary vector $\vec{v}$ is represented as 
\begin{eqnarray}
\vec{v}=\sum_{q=-1}^1\,(\vec{v})_q\,\vec{\varepsilon}_{q}\,\,,
\end{eqnarray}
where $\vec{\varepsilon}_{0}=\hat{z},\,\vec{\varepsilon}_{\pm}=\tfrac{1}{\sqrt{2}}\left(\mp\hat{x}+i\hat{y}\right)$ and $(\vec{v})_q=\vec{v}\cdot\vec{\varepsilon}^{*}_q$. We start with the matrix element $\left\langle l',m'\right|Y_{l_\gamma}^{m_\gamma}\left(\hat{r}\right)\,\left(\hat{r}\right)_{q}\left|l,l\right\rangle $.
We can calculate this by adding first the angular momentum of $Y_{l_\gamma}^{m_\gamma}$
with that of $\left| l,l\right\rangle$ using the contraction formula,
thereby transforming the element to 
\begin{equation}
\label{SphericaltimesrElem}
\begin{gathered}
\left\langle l',m'\right|Y_{l_{\gamma}}^{m_\gamma}\left(\hat{r}\right)_{q}\left|l,l\right\rangle =\left(-1\right)^{l+m_{\gamma}}\sqrt{\frac{\left(2l_{\gamma}+1\right)\left(2l+1\right)}{4\pi}}\\
\times\sum_{c=m_{\gamma}}^{l_{\gamma}}\sqrt{2\left(l+c\right)+1}\left(\begin{array}{ccc}
l+c & l & l_{\gamma}\\
0 & 0 & 0
\end{array}\right)\left(\begin{array}{ccc}
l+c & l & l_{\gamma}\\
-\left(l+m_{\gamma}\right) & l & m_{\gamma}
\end{array}\right)\left\langle l',m'\right|\left(\hat{r}\right)_{q}\left|l+c,l+m_{\gamma}\right\rangle .
\end{gathered}
\end{equation}
We then use the relation \cite{Khersonskii1988}
\begin{eqnarray}
&&\left\langle l',m'\right|\left(\hat{r}\right)_{q}\left|l_{c},m_{c}\right\rangle =\nonumber\\[5pt]
&&\delta_{m',m_{c}+q}\left\{ \left(\sqrt{\frac{\left(l_{c}-m_{c}+1\right)\left(l_{c}+m_{c}+1\right)}{\left(2l_{c}+1\right)\left(2l_{c}+3\right)}}\delta_{q,0}+\sqrt{\frac{\left(l_{c}\pm m_{c}+1\right)\left(l_{c}\pm m_{c}+2\right)}{2\left(2l_{c}+1\right)\left(2l_{c}+3\right)}}\delta_{q,\pm1}\right)\delta_{l',l_{c}+1}\right.\nonumber\\[5pt]
&&~~~~~~~~~~+\left.\left(\sqrt{\frac{\left(l_{c}-m_{c}\right)\left(l_{c}+m_{c}\right)}{\left(2l_{c}+1\right)\left(2l_{c}-1\right)}}\delta_{q,0}-\sqrt{\frac{\left(l_{c}\mp m_{c}-1\right)\left(l_{c}\mp m_{c}\right)}{2\left(2l_{c}+1\right)\left(2l_{c}-1\right)}}\delta_{q,\pm1}\right)\delta_{l',l_{c}-1}\right\} .
\end{eqnarray}
\subsubsection{Spherical harmonic times gradient}
Finally, the matrix element $\left\langle l',m'\right|Y_{l_\gamma}^{m_\gamma}\left(\hat{r}\right)\left(\vec{\nabla}_{\Omega}\right)_{q}\left|l,l\right\rangle $
can be calculated using the expression \cite{Khersonskii1988}
\begin{equation}
\label{SphericaltimesGradientElem}
\begin{gathered} 
\left\langle l',m'\right|Y_{l_\gamma}^{m_\gamma}\left(\hat{r}\right)\left(\vec{\nabla}_{\Omega}\right)_{q}\left|l,l\right\rangle =\left(-1\right)^{m_\gamma+q}\sqrt{\frac{\left(2l'+1\right)\left(2l_\gamma+1\right)}{4\pi\left(2l+1\right)}} \\
\times\sum_{s=\pm1}\left(l+\frac{1+3s}{2}\right)\sqrt{\frac{l+\frac{1+s}{2}}{2l+1+2s}}C_{l'\,0\,l_\gamma\,0}^{l+s\,0}C_{l'\,m'\,l_\gamma\,-m_\gamma}^{l+s\,m'-m_\gamma}C_{l+s\,m'-m_\gamma\,1\,-q}^{l\,l},
\end{gathered} 
\end{equation}
where the $C$ coefficients are the Clebsch-Gordan coefficients.
\subsection{Radial matrix elements}
For hydrogen-like atoms, we can express the radial wave-function using the Kummer confluent hypergeometric function $\,_{1}F_{1}$ as follows

\begin{equation} \label{HydrogenKummer}
R_{n,l}\left(r\right)=\frac{1}{\left(2l+1\right)!}\sqrt{\frac{\left(n+l\right)!}{\left(n-l-1\right)!2n}}\left(\frac{2Z}{na_{0}}\right)^{l+3/2}\exp\left(-\frac{Zr}{na_{0}}\right)r^{l}\, \,_{1}F_{1}\left(-n+l+1;2l+2;\frac{2Zr}{na_{0}}\right).
\end{equation}
where $a_{0}=\frac{4\pi \hbar^{2}}{\mu  q^{2}}$ is the Bohr radius, and $Z$ is the atomic number. By using Gordon's integral \cite{Gordon1929,Matsumoto1991},
\begin{equation} \label{IntIdenKummer}
\int_{0}^{\infty}e^{-sr}r^{\rho-1}\,_{1}F_{1}\left(a;b;pr\right)\,_{1}F_{1}\left(c;d;qr\right)\,dr =s^{-\rho}\,\Gamma\left(\rho\right)F_{2}\left(\rho,a,c,b,d;\frac{p}{s},\frac{q}{s}\right),
\end{equation}
one can explicitly calculate matrix elements of the form $\left\langle n',l'\right|r^{j}\left|n,l\right\rangle$ and $\left\langle n',l'\right|r^{j}\partial_r\left|n,l\right\rangle$.

\subsubsection{General transition}
\begin{equation} \label{RadialMat1}
\begin{gathered} 
\left\langle n',l'\right|r^{j}\left|n,l\right\rangle =\left(\frac{\hbar^{2}}{\mu  K}\right)^{j}\sqrt{\frac{\left(n+l\right)!\left(n'+l'\right)!}{\left(n-l-1\right)!\left(n'-l'-1\right)!}}
2^{l+l'+2}\frac{\left(l+l'+j+2\right)!}{\left(2l+1\right)!\left(2l'+1\right)!}\frac{n^{l'+j+1}\left(n'\right)^{l+j+1}}{\left(n+n'\right)^{l+l'+j+3}}\\
\times F_{2}\left(l+l'+j+3,-n+l+1,-n'+l'+1,2l+2,2l'+2;\frac{2n'}{n+n'},\frac{2n}{n+n'}\right),
\end{gathered}
\end{equation} 
where $F_{2}$ is the Appell hypergeometric function of two variables, defined as:
\begin{equation}
F_{2}\left(a,b_{1},b_{2},c_{1},c_{2};x,y\right)=\sum_{m,n=0}^{\infty}\frac{\left(a\right)_{m+n}\left(b_{1}\right)_{m}\left(b_{2}\right)_{n}}{\left(c_{1}\right)_{m}\left(c_{2}\right)_{n}m!n!}x^{m}y^{n}.
\end{equation}

Next, we wish to calculate $\left\langle n',l'\right|r^{j}\partial_{r}\left|n,l\right\rangle$. Using Eq. \ref{HydrogenKummer}, it is straightforward to show that

\begin{equation}
\begin{gathered}
\partial_{r}R_{n,l}\left(r\right)=-\frac{Z}{na_{0}}R_{n,l}\left(r\right)+\frac{l}{r}R_{n,l}\left(r\right)
-\frac{2Z}{a_{0}n}\frac{n-l-1}{2l+2}\frac{1}{\left(2l+1\right)!}\sqrt{\frac{\left(n+l\right)!}{2n\left(n-l-1\right)!}}\left(\frac{2Z}{a_{0}n}\right)^{l+3/2}\\
\times\exp\left(-\frac{Zr}{a_{0}n}\right)r^{l}\,\,_{1}F_{1}\left(-n+l+2;2l+3;\frac{2Zr}{a_{0}n}\right).
\end{gathered}
\end{equation}
Hence, we have that
\begin{equation}\label{eq:partialrME}
\begin{gathered}
\hbar\left\langle n',l'\right|r^{j}\partial_{r}\left|n,l\right\rangle =-\frac{\mu K}{ N}\left\langle n',l'\right|r^{j}\left|n,l\right\rangle +L\left\langle n',l'\right|r^{j-1}\left|n,l\right\rangle
-\frac{2\mu K}{N}\frac{n-l-1}{2l+2}\left(\frac{a_{0}}{Z}\right)^{j}\\
\times\sqrt{\frac{\left(n+l\right)!\left(n'+l'\right)!}{\left(n-l-1\right)!\left(n'-l'-1\right)!}} 2^{l+l'+2}\frac{\left(l+l'+j+2\right)!}{\left(2l+1\right)!\left(2l'+1\right)!}\frac{n^{l'+j+1}\left(n'\right)^{l+j+1}}{\left(n+n'\right)^{l+l'+j+3}}\\
\times F_{2}\left(l+l'+j+3,-n+l+2,-n'+l'+1,2l+3,2l'+2;\frac{2n'}{n+n'},\frac{2n}{n+n'}\right).
\end{gathered}
\end{equation}
\subsubsection{Bessel matrix elements}
Using the recursion relations for spherical Bessel functions:
\begin{equation}\label{eq:rad2Q}  
\left\langle n',l'\right|r^{-1}j_{l_\gamma}\left(\omega_{\Delta n}\,r\right)\left|n,l\right\rangle=\frac{\omega_{\Delta n}}{2l_\gamma+1}~\left\{\,\left\langle n',l'\right|j_{l_\gamma+1}\left(\omega_{\Delta n} \,r\right)\left|n,l\right\rangle+\left\langle n',l'\right|j_{l_\gamma-1}\left(\omega_{\Delta n} \,r\right)\left|n,l\right\rangle\,\right\}\,.
\end{equation}
On the other hand, Taylor expanding the spherical Bessel gives
\begin{eqnarray} \label{eq:rad1Q}  
\left\langle n',l'\right|j_{l_\gamma}\left(\omega_{\Delta n} \,r\right)\left|n,l\right\rangle&=&2^{l_\gamma}\sum_{j=0}^{\infty}\frac{\left(-1\right)^{j}\left(j+l_\gamma\right)!}{j!\left(2j+2l_\gamma+1\right)!}\,\omega_{\Delta n}^{2j+l_\gamma} \left\langle n',l'\right|r^{2j+l_{\gamma}}\left|n,l\right\rangle\nonumber\\[5pt]
\left\langle n',l'\right|j_{l_\gamma}\left(\omega_{\Delta n} \,r\right)\partial_r\left|n,l\right\rangle&=&2^{l_\gamma}\sum_{j=0}^{\infty}\frac{\left(-1\right)^{j}\left(j+l_\gamma\right)!}{j!\left(2j+2l_\gamma+1\right)!}\,\omega_{\Delta n}^{2j+l_\gamma} \left\langle n',l'\right|r^{2j+l_{\gamma}}\partial_r\left|n,l\right\rangle\,.\nonumber \\
\end{eqnarray}
\section{Classical limit of quantum matrix elements}\label{app:clas}
In this appendix, we evaluate the classical limit of the quantum matrix elements given in section \ref{app:quant}, starting from the spherical
elements then moving on to the radial elements.

\subsection{Angular matrix elements}\label{app:angc}
\subsubsection{Spherical harmonic}
By explicit calculation, the classical limit of \eqref{SphericalScalarElem} is given by
\begin{equation}\label{eq:ang1}
\begin{gathered} 
\lim_{\hbar\rightarrow 0}\left\langle l',m'\right|Y_{l_\gamma}^{m_\gamma}\left(\hat{r}\right)\left|l,l\right\rangle =\delta_{l',m'}\delta_{-\Delta l,m_\gamma}\,f_{l_\gamma,m_\gamma}\,,
\end{gathered} 
\end{equation} 
where
\begin{eqnarray} 
f_{l_\gamma,m_\gamma}&\equiv&\,Y_{l_\gamma}^{-m_\gamma}\left(\pi/2,0\right)\,,
\end{eqnarray}
or explicitly 
\begin{eqnarray} 
f_{l_\gamma,m_\gamma}=
\frac{\cos\left[\tfrac{\pi(l_\gamma-m_\gamma)}{2}\right]}{2\pi}\sqrt{\frac{\left(2 l_\gamma+1\right) \Gamma\left(\frac{l_\gamma+m_\gamma+1}{2}\right) \Gamma\left(\frac{l_\gamma-m_\gamma+1}{2}\right)}{\Gamma\left(\frac{l_\gamma+m_\gamma}{2}+1\right) \Gamma\left(\frac{l_\gamma-m_\gamma}{2}+1\right)}}
\end{eqnarray}
\subsubsection{Spherical harmonic times $\hat{r}$}
Again by explicit calculation, the classical limit of \eqref{SphericaltimesrElem} is
\begin{equation}\label{eq:ang2}
\begin{gathered} 
\lim_{\hbar\rightarrow 0}\left\langle l',m'\right|Y_{l_\gamma}^{m_\gamma}\left(\hat{r}\right)\left(\hat{r}\right)_{q}\left|l,l\right\rangle =\delta_{l',m'}\delta_{-\Delta l,m_\gamma+q}\,\frac{1}{\sqrt{2}}\left(\delta_{q,1}-\delta_{q,-1}\right)\,f_{l_\gamma,m_\gamma}\,.
\end{gathered} 
\end{equation} 
\subsubsection{Spherical harmonic times gradient}
Finally, the classical limit of \eqref{SphericaltimesGradientElem} is
\begin{equation}\label{eq:ang3} 
\begin{gathered} 
\lim_{\hbar\rightarrow 0}\,\hbar\left\langle l',m'\right|Y_{l_\gamma}^{m_\gamma}\left(\hat{r}\right)\left(\vec{\nabla}_{\Omega}\right)_{q}\left|l,l\right\rangle =-\frac{L}{\sqrt{2}}\delta_{l',m'}\delta_{-\Delta l,m_\gamma+q}\,\left(\delta_{q,1}+\delta_{q,-1}\right)\,f_{l_\gamma,m_\gamma}\,.
\end{gathered} 
\end{equation} 
\subsection{Radial matrix elements}
\subsubsection{General transition}
To calculate $\lim_{\hbar\to0}\left\langle n',l'\right|r^{j}\left|n,l\right\rangle$, from \eqref{RadialMat1}, we make use of the Taylor expanded Appell $F_2$, \eqref{eq:F2fin}, with the parameters
\begin{eqnarray}
&&a=\frac{2L}{\hbar}-\Delta l+j+3~~~,~~~b_1=\frac{N-L}{\hbar}-1
~~~,~~~b_2=\frac{N-L}{\hbar}+\Delta l-\Delta n-1\nonumber\\[5pt]
&&c_1=\frac{2L}{\hbar}+2~~~,~~~c_2=\frac{2L}{\hbar}-2\Delta l+2~~~,~~~\Delta=\hbar \frac{\Delta n}{2N}\,.
\end{eqnarray}
The $\hbar\rightarrow 0$ limit is then straightforward, using the limit of the Pochhammer symbol \eqref{eq:Poclim}.
In the physical cases that we encounter in this work, at least one of $j\pm \Delta l+1\ge 0$ is positive, so we specialize to such cases. Assuming $j\pm \Delta l+1\ge 0$, the limit is then
\begin{eqnarray} 
\left\langle n',l'\right|r^{j}\left|n,l\right\rangle =&&\left(\frac{p}{1-e^2}\right)^{j}\left(-\eta\right)^{-\Delta n-\Delta l}\left(\frac{\eta e}{2}\right)^{j+1}\sum_{m=0}^{\infty}\,\eta^{-2m}\times\nonumber\\[5pt]
&&\left[\sum_{s=0}^\infty\,\tfrac{1}{s!}\left(\begin{array}{c}
j-\Delta l+1+s\\
m
\end{array}\right)\left(\frac{\eta e\Delta n}{2}\right)^s\right]\,\left[\sum_{r=0}^\infty\,\tfrac{1}{r!}\left(\begin{array}{c}
j+\Delta l+1+r\\
\Delta l+\Delta n + m
\end{array}\right)\left(-\frac{\eta e\Delta n}{2}\right)^r\right]\,\nonumber\\
\end{eqnarray}
where 
\begin{eqnarray}
{\eta\equiv\frac{1-\sqrt{1-e^2}}{e}}\,,
\end{eqnarray}
and also $\Delta n=n-n'$ and $\Delta l=l-l'$. The $r$ and $s$ sums can be carried analytically, giving rise to our final expression
\begin{eqnarray} \label{ClassLimRadialLaguerre2}
\left\langle n',l'\right|r^{j}\left|n,l\right\rangle =&&\left(\frac{p}{1-e^2}\right)^{j}\left(-\eta\right)^{-\Delta n-\Delta l}\left(\frac{\eta e}{2}\right)^{j+1}\times\nonumber\\[5pt]
&&\sum_{m=0}^{\infty}L_{m+\Delta l+\Delta n }^{j+1-m-\Delta n}\left(\frac{\eta e\Delta n}{2}\right)L_{m}^{j+1-m-\Delta l}\left(-\frac{\eta e \Delta n}{2}\right)\eta^{-2m}\,.
\end{eqnarray}
Otherwise, if $(j+\Delta l+1)(j-\Delta l+1)< 0$, then the classical limit is obtained similarly using \eqref{eq:F2fin}, albeit a less compact expression is obtained.
Next, we calculate $\lim_{\hbar\to0}\left\langle n',l'\right|r^{j}\partial_r\left|n,l\right\rangle$, from \eqref{eq:partialrME}, using the same asymptotic form of the Appell $F_2$ function from \eqref{eq:F2fin}. Assuming $j \ge \rm{Max}$ $\left\{\Delta l,-\Delta l-1\right\} $, we get
\begin{eqnarray} \label{RadialDerivativeClassLim}   
&&\lim_{\hbar\to0}\hbar\left\langle n',l'\right|r^{j}\partial_{r}\left|n,l\right\rangle =\nonumber\\[5pt]
&&L\left\{\frac{2E}{K\sqrt{1-e^2}}\left\langle n',l'\right|r^{j}\left|n,l\right\rangle +\left\langle n',l'\right|r^{j-1}\left|n,l\right\rangle
 +\frac{2}{\sqrt{1-e^2}}\left(\frac{p}{1-e^2}\right)^{j-1}\left(-\eta\right)^{-\Delta n-\Delta l}\left(\frac{\eta e}{2}\right)^{j+1}\right.\nonumber\\[5pt]
 &&~~~\left.\times\sum_{m=0}^{\infty}L_{m+\Delta l+\Delta n }^{j+1-m-\Delta n}\left(\frac{\eta e\Delta n}{2}\right)L_{m}^{j-m-\Delta l}\left(-\frac{\eta e \Delta n}{2}\right)\eta^{-2m} \right\}\,.
\end{eqnarray}
Otherwise, the resulting expression is less compact.
\subsubsection{Bessel matrix elements}
Since the prefactors in \eqref{eq:rad2Q}-\eqref{eq:rad1Q} are finite in the classical limit, we can commute the $\hbar\rightarrow 0$ limit past them and get 
\begin{equation}\label{eq:rad2}  
\lim_{\hbar\rightarrow 0}\left\langle n',l'\right|r^{-1}j_{l_\gamma}\left(\omega_{\Delta n}\,r\right)\left|n,l\right\rangle=\frac{\omega_{\Delta n}}{2l_\gamma+1}~\left\{\,\lim_{\hbar\rightarrow 0}\left\langle n',l'\right|j_{l_\gamma+1}\left(\omega_{\Delta n} \,r\right)\left|n,l\right\rangle+\lim_{\hbar\rightarrow 0}\left\langle n',l'\right|j_{l_\gamma-1}\left(\omega_{\Delta n} \,r\right)\left|n,l\right\rangle\,\right\}\,,
\end{equation}
as well as
\begin{eqnarray}\label{eq:rad1}  
\lim_{\hbar\rightarrow 0}\left\langle n',l'\right|j_{l_\gamma}\left(\omega_{\Delta n} \,r\right)\left|n,l\right\rangle&=&2^{l_\gamma}\sum_{j=0}^{\infty}\frac{\left(-1\right)^{j}\left(j+l_\gamma\right)!}{j!\left(2j+2l_\gamma+1\right)!}\,\omega_{\Delta n}^{2j+l_\gamma} \lim_{\hbar\rightarrow 0}\left\langle n',l'\right|r^{2j+l_\gamma}\left|n,l\right\rangle\nonumber\\
\end{eqnarray}
\begin{eqnarray}\label{eq:rad3} 
\lim_{\hbar\rightarrow 0}\,\hbar\,\left\langle n',l'\right|j_{l_\gamma}\left(\omega_{\Delta n} \,r\right)\partial_r\left|n,l\right\rangle&=&2^{l_\gamma}\sum_{j=0}^{\infty}\frac{\left(-1\right)^{j}\left(j+l_\gamma\right)!}{j!\left(2j+2l_\gamma+1\right)!}\,\omega_{\Delta n}^{2j+l_\gamma} \lim_{\hbar\rightarrow 0}\,\hbar\,\left\langle n',l'\right|r^{2j+l_\gamma}\partial_r\left|n,l\right\rangle\,.\nonumber\\
\end{eqnarray}

\section{Auxiliary calculations}\label{app:aux}
Here we derive for completeness various asymptotic formulae  relevant for the classical limit of the radial matrix elements. These include asymptotic forms of the Pochhammer symbol, the Gauss hypergeometric function ${}_2F_1$, and the Appell function $F_2$.
\subsection{Pochhammer symbol}
As a warm-up, we calculate the classical limit of the Pochhammer Symbol,
\begin{eqnarray}\label{eq:Poclim}
\lim_{\hbar \rightarrow 0}\hbar^m(A\hbar^{-1})_m=A^m\,,
\end{eqnarray}
where $(a)_m$ is the Pochhammer symbol.
\subsection{Appell $F_2$}
We are looking to expand
\begin{equation}
F_{2}\left(a,-b_1,-b_2,c_1,c_2;1-\Delta,1+\Delta\right),
\end{equation}
as a power series in $\Delta$. For this purpose we verify by explicit calculation that
\begin{eqnarray}
&&\frac{\partial^{r+s}}{\partial x^r\partial y^s}F_{2}\left(a,-b_1,-b_2,c_1,c_2;x,y\right)=\nonumber\\[5pt]
&&\frac{(a)_{r+s}(-b_1)_r(-b_2)_s}{(c_1)_r(c_2)_s}\,F_{2}\left(a+r+s,-b_1+r,-b_2+s,c_1+r,c_2+s;x,y\right)\,.
\end{eqnarray}
Using this expression for our Taylor expansion in $\Delta$, we have
\begin{eqnarray}\label{eq:F2Del}
&&F_{2}\left(a,-b_1,-b_2,c_1,c_2;1-\Delta,1+\Delta\right)=\nonumber\\[5pt]
&&\sum_{r,s=0}^\infty\,\frac{(a)_{r+s}(-b_1)_r(-b_2)_s}{(c_1)_r(c_2)_s\Gamma(r+1)\Gamma(s+1)}\,F_{2}\left(a+r+s,-b_1+r,-b_2+s,c_1+r,c_2+s;1,1\right)(-\Delta)^r\Delta^s\,.\nonumber\\
\end{eqnarray}
To compute $F_2$ at $x=1,y=1$ we use the representation of $F_2$ as the sum of products of Gauss ${}_2F_1$:
\begin{eqnarray}
&&F_{2}\left(A,-B_1,-B_2,C_1,C_2;1,1\right)=\nonumber\\[5pt]
&&\sum_m\frac{(A)_m(-B_1)_m(-B_2)_m}{(C_1)_m(C_2)_mm!}{}_2F_1(A+m,-B_1+m,C_1+m,1){}_2F_1(A+m,-B_2+m,C_2+m,1)\,.\nonumber\\
\end{eqnarray}
In the cases of interest in this work, $A-C_1$ and $A-C_2$ can not be both negative simultaneously. Without loss of generality, assume that $A-C_1\ge0$.  After a little massaging we get
\begin{equation}
\begin{gathered} 
F_{2}\left(A,-B_{1},-B_{2},C_{1},C_{2};1,1\right)=\frac{\left(-1\right)^{B_{2}-B_{1}}B_{1}!B_{2}!}{\left(C_{1}\right)_{B_{1}}\left(C_{2}\right)_{B_{2}}\left(A-1\right)!}\\[5pt]
\times\sum_{m=0}^{A-C_{1}+B_{2}-B_{1}}\left(\begin{array}{c}
A-C_{2}\\
C_{1}-C_{2}+B_{1}-B_{2}+m
\end{array}\right)\left(\begin{array}{c}
A-C_{1}\\
m
\end{array}\right)\frac{\left(B_{1}+C_{1}+m-1\right)!}{\left(B_{1}+C_{1}-A+m\right)!}.
\end{gathered}
\end{equation} 
Using this expression in \eqref{eq:F2Del}, we finally get
\begin{eqnarray}\label{eq:F2fin}
&&F_{2}\left(a,-b_1,-b_2,c_1,c_2;1-\Delta,1+\Delta\right)=\nonumber\\[5pt]
&&
\sum_{r,s=0}^\infty\,\frac{(a)_{r+s}(-b_1)_r(-b_2)_s}{(c_1)_r(c_2)_s\Gamma(r+1)\Gamma(s+1)}\frac{\left(-1\right)^{b_{2}-b_{1}+r-s}(b_{1}-r)!(b_{2}-s)!}{\left(c_{1}+r\right)_{b_{1}-r}\left(c_{2}+s\right)_{b_{2}-s}\left(a+r+s-1\right)!}\times\nonumber\\[5pt]
&&\,\,\sum_{m=0}^{a-c_{1}+b_{2}-b_{1}+r}\left(\begin{array}{c}
a+r-c_{2}\\
c_{1}-c_{2}+b_{1}-b_{2}+m
\end{array}\right)\left(\begin{array}{c}
a-c_{1}+s\\
m
\end{array}\right)\frac{\left(b_{1}+c_{1}+m-1\right)!}{\left(b_{1}+c_{1}-a-r-s+m\right)!}(-\Delta)^r\Delta^s\,.\nonumber\\
\end{eqnarray}
This expression might seem cumbersome, but it simplifies greatly upon taking the classical limit.
\bibliographystyle{JHEP}

\providecommand{\href}[2]{#2}\begingroup\raggedright\endgroup

\end{document}